\documentclass[oldversion]{aa}

\usepackage{graphicx}
\usepackage[varg]{txfonts}
\usepackage{natbib}

\bibpunct{(}{)}{;}{a}{}{,}

\newcommand{\ngc}{NGC~}

\newcommand{\Mv}{\mbox{$M_{V}$}}
\newcommand{\Av}{\mbox{$A_{V}$}}
\newcommand{\Rv}{\mbox{$R_{V}$}}

\newcommand{\fastwind}{{\sc fastwind}}
\newcommand{\pikaia}{{\sc pikaia}}
\newcommand{\cmfgen}{{\sc cmfgen}}

\newcommand{\ha}{H$\alpha$}

\newcommand{\hg}{H$\gamma$}
\newcommand{\hd}{H$\delta$}

\newcommand{\teff}{\mbox{$T_{\rm eff}$}}
\newcommand{\logg}{\mbox{$\log{{g}}$}}
\newcommand{\loggc}{\mbox{$\log{{g}}_{\rm c}$}}
\newcommand{\mdot}{\mbox{$\dot{M}$}}
\newcommand{\yhe}{\mbox{$Y_{\rm He}$}}
\newcommand{\rstar}{\mbox{$R_{\star}$}}
\newcommand{\rsun}{\mbox{$R_{\sun}$}}
\newcommand{\lstar}{\mbox{$L_{\star}$}}
\newcommand{\lsun}{\mbox{$L_{\sun}$}}
\newcommand{\Ms}{\mbox{$M_{\rm s}$}}
\newcommand{\Mev}{\mbox{$M_{\rm ev}$}}
\newcommand{\msun}{\mbox{$M_{\sun}$}}
\newcommand{\zsun}{\mbox{$Z_{\sun}$}}
\newcommand{\kmsec}{\mbox{km\,s$^{-1}$}}
\newcommand{\cmsecsec}{\mbox{cm\,s$^{-2}$}}
\newcommand{\msunyr}{\mbox{$M_{\sun}{\rm yr}^{-1}$}}
\newcommand{\Dmom}{\mbox{$D_{\rm mom}$}}

\newcommand{\vturb}{\mbox{$v_{\rm turb}$}}

\newcommand{\vrot}{\mbox{$v_{\rm rot}$}}

\newcommand{\vsini}{\mbox{$v_{\rm r}\sin i$}}
\newcommand{\vinf}{\mbox{$v_{\infty}$}}
\newcommand{\vesc}{\mbox{$v_{\rm esc}$}}
\newcommand{\vcrit}{\mbox{$v_{\rm crit}$}}

\newcommand{\hii}{\ion{H}{ii}}
\newcommand{\hei}{\ion{He}{i}}
\newcommand{\heii}{\ion{He}{ii}}

\newcommand{\niii}{\ion{N}{iii}} 
\newcommand{\niv}{\ion{N}{iv}}
\newcommand{\nv}{\ion{N}{v}}
\newcommand{\oiii}{\ion{O}{iii}} 
\newcommand{\oiv}{\ion{O}{iv}}
\newcommand{\siiii}{\ion{Si}{iii}}

\newcommand{\heil}{\ion{He}{i}~$\lambda$}
\newcommand{\heiil}{\ion{He}{ii}~$\lambda$}

\defcitealias{mokiem05}{Paper~I}
\defcitealias{mokiem06}{Paper~II}

\begin{document}

\title{The VLT-FLAMES survey of massive stars: \\Wind properties and
evolution of hot massive stars in the LMC}

\titlerunning{Wind properties and evolution of hot massive stars in the LMC}

\author{M.\,R.~Mokiem\inst{1}
   \and A.~de~Koter\inst{1}
   \and C.\,J.~Evans\inst{2}
   \and J.~Puls\inst{3}
   \and S.\,J.~Smartt\inst{4}
   \and P.~A.~Crowther\inst{5}
   \and A.~Herrero\inst{6,7}
   \and N.~Langer\inst{8}
   \and D.\,J.~Lennon\inst{9,6}
   \and F.~Najarro\inst{10}
   \and M.\,R.~Villamariz\inst{11,6}
   \and J.\,S.~Vink\inst{12}
}


\institute{
  Astronomical Institute Anton Pannekoek, University of Amsterdam,
  Kruislaan 403, 1098~SJ Amsterdam, The Netherlands
  \and
  UK Astronomy Technology Centre, Royal Observatory, Blackford Hill,
  Edinburgh, EH9 3HJ, UK
  \and
  Universit\"ats-Sternwarte M\"unchen, Scheinerstr. 1,
  D-81679 M\"unchen, Germany
  \and
  The Department of Pure and Applied Physics,
  The Queen's University of Belfast,
  Belfast BT7 1NN, Northern Ireland, UK
  \and
  Department of Physics and
  Astronomy, University of Sheffield, Hicks Building, Hounsfield Rd,
  Shefffield, S3 7RH,  UK
  \and
  Instituto de Astrof\'{\i}sica de Canarias, E-38200, La Laguna,
  Tenerife, Spain
  \and
  Departamento de Astrof\'{\i}sica, Universidad de La Laguna,
  Avda.\ Astrof\'{\i}sico Francisco S\'anchez, s/n, E-38071
  La Laguna, Spain
  \and
  Astronomical Institute, Utrecht University, Princetonplein 5,
  3584 CC Utrecht, The Netherlands
  \and
  The Isaac Newton Group of Telescopes,
  Apartado de Correos 321, E-38700,
  Santa Cruz de La Palma, Canary Islands, Spain
  \and 
  Instituto de Estructura de la Materia, Consejo Superior de
  Investigaciones Cient\'{\i}ficas, CSIC, Serrano 121, E-28006
  Madrid, Spain
  \and
  Grantecan S.A., E-38200, La Laguna,
  Tenerife, Spain
  \and
  Astrophysics Group, Lennard-Jones Laboratories, Keele University,
  Staffordshire, ST55BG, UK
}

\date{Accepted: 24 January 2007}

\abstract{We have studied the optical spectra of
a sample of 28 O- and early B-type stars in the Large Magellanic
Cloud, 22 of which are associated with the young star forming region
N11. Our observations sample the central associations of LH9 and LH10,
and the surrounding regions. Stellar parameters are determined using
an automated fitting method (Mokiem et al.  2005), which combines the
stellar atmosphere code \fastwind\ (Puls et al. 2005) with the
genetic algorithm based optimisation routine \pikaia\ (Charbonneau
1995). We derive an age of $7.0 \pm 1.0$ and $3.0 \pm 1.0$~Myr for LH9
and LH10, respectively. The age difference and relative distance of
the associations are consistent with a sequential star formation
scenario in which stellar activity in LH9 triggered the formation of
LH10. Our sample contains four stars of spectral type O2. From helium
and hydrogen line fitting we find the hottest three of these stars to
be $\sim$$49-54$~kK (compared to $\sim$$45-46$~kK for O3
stars). Detailed determination of the helium mass fraction reveals
that the masses of helium enriched dwarfs and giants derived in our
spectroscopic analysis are systematically lower than those implied by
non-rotating evolutionary tracks. We interpret this as evidence for
efficient rotationally enhanced mixing leading to the surfacing of
primary helium and to an increase of the stellar luminosity. This
result is consistent with findings for SMC stars by
Mokiem et al. (2006). For bright giants and supergiants no such mass
discrepancy is found; these stars therefore appear to follow tracks of
modestly or non-rotating objects.  The set of programme stars was
sufficiently large to establish the mass loss
rates of OB stars in this $Z \sim 1/2 \,\zsun$ environment
sufficiently accurate to allow for a {\em quantitative} comparison
with similar objects in the Galaxy and the SMC. The mass loss
properties are found to be intermediate to massive stars in the
Galaxy and SMC.  Comparing the derived modified wind momenta \Dmom\
as a function of luminosity with predictions for LMC metallicities by
Vink et al.\ (2001) yields good agreement in the entire luminosity
range that was investigated, i.e. $5.0 < \log L/\lsun < 6.1$.
}

\keywords{Magellanic Clouds -- stars:atmospheres -- stars:
  early-type -- stars: fundamental parameters -- stars: mass loss}

\maketitle

\section{Introduction}

Massive stars play an intricate role in the evolution of galaxies.
Because of the large energies associated with their stellar winds,
ionising radiation, and life-ending supernova explosions, they dictate
galactic structuring processes such as star formation and the creation
and evolution of supperbubbles \citep[e.g.][]{oey99}.  Mounting
evidence also points to a direct link between massive stars and exotic
phenomena such as $\gamma$-ray bursts \citep[e.g.][]{hjorth03} and the
reionisation of the early universe \citep{bromm01}. Accordingly,
understanding the properties of these stars, both in terms of their
fundamental parameters as well as their evolution, is fundamental.

The initial metal composition ($Z$) of the gas out of which massive
stars form has a strong impact on their global properties and
characteristics. Many studies have shown that parameters such as the
effective temperature and ionising fluxes are strong functions of $Z$
\citep[e.g.][]{kudritzki02, mokiem04, massey05, mokiem06}, adding an
extra dimension to the conversion of morphological properties such as
spectral type to physical quantities. Theoretical and observational
arguments \citep[e.g.][]{kudritzki00, vink01} also point to a relation
between the strengths of the stellar winds of these objects and
metallicity. As wind mass loss leads to partial evaporation of the
star \citep[e.g.][]{chiosi86} -- with possible consequences for the
nature of the compact object that is left behind after the final
supernova explosion -- and to loss of angular momentum
\citep[e.g.][]{meynet00}, it also dictates to a large extent its
evolutionary path and fate \citep[e.g.][]{yoon05,
woosley06}. Quantifying the mass loss versus metallicity dependence
$\mdot(Z)$, therefore, is an important quest in astrophysics.

Due to their proximity and low metal content the Magellanic Clouds
provide us with unparallelled laboratories to test and enlarge our
knowledge of massive stars. These galaxies, therefore, have been in
the focal point of many studies analysing their massive star
content. Early studies \citep[e.g.][]{conti86, garmany87, massey89,
parker92, walborn99} predominantly relied on photometric data and
spectral type calibrations. Only relatively recent the advent of large
telescopes and the development of sophisticated stellar atmosphere
models has allowed for more detailed analyses of individual stars
\citep[e.g.][]{puls96, hillier99, crowther02, bouret03,
martins04}. Though all these studies have contributed enormously to
our understanding of massive stars, the samples analysed so far have
been rather limited in size (a few objects at a time) and have been
focused predominantly on objects in the Small Magellanic Cloud (SMC)
-- because its metal deficiency is more extreme than that of the Large
Magellanic Cloud (LMC) \cite[but see][]{massey04, massey05}.

The limited sizes of the samples that have so far been studied are at
the root cause of the perhaps somewhat disheartening conclusion that
-- in spite of all the progress that has been made -- we still cannot
provide robust and sound answers to the question: what is
the role of metal content, stellar winds and rotation in the evolution
of massive stars?
To help attack this problem,
our research group has conducted a {\em VLT-FLAMES Survey of Massive
Stars} \citep[see][]{evans05}. In this ESO Large Program the {\em
Fibre Large Array Multi-Element Spectrograph} at the {\em Very Large
Telescope} was used to obtain optical spectra of more than 50 O- and
early B-type stars in the Magellanic Clouds.

Here we present the homogeneous analysis, employing automated spectral
fitting methods, of a sample of 28 O-type and early B-type stars in
the Large Magellanic Cloud; 22 targets from the FLAMES survey and 6
from other sources. This is so far the largest sample of massive stars
studied in the LMC and it almost doubles the amount of massive objects
in this galaxy for which parameters have been derived from
quantitative spectroscopy. Specifically, we will try to establish the
mass loss rates of OB stars in this $Z \sim 1/2 \,\zsun$ environment
to a level of precision that allows for a {\em quantitative}
comparison with similar objects in the SMC and our Galaxy.
This will provide a new $Z$-point in testing the
fundamental prediction provided by radiation driven wind theory for
the mass-loss -- metallicity dependence: $\mdot(Z)\propto
Z^{0.5 - 0.7}$ \citep[e.g.][]{kudritzki87, puls00, vink01}. 

The majority of our LMC sample is associated with the spectacular star
forming region N11 \citep[][]{henize56}. It has a \ha\ luminosity only
surpassed by that of 30~Doradus \citep{kennicutt86},
ranking it as the second largest \hii\ region in the Magellanic
Clouds. N11 is host to several OB associations of apparently different
ages, the formation of which is believed to have been triggered by
stellar activity in the central OB cluster \citep{parker92, walborn92,
walborn99}. Our observations sample both the central cluster LH9 as
well as the younger cluster LH10, allowing for an investigation of a
possible sequential star formation scenario. We will use our N11 stars
to test predictions of massive star evolution, including the role of
rotation, and the star formation history.

This paper is organised as follows: in Sect.~\ref{sec:data} we
describe the LMC data set that was analysed using our automated
genetic algorithm based fitting method. A short description of this
method is given in Sect.~\ref{sec:method} and the results obtained
are presented in Sect.~\ref{sec:fun_par}, with fits and comments
on individual objects given in the appendix. In
Sect.~\ref{sec:mdisc} we investigate the discrepancy between
spectroscopically determined masses and those derived from
evolutionary tracks. The evolutionary status of N11 is discussed in
Sect.~\ref{sec:age}. Finally, Sect.~\ref{sec:sum} summarises
and lists our most important findings.

\section{Data description}
\label{sec:data}

Our OB-type star sample is mainly drawn from the targets observed in
the LMC within the context of the VLT-FLAMES survey of massive stars
\citep[see][]{evans05}. Two fields in the LMC centred on the clusters
N11 and \ngc2004 were observed in the survey. Here we analyse a subset
of the objects observed in the N11 field. This set consists of all
O-type spectra obtained, excluding those that correspond to confirmed
binaries, and five early B-type spectra of luminous giant and
supergiant stars.

To improve the sampling in luminosity and temperature, we supplemented
the FLAMES targets with six relatively bright O-type field
stars. These objects are part of the \citet[][ hereafter
Sk]{sanduleak70} and \citet[][ hereafter BI]{brunet75} catalogues, and
were observed as part of the programs 67.D-0238, 70.D-0164 and
074.D-0109 (P.I.\ Crowther) using the Ultraviolet and Visual Echelle
Spectrograph (UVES) at the VLT.

The observations of the FLAMES targets are described extensively by
\cite{evans06}, to which we refer for full details. Here we only
summarise the most important observational parameters. Basic
observational properties of the programme stars together with common
aliases are given in Tab.~\ref{tab:data}. Note that N11-031,
BI~237, BI~253 and Sk$-67$~166 were studied recently using line
blanketed stellar atmosphere models. In the appendix a comparison with
these analyses is provided. The FLAMES targets were observed with the
Giraffe spectrograph mounted at UT2. For six wavelength settings a
spectrum was acquired six times for each object with an effective
resolving power of $R \simeq 20\,000$. These multiple exposures, often
at different epochs, allowed for the detection of variable radial
velocities. As a result, a considerable number of binaries could be
detected \citep{evans06}, which we subsequently excluded from our
analysis.

\begin{table*}
  \caption{Basic parameters. Primary identification numbers for N11
  are from \cite{evans06}. Identifications starting with ``Sk'',
  ``PGMW'' and ``BI'' are from, respectively, \cite{sanduleak70},
  \cite{parker92} and \cite{brunet75}.  Photometric data for these
  objects are from \cite{evans06} and \cite{parker92}, the latter are
  flagged with an asterisk. For non-N11 objects these are from
  \cite{ardeberg72}, \cite{issersted79} and \cite{massey02}. Wind
  velocities given without brackets are from \cite{crowther02},
  \cite{massa03} and \cite{massey05}. For Sk~$-66$~18 the wind
  velocity was measured from \oiv\ (1031-1037~\AA) in its FUSE
  spectrum. Values between brackets are calculated from the escape
  velocity at the stellar surface. For N11-031 the value of
  \vinf\ is from \cite{walborn04}, though this is actually based on the
  value obtained for Sk-68\,137 by \cite{prinja98}.}
  \label{tab:data}
  \begin{center} \begin{tabular}{llllccc} \hline\\[-9pt] \hline
  \\[-7pt] Primary ID & Cross-IDs & Spectral & \multicolumn{1}{c}{$V$} & \Av & \Mv & \vinf
  \\[2pt] &  & Type & & & & [\kmsec]\\[1pt]
\hline\\[-9pt]
N11-004      & Sk~$-66$~16 & OC9.7 Ib & 12.56 & 0.74 & $-6.68$ & [2387]\\[3.5pt]
N11-008      & Sk~$-66$~15 & B0.7 Ia & 12.77 & 0.84 & $-6.57$ & [1619]\\[3.5pt]
N11-026      & ...         & O2 III(f*) & 13.51 & 0.47 & $-5.46$ & [3116]\\[3.5pt]
N11-029      & ...         & O9.7 Ib & 13.63 & 0.56 & $-5.43$ & [1576]\\[3.5pt]
N11-031      & PGMW~3061   & ON2 III(f*) & 13.68$^*$ & 0.96 & $-5.78$ & 3200\\[3.5pt]
N11-032      & PGMW~3168   & O7 II(f) & 13.68$^*$ & 0.65 & $-5.47$ & [1917]\\[3.5pt]
N11-033      & PGMW~1005   & B0 IIIn & 13.68 & 0.43 & $-5.25$ & [1536]\\[3.5pt]
N11-036      & ...         & B0.5 Ib & 13.72 & 0.40 & $-5.18$ & [1714]\\[3.5pt]
N11-038      & PGMW~3100   & O5 II(f+) & 13.81$^*$ & 0.99 & $-5.68$ & [2601]\\[3.5pt]
N11-042      & PGMW~1017   & B0 III & 13.93 & 0.22 & $-4.79$ & [2307]\\[3.5pt]
N11-045      & ...         & O9 III & 13.97 & 0.50 & $-5.03$ & [1548]\\[3.5pt]
N11-048      & PGMW~3204   & O6.5 V((f)) & 14.02$^*$ & 0.47 & $-4.95$ & [3790]\\[3.5pt]
N11-051      & ...         & O5 Vn((f)) & 14.03 & 0.19 & $-4.66$ & [2108]\\[3.5pt]
N11-058      & ...         & O5.5 V((f)) & 14.16 & 0.28 & $-4.62$ & [2472]\\[3.5pt]
N11-060      & PGMW~3058   & O3 V((f*)) & 14.24$^*$ & 0.81 & $-5.07$ & [2738]\\[3.5pt]
N11-061      & ...         & O9 V & 14.24 & 0.78 & $-5.04$ & [1898]\\[3.5pt]
N11-065      & PGMW~1027   & O6.5 V((f)) & 14.40 & 0.25 & $-4.35$ & [2319]\\[3.5pt]
N11-066      & ...         & O7 V((f)) & 14.40 & 0.25 & $-4.35$ & [2315]\\[3.5pt]
N11-068      & ...         & O7 V((f)) & 14.55 & 0.28 & $-4.23$ & [3030]\\[3.5pt]
N11-072      & ...         & B0.2 III & 14.61 & 0.09 & $-3.98$ & [2098]\\[3.5pt]
N11-087      & PGMW~3042   & O9.5 Vn & 14.76$^*$ & 0.62 & $-4.36$ & [3025]\\[3.5pt]
N11-123      & ...         & O9.5 V & 15.29 & 0.16 & $-3.37$ & [2890]\\[1pt]
\hline\\[-9pt]
BI 237       & ...         & O2 V((f*)) & 13.89 & 0.62 & $-5.23$ & 3400\\[3.5pt]
BI 253       & ...         & O2 V((f*)) & 13.76 & 0.71 & $-5.45$ & 3180\\[3.5pt]
Sk $-66$ 18  & ...         & O6 V((f)) & 13.50 & 0.37 & $-5.37$ & 2200\\[3.5pt]
Sk $-66$ 100 & ...         & O6 II(f) & 13.26 & 0.34 & $-5.58$ & 2075\\[3.5pt]
Sk $-67$ 166 & HD~269698   & O4 Iaf+ & 12.27 & 0.31 & $-6.54$ & 1750  \\[3.5pt]
Sk $-70$ 69  & ...         & O5 V & 13.95 & 0.28 & $-4.83$ & 2750\\[1pt]
\hline
  \end{tabular}
  \end{center}
\end{table*}

To allow for a sky subtraction a master sky-spectrum was created from
combining the sky fibres in the Giraffe spectrograph (typically 15),
individually scaled by their relative fibre throughput. Even though in
general the sky background is low and this approach successfully
removes the background contribution, in crowded regions such as N11
accurate subtraction of nebular features remains very difficult. As a
result of this, the line profiles of many of our programme stars still
suffer from nebular contamination. This in principle does not hamper
our analysis. For most stars the core nebular emission is
well-resolved and we simply disregard this contaminated part of the
profile in the automated line fits. \cite{mokiem06} showed by
performing tests using synthetic data that with this reduced amount of
information, the automated method can still accurately determine the
correct fit parameters. Also tests assessing the impact of possible
residual nebular contamination in the line wings or over-subtraction
of sky components showed that its effect is negligible.

For each wavelength range the individual sky-subtracted spectra were
co-added and then normalised using a cubic-spline fit to the
continuum. A final spectrum covering 3850-4750 and 6300-6700~\AA\ was
obtained by merging the normalised data. Depending on the magnitude of
the target these combined spectra have typical signal-to-noise ratios
of 100-400.

The first four field stars listed in Tab.~\ref{tab:data} were observed
with the VLT-UVES spectrograph in service mode on 29 and 30 November
2004 under program 74.D-0109. UVES is a two armed cross-dispersed
echelle spectrograph allowing for simultaneous observations in the
blue and red part of the spectrum. In the blue arm standard settings
with central wavelengths of 390 and 437~nm were used to observe the
spectral ranges 3300-4500 and 3730-5000~\AA. For the red arm standard
settings with central wavelengths of 564 and 860~nm provided coverage
between 4620-5600 and 6600-10400~\AA. A 1.2$''$ wide slit was used,
providing a spectral resolution of 0.1~\AA\ at \hg, corresponding to
an effective resolving power of $R \simeq 40\,000$, a value which
applies to all UVES setups. Individual exposure times ranged from 1500
to 2200 seconds.

Sk~$-67$~166 was observed on 27 and 29 September 2001 under program
67.D-0238 with UVES using a 1$''$ slit. A standard blue setting with
central wavelength 437~nm provided continuous coverage between
3730-5000~\AA. A non-standard red setup with central wavelength 830~nm
was used to observe the range between 6370-10250~\AA. Three exposures,
each of 1000~s were obtained for each setup. Note that the data for
this target was previously presented by \cite{crowther02}. Finally,
Sk~$-70$~69 was observed on 1 and 2 December 2002 using UVES under
program 70.D-0164. Standard blue and red settings with central
wavelengths 390 and 564~nm were used in simultaneous 2400~s exposures
using dichroic. A second non-standard red setup with central
wavelength 520~nm (4170-6210~\AA) without dichroic was used in a
1500~s exposure. The typical two pixel S/N ratios obtained for all
spectra are 80 at \hg\ and 60 at \ha.

Spectral types for FLAMES stars were determined by visual inspection
of the spectra, using published standards. In particular the atlas of
\cite{walborn90} was used while considering the lower metallicity
environment of the LMC. The classifications given in
Tab.~\ref{tab:data} are from \cite{evans06}. A comparison with
previously published spectral types is also given in the latter
paper. For the field stars we adopted the classifications given by
\cite{walborn77}, \cite{walborn95, walborn02b} and \cite{massey95,
massey05}.

Photometric data for the FLAMES targets was obtained predominantly
from $B$ and $V$ images of the N11 field taken with the Wide Field
Imager at the 2.2-m Max Planck Gesellschaft/ESO telescope on 2003
April 2004 \citep{evans06}. The photometry for stars flagged with an
asterisk was adopted from \cite{parker92}. The caption of
Tab.~\ref{tab:data} lists the references to the various sources of the
photometric data of the field stars.

To calculate the interstellar extinction (\Av) given in
Tab.~\ref{tab:data} we adopted intrinsic colours from \citet[][ and
references therein]{johnson66} and a ratio of total to selective
extinction of $\Rv = 3.1$. With these \Av\ values, extinction
corrected visual magnitudes ($V_0$) were calculated from the observed
$V$-band magnitudes. Finally, we calculated the absolute visual
magnitude \Mv, while adopting a distance modulus of 18.5 for the LMC
\citep{panagia91, mitchell02}.

\section{Analysis method}
\label{sec:method}

All optical spectra are analysed using the automated fitting
method developed by \citet[][hereafter referred to as
\citetalias{mokiem05}]{mokiem05}. Here we will suffice with a short
description of the method and refer to the before mentioned paper and
to \citet[][ hereafter \citetalias{mokiem06}]{mokiem06} for the
details.

In short the automated fitting method uses the genetic algorithm based
optimisation routine \pikaia\ \citep{charbonneau95} to determine the
set of input parameters for the stellar atmosphere code \fastwind\
\citep{puls05} which fit an observed spectrum the best. This best
fitting model is constructed by evolving a population of \fastwind\
models over a course of generations. At the end of every generation
the parameters of the models which relatively fit the observed
spectrum the best are used to construct a new population of models. By
repeating this procedure a natural optimisation is obtained and after
a number of generations (see below) the best fitting model, i.e.\
global optimum in parameter space, is found.

Using the concept of a unified model atmosphere the fast performance
code \fastwind\ incorporates non-LTE and an approximate approach to
line blanketing to synthesise hydrogen and helium line
profiles. Consequently, given the spectral range of the observed data
set in this study we will focus on the modelling of the optical
hydrogen and helium lines. To account for the accuracy with which each
individual line can be reproduced by
\fastwind\ we adopt the line weighting scheme as described in
\citetalias{mokiem05}.

In the fitting of the spectra we allow for seven free
parameters. These are the effective temperature \teff, the surface
gravity $g$, the helium number density defined as $\yhe \equiv N({\rm
He})/N({\rm H})$, the microturbulent velocity \vturb, the projected
rotational velocity \vsini, the mass loss rate \mdot\ and the exponent
of the beta-type velocity law describing the supersonic regime of the
stellar wind. For the terminal velocity of the wind \vinf, which
cannot be accurately determined from the optical spectrum, we adopt
values determined from the analysis of ultraviolet (UV) wind lines. If
no UV determination of \vinf\ is available a scaling relation of
\vinf\ with the escape velocity (\vesc) defined at the stellar surface
is used throughout the fitting process. For our programme stars we
adopt the ratio as determined by \cite{lamers95} of $\vinf/\vesc=2.6$.
Similar as in \citetalias{mokiem06} we adopt fixed values for the
atmospheric abundances of the background metals. These were scaled
with respect to mass ratios based on the Solar abundances of \citet[][
and references therein]{grevesse98}. The metallicity scaling factor
was set equal to the mean metal deficiency of 0.5 as found for the LMC
\citep{russell89, rolleston02}.

For our current data set the spectral range and quality, with
exception of the signal-to-noise ratio, is similar to the set analysed
in \citetalias{mokiem06}. Consequently, we adopt the same minimum
number of generations that have to be calculated to determine the best
fit, i.e.\ to assure that the global optimum in parameter space is
found. In \citetalias{mokiem06}, using formal tests that accounted for
nebular contamination of the line profiles and a signal-to-noise ratio
of 50, this number was determined to be 150.

The uncertainties of the fit parameters are determined using so-called
optimum width based error estimates. In \citetalias{mokiem05} we have
argued that an error estimate for a parameter can be defined as the
maximum variation of this parameter within the global optimum in
parameter space. By measuring the width of this optimum in terms of
fit quality one can determine which models, based on their fit
quality, are associated with the global optimum. The maximum variation
of the individual fit parameters within this group of models then gives
the error estimates \citepalias[see Sect. 4 in][]{mokiem05}.

In Tab.~\ref{tab:errors} the optimum width based error estimates of
the fit parameters for our programme stars are listed. The
uncertainties in \rstar, \lstar, \Ms\ and \Mev, that are derived from
the fit parameters, were calculated using the same approach as in
\citetalias{mokiem06}. The adopted uncertainty in the visual magnitude
is $\Delta\Mv = 0.13$ \citep{panagia91} for all our programme stars.

\begin{table*}
  \caption{Fundamental parameters of the LMC sample determined using
  GA optimised spectral fits, with \teff\ in kK, \logg\ and \loggc\ in
  \cmsecsec, \rstar\ in \rsun, \lstar\ in \lsun, \vturb\ and \vsini\
  in \kmsec, \mdot\ in \msunyr and \Ms\ and \Mev\ in \msun. Results
  were obtained using a population of 72 \fastwind\ models evolved
  over a minimum of 150 generations. Gravities corrected for
  centrifugal acceleration (\loggc) were used to calculate the
  spectroscopic masses (\Ms). Evolutionary masses (\Mev) were derived
  from the tracks of \cite{schaerer93}.}
  \label{tab:results}
  \begin{center}
  \begin{tabular}{llcccrccrrclrr}
  \hline\\[-9pt] \hline \\[-7pt]
  ID & ST & \teff & \logg & \loggc & \multicolumn{1}{c}{\rstar} & $\log \lstar$ & \yhe
  & \multicolumn{1}{c}{\vturb} & \multicolumn{1}{c}{\vsini} & \mdot & \multicolumn{1}{c}{$\beta$}
  & \multicolumn{1}{c}{\Ms} & \multicolumn{1}{c}{\Mev} \\[1pt]
 \hline \\[-9pt]
N11-004 & OC9.7 Ib &   31.6 & 3.36 & 3.37 & 26.5 & 5.80 & 0.10 & 5.7 & 81 & 1.78$\cdot10^{-6}$ & 1.18 & 59.9        & 47.5\\[3.5pt]
N11-008 & B0.7 Ia &    26.0 & 2.98 & 2.99 & 29.6 & 5.55 & 0.10 & 18.0 & 83 & 4.96$\cdot10^{-7}$ & 1.87 & 31.4       & 31.7\\[3.5pt]
N11-026 & O2 III(f*) & 53.3 & 4.00 & 4.00 & 10.7 & 5.92 & 0.11 & 19.0 & 109 & 1.81$\cdot10^{-6}$ & 1.08 & 42.4      & 81.9\\[3.5pt]
N11-029 & O9.7 Ib &    29.4 & 3.23 & 3.24 & 15.7 & 5.21 & 0.07 & 19.1 & 77 & 1.73$\cdot10^{-7}$ & 1.63 & 15.5       & 24.9\\[3.5pt]
N11-031 & ON2 III(f*) &45.0 & 3.85 & 3.86 & 13.7 & 5.84 & 0.10 & 20.0 & 116 & 3.88$\cdot10^{-6}$ & 0.89 & 49.3      & 60.8\\[3.5pt]
N11-032 & O7 II(f) &   35.2 & 3.45 & 3.46 & 14.0 & 5.43 & 0.09 & 11.3 & 96 & 8.06$\cdot10^{-7}$ & 1.03 & 20.6       & 33.8\\[3.5pt]
N11-033 & B0 IIIn &    27.2 & 3.21 & 3.35 & 15.6 & 5.07 & 0.08 & 17.9 & 256 & 2.44$\cdot10^{-7}$ & 1.03 & 19.8      & 21.0\\[3.5pt]
N11-036 & B0.5 Ib &    26.3 & 3.31 & 3.32 & 15.6 & 5.02 & 0.08 & 13.6 & 54 & 1.06$\cdot10^{-7}$ & 0.80$^{a)}$       & 18.4 & 20.0\\[3.5pt]
N11-038 & O5 II(f+) &  41.0 & 3.72 & 3.74 & 14.0 & 5.69 & 0.10 & 9.2 & 145 & 1.52$\cdot10^{-6}$ & 0.98 & 38.8       & 48.3\\[3.5pt]
N11-042 & B0 III &     30.2 & 3.69 & 3.70 & 11.8 & 5.01 & 0.10 & 4.0 & 42 & 1.89$\cdot10^{-7}$ & 1.19 & 25.1        & 21.4\\[3.5pt]
N11-045 & O9 III &     32.3 & 3.32 & 3.35 & 12.0 & 5.15 & 0.07 & 16.8 & 105 & 5.48$\cdot10^{-7}$ & 0.80$^{a)}$      & 11.8 & 24.6\\[3.5pt]
N11-048 & O6.5 V((f) & 40.7 & 4.19 & 4.20 & 9.9 & 5.38 & 0.06 & 1.0 & 130 & 1.67$\cdot10^{-7}$ & 0.80$^{a)}$        & 56.6 & 36.6\\[3.5pt]
N11-051 & O5 Vn((f)) & 42.4 & 3.75 & 3.88 & 8.4 & 5.31 & 0.08 & 19.7 & 333 & 1.01$\cdot10^{-6}$ & 0.60 & 19.5       & 36.4\\[3.5pt]
N11-058 & O5.5 V((f) & 41.3 & 3.89 & 3.90 & 8.4 & 5.27 & 0.10 & 14.5 & 85 & 1.52$\cdot10^{-7}$ & 1.42 & 20.3        & 34.4\\[3.5pt]
N11-060 & O3 V((f*)) & 45.7 & 3.92 & 3.93 & 9.7 & 5.57 & 0.12 & 19.3 & 106 & 5.22$\cdot10^{-7}$ & 1.26 & 29.2       & 49.4\\[3.5pt]
N11-061 & O9 V &       33.6 & 3.51 & 3.52 & 11.7 & 5.20 & 0.09 & 19.8 & 87 & 2.14$\cdot10^{-7}$ & 1.80 & 16.7       & 26.6\\[3.5pt]
N11-065 & O6.5 V((f) & 41.7 & 3.89 & 3.90 & 7.4 & 5.17 & 0.17 & 9.4 & 83 & 3.63$\cdot10^{-7}$ & 0.80$^{a)}$ & 15.8  & 32.9\\[3.5pt]
N11-066 & O7 V((f))  & 39.3 & 3.87 & 3.88 & 7.7 & 5.10 & 0.11 & 4.8 & 71 & 4.08$\cdot10^{-7}$ & 0.80$^{a)}$ & 16.2  & 29.5\\[3.5pt]
N11-068 & O7 V((f))  & 39.9 & 4.13 & 4.13 & 7.1 & 5.06 & 0.10 & 15.6 & 54 & 3.43$\cdot10^{-7}$ & 1.12 & 25.2        & 29.2\\[3.5pt]
N11-072 & B0.2 III &   30.8 & 3.78 & 3.78 & 7.9 & 4.70 & 0.12 & 7.6 & 14 & 2.35$\cdot10^{-7}$ & 0.84 & 13.8         & 17.7\\[3.5pt]
N11-087 & O9.5 Vn &    32.7 & 4.04 & 4.09 & 8.9 & 4.91 & 0.10 & 15.5 & 276 & 1.38$\cdot10^{-7}$ & 0.80$^{a)}$       & 35.6 & 20.9\\[3.5pt]
N11-123 & O9.5 V &     34.8 & 4.22 & 4.23 & 5.4 & 4.58 & 0.09 & 9.0 & 110 & 7.62$\cdot10^{-8}$ & 0.80$^{a)}$ & 17.8 & 18.8\\[1pt]
\hline\\[-9pt]
BI 237       & O2 V((f*)) & 53.2 & 4.11 & 4.11 & 9.7 & 5.83 & 0.10 & 12.8 & 126 & 7.81$\cdot10^{-7}$ & 1.26 & 44.6 & 75.0\\[3.5pt]
BI 253       & O2 V((f*)) & 53.8 & 4.18 & 4.19 & 10.7 & 5.93 & 0.09 & 18.6 & 191 & 1.92$\cdot10^{-6}$ & 1.21 & 64.6 & 84.1\\[3.5pt]
Sk $-66$ 18  & O6 V((f))  & 40.2 & 3.76 & 3.76 & 12.2 & 5.55 & 0.14 & 10.8 & 82 & 1.07$\cdot10^{-6}$ & 0.94 & 31.5 & 40.7\\[3.5pt]
Sk $-66$ 100 & O6 II(f)   & 39.0 & 3.70 & 3.71 & 13.6 & 5.58 & 0.19 & 8.7 & 84 & 8.81$\cdot10^{-7}$ & 1.27 & 34.7 & 41.4\\[3.5pt]
Sk $-67$ 166 & O4 Iaf+    & 40.3 & 3.65 & 3.66 & 21.3 & 6.03 & 0.28 & 20.0 & 97 & 9.28$\cdot10^{-6}$ & 0.94 & 75.0 & 70.4\\[3.5pt]
Sk $-70$ 69  & O5 V       & 43.2 & 3.87 & 3.88 & 9.0 & 5.41 & 0.17 & 16.1 & 131 & 1.03$\cdot10^{-6}$ & 0.78 & 22.7 & 39.7\\[1pt]
\hline
\end{tabular}
  \end{center}
$^{a)}$ assumed fixed value
\end{table*}

\begin{table*}
  \caption{Optimum width based error estimates for the seven fit
  parameters. The ND entries correspond to error in \vturb\ that reach
  up to the maximum allowed value of \vturb\ and, therefore, are
  formally not defined. See text for details on the calculation of the
  uncertainties in the derived parameters. Units: \teff\ in kK, \logg\
  and \loggc\ in \cmsecsec, \rstar\ in \rsun, \lstar\ in \lsun,
  \vturb\ and \vsini\ in \kmsec, \mdot\ in \msunyr and \Ms\ and \Mev\
  in \msun.}
  \label{tab:errors}
  \begin{center}
    \begin{tabular}{lcccccllclll}
  \hline\\[-9pt] \hline \\[-7pt]
  ID & $\Delta$\teff & $\Delta$\loggc & $\Delta$\rstar & $\Delta \log
  \lstar$ & $\Delta$\yhe & \multicolumn{1}{c}{$\Delta$\vturb} &
  \multicolumn{1}{c}{$\Delta$\vsini} &
  $\log \Delta$\mdot & \multicolumn{1}{c}{$\Delta$$\beta$} &
  \multicolumn{1}{c}{$\Delta$\Ms} & \multicolumn{1}{c}{$\Delta$\Mev}\\[1pt]
  \hline \\[-9pt]
N11-004 & $^{-0.5}_{+0.5}$ & $^{-0.06}_{+0.06}$ & $\pm$1.7 & $\pm$0.06 & $^{-0.01}_{+0.02}$ & \hspace{6pt}$^{-2.5}_{+8.2}$ & \hspace{10pt}$^{-6}_{+7}$ & $^{-0.14}_{+0.05}$ & $^{-0.00}_{+0.19}$ & \hspace{4pt}$^{-10}_{+11}$ & \hspace{4pt}$^{-4}_{+4}$ \\[3.5pt]
N11-008 & $^{-0.6}_{+0.5}$ & $^{-0.06}_{+0.05}$ & $\pm$1.9 & $\pm$0.07 & $^{-0.01}_{+0.04}$ & \hspace{6pt}$^{-4.2}_{+\rm ND}$ & \hspace{10pt}$^{-8}_{+9}$ & $^{-0.20}_{+0.16}$ & $^{-0.31}_{+1.08}$ & \hspace{4pt}$^{-5}_{+8}$ & \hspace{4pt}$^{-2}_{+2}$ \\[3.5pt]
N11-026 & $^{-3.9}_{+0.8}$ & $^{-0.05}_{+0.05}$ & $\pm$0.8 & $\pm$0.14 & $^{-0.01}_{+0.02}$ & \hspace{6pt}$^{-4.8}_{+\rm ND}$ & \hspace{10pt}$^{-11}_{+13}$ & $^{-0.05}_{+0.10}$ & $^{-0.12}_{+0.03}$ & \hspace{4pt}$^{-7}_{+9}$ & \hspace{4pt}$^{-16}_{+21}$ \\[3.5pt]
N11-029 & $^{-0.6}_{+0.8}$ & $^{-0.05}_{+0.06}$ & $\pm$1.0 & $\pm$0.07 & $^{-0.01}_{+0.02}$ & \hspace{6pt}$^{-3.9}_{+\rm ND}$ & \hspace{10pt}$^{-6}_{+7}$ & $^{-0.13}_{+0.16}$ & $^{-0.42}_{+0.31}$ & \hspace{4pt}$^{-2}_{+4}$ & \hspace{4pt}$^{-2}_{+2}$ \\[3.5pt]
N11-031 & $^{-1.6}_{+2.2}$ & $^{-0.06}_{+0.07}$ & $\pm$0.9 & $\pm$0.10 & $^{-0.02}_{+0.02}$ & \hspace{6pt}$^{-7.6}_{+\rm ND}$ & \hspace{10pt}$^{-22}_{+21}$ & $^{-0.05}_{+0.06}$ & $^{-0.04}_{+0.06}$ & \hspace{4pt}$^{-9}_{+13}$ & \hspace{4pt}$^{-8}_{+11}$ \\[3.5pt]
N11-032 & $^{-0.7}_{+0.4}$ & $^{-0.07}_{+0.06}$ & $\pm$0.9 & $\pm$0.06 & $^{-0.02}_{+0.02}$ & \hspace{6pt}$^{-6.1}_{+6.6}$ & \hspace{10pt}$^{-8}_{+11}$ & $^{-0.16}_{+0.09}$ & $^{-0.11}_{+0.28}$ & \hspace{4pt}$^{-4}_{+5}$ & \hspace{4pt}$^{-2}_{+2}$ \\[3.5pt]
N11-033 & $^{-0.9}_{+1.0}$ & $^{-0.06}_{+0.09}$ & $\pm$1.0 & $\pm$0.08 & $^{-0.01}_{+0.03}$ & \hspace{6pt}$^{-5.9}_{+\rm ND}$ & \hspace{10pt}$^{-20}_{+20}$ & $^{-0.26}_{+0.10}$ & $^{-0.20}_{+0.74}$ & \hspace{4pt}$^{-3}_{+5}$ & \hspace{4pt}$^{-2}_{+2}$ \\[3.5pt]
N11-036 & $^{-0.3}_{+1.1}$ & $^{-0.09}_{+0.06}$ & $\pm$1.0 & $\pm$0.09 & $^{-0.01}_{+0.02}$ & \hspace{6pt}$^{-4.2}_{+1.4}$ & \hspace{10pt}$^{-4}_{+6}$ & $^{-1.99}_{+0.30}$ & $-$ & \hspace{4pt}$^{-4}_{+6}$ & \hspace{4pt}$^{-2}_{+2}$ \\[3.5pt]
N11-038 & $^{-1.0}_{+0.8}$ & $^{-0.06}_{+0.07}$ & $\pm$0.9 & $\pm$0.07 & $^{-0.01}_{+0.03}$ & \hspace{6pt}$^{-9.0}_{+8.3}$ & \hspace{10pt}$^{-19}_{+18}$ & $^{-0.16}_{+0.10}$ & $^{-0.13}_{+0.22}$ & \hspace{4pt}$^{-6}_{+16}$ & \hspace{4pt}$^{-4}_{+4}$ \\[3.5pt]
N11-042 & $^{-0.8}_{+0.6}$ & $^{-0.07}_{+0.09}$ & $\pm$0.7 & $\pm$0.07 & $^{-0.01}_{+0.03}$ & \hspace{6pt}$^{-3.8}_{+4.2}$ & \hspace{10pt}$^{-6}_{+4}$ & $^{-0.47}_{+0.40}$ & $^{-0.47}_{+0.77}$ & \hspace{4pt}$^{-5}_{+14}$ & \hspace{4pt}$^{-1}_{+2}$ \\[3.5pt]
N11-045 & $^{-1.2}_{+0.6}$ & $^{-0.09}_{+0.21}$ & $\pm$0.8 & $\pm$0.08 & $^{-0.02}_{+0.03}$ & \hspace{6pt}$^{-7.6}_{+\rm ND}$ & \hspace{10pt}$^{-15}_{+17}$ & $^{-0.20}_{+0.14}$ & $-$ & \hspace{4pt}$^{-2}_{+5}$ & \hspace{4pt}$^{-2}_{+3}$ \\[3.5pt]
N11-048 & $^{-2.0}_{+1.6}$ & $^{-0.21}_{+0.06}$ & $\pm$0.7 & $\pm$0.10 & $^{-0.01}_{+0.02}$ & \hspace{6pt}$^{-0.8}_{+13.2}$ & \hspace{10pt}$^{-27}_{+38}$ & $^{-1.48}_{+0.74}$ & $-$ & \hspace{4pt}$^{-22}_{+30}$ & \hspace{4pt}$^{-4}_{+5}$ \\[3.5pt]
N11-051 & $^{-1.9}_{+1.1}$ & $^{-0.06}_{+0.06}$ & $\pm$0.6 & $\pm$0.09 & $^{-0.02}_{+0.03}$ & \hspace{6pt}$^{-8.4}_{+\rm ND}$ & \hspace{10pt}$^{-38}_{+23}$ & $^{-0.28}_{+0.11}$ & $^{-0.09}_{+0.52}$ & \hspace{4pt}$^{-3}_{+5}$ & \hspace{4pt}$^{-4}_{+5}$ \\[3.5pt]
N11-058 & $^{-0.5}_{+0.7}$ & $^{-0.06}_{+0.05}$ & $\pm$0.5 & $\pm$0.06 & $^{-0.01}_{+0.01}$ & \hspace{6pt}$^{-5.9}_{+3.5}$ & \hspace{10pt}$^{-8}_{+7}$ & $^{-0.36}_{+0.08}$ & $^{-0.20}_{+0.60}$ & \hspace{4pt}$^{-3}_{+4}$ & \hspace{4pt}$^{-2}_{+2}$ \\[3.5pt]
N11-060 & $^{-1.0}_{+2.3}$ & $^{-0.05}_{+0.09}$ & $\pm$0.7 & $\pm$0.10 & $^{-0.02}_{+0.03}$ & \hspace{6pt}$^{-13.1}_{+\rm ND}$ & \hspace{10pt}$^{-16}_{+16}$ & $^{-0.22}_{+0.06}$ & $^{-0.21}_{+0.20}$ & \hspace{4pt}$^{-5}_{+9}$ & \hspace{4pt}$^{-7}_{+8}$ \\[3.5pt]
N11-061 & $^{-0.6}_{+0.6}$ & $^{-0.09}_{+0.07}$ & $\pm$0.7 & $\pm$0.06 & $^{-0.01}_{+0.03}$ & \hspace{6pt}$^{-5.4}_{+\rm ND}$ & \hspace{10pt}$^{-9}_{+11}$ & $^{-0.21}_{+0.14}$ & $^{-0.49}_{+0.38}$ & \hspace{4pt}$^{-4}_{+6}$ & \hspace{4pt}$^{-2}_{+2}$ \\[3.5pt]
N11-065 & $^{-0.9}_{+0.7}$ & $^{-0.07}_{+0.15}$ & $\pm$0.5 & $\pm$0.06 & $^{-0.03}_{+0.04}$ & \hspace{6pt}$^{-9.2}_{+6.3}$ & \hspace{10pt}$^{-8}_{+8}$ & $^{-1.06}_{+0.14}$ & $-$ & \hspace{4pt}$^{-3}_{+3}$ & \hspace{4pt}$^{-2}_{+2}$ \\[3.5pt]
N11-066 & $^{-1.3}_{+1.8}$ & $^{-0.15}_{+0.19}$ & $\pm$0.5 & $\pm$0.09 & $^{-0.03}_{+0.05}$ & \hspace{6pt}$^{-4.5}_{+\rm ND}$ & \hspace{10pt}$^{-15}_{+22}$ & $^{-0.49}_{+0.22}$ & $-$ & \hspace{4pt}$^{-5}_{+12}$ & \hspace{4pt}$^{-4}_{+3}$ \\[3.5pt]
N11-068 & $^{-1.0}_{+0.8}$ & $^{-0.19}_{+0.07}$ & $\pm$0.4 & $\pm$0.07 & $^{-0.02}_{+0.03}$ & \hspace{6pt}$^{-14.8}_{+\rm ND}$ & \hspace{10pt}$^{-13}_{+13}$ & $^{-0.79}_{+0.16}$ & $^{-0.24}_{+0.36}$ & \hspace{4pt}$^{-9}_{+14}$ & \hspace{4pt}$^{-2}_{+2}$ \\[3.5pt]
N11-072 & $^{-0.6}_{+0.6}$ & $^{-0.07}_{+0.09}$ & $\pm$0.5 & $\pm$0.06 & $^{-0.02}_{+0.02}$ & \hspace{6pt}$^{-2.8}_{+2.5}$ & \hspace{10pt}$^{-6}_{+8}$ & $^{-1.29}_{+0.24}$ & $^{-0.32}_{+0.31}$ & \hspace{4pt}$^{-3}_{+5}$ & \hspace{4pt}$^{-1}_{+1}$ \\[3.5pt]
N11-087 & $^{-0.6}_{+1.0}$ & $^{-0.09}_{+0.11}$ & $\pm$0.6 & $\pm$0.07 & $^{-0.03}_{+0.02}$ & \hspace{6pt}$^{-6.7}_{+\rm ND}$ & \hspace{10pt}$^{-20}_{+21}$ & $^{-2.82}_{+0.27}$ & $-$ & \hspace{4pt}$^{-8}_{+14}$ & \hspace{4pt}$^{-1}_{+2}$ \\[3.5pt]
N11-123 & $^{-0.7}_{+0.6}$ & $^{-0.11}_{+0.00}$ & $\pm$0.3 & $\pm$0.06 & $^{-0.01}_{+0.02}$ & \hspace{6pt}$^{-5.1}_{+6.6}$ & \hspace{10pt}$^{-11}_{+13}$ & $^{-2.03}_{+0.35}$ & $-$ & \hspace{4pt}$^{-4}_{+7}$ & \hspace{4pt}$^{-1}_{+1}$  \\[1pt]  
  \hline\\[-9pt]
BI 237 & $^{-3.8}_{+5.2}$ & $^{-0.08}_{+0.15}$ & $\pm$0.8 & $\pm$0.18 & $^{-0.01}_{+0.03}$ & \hspace{6pt}$^{-12.6}_{+\rm ND}$ & \hspace{10pt}$^{-15}_{+23}$ & $^{-0.16}_{+0.12}$ & $^{-0.17}_{+0.16}$ & \hspace{4pt}$^{-9}_{+22}$ & \hspace{4pt}$^{-18}_{+25}$ \\[3.5pt]
BI 253 & $^{-5.5}_{+5.8}$ & $^{-0.15}_{+0.14}$ & $\pm$0.9 & $\pm$0.19 & $^{-0.02}_{+0.03}$ & \hspace{6pt}$^{-18.0}_{+\rm ND}$ & \hspace{10pt}$^{-21}_{+18}$ & $^{-0.05}_{+0.10}$ & $^{-0.21}_{+0.07}$ & \hspace{4pt}$^{-21}_{+39}$ & \hspace{4pt}$^{-23}_{+31}$ \\[3.5pt]
Sk $-$66 18 & $^{-1.1}_{+0.6}$ & $^{-0.13}_{+0.09}$ & $\pm$0.8 & $\pm$0.07 & $^{-0.03}_{+0.04}$ & \hspace{6pt}$^{-10.4}_{+6.4}$ & \hspace{10pt}$^{-12}_{+15}$ & $^{-0.05}_{+0.06}$ & $^{-0.10}_{+0.11}$ & \hspace{4pt}$^{-9}_{+9}$ & \hspace{4pt}$^{-3}_{+4}$ \\[3.5pt]
Sk $-$66 100 & $^{-1.0}_{+0.8}$ & $^{-0.09}_{+0.08}$ & $\pm$0.9 & $\pm$0.07 & $^{-0.03}_{+0.05}$ & \hspace{6pt}$^{-8.1}_{+8.3}$ & \hspace{10pt}$^{-10}_{+13}$ & $^{-0.12}_{+0.12}$ & $^{-0.19}_{+0.18}$ & \hspace{4pt}$^{-8}_{+12}$ & \hspace{4pt}$^{-3}_{+4}$ \\[3.5pt]
Sk $-$67 166 & $^{-0.8}_{+0.9}$ & $^{-0.08}_{+0.00}$ & $\pm$1.3 & $\pm$0.07 & $^{-0.04}_{+0.06}$ & \hspace{6pt}$^{-8.0}_{+\rm ND}$ & \hspace{10pt}$^{-23}_{+15}$ & $^{-0.05}_{+0.02}$ & $^{-0.04}_{+0.09}$ & \hspace{4pt}$^{-15}_{+35}$ & \hspace{4pt}$^{-7}_{+7}$ \\[3.5pt]
Sk $-$70 69 & $^{-1.4}_{+1.0}$ & $^{-0.14}_{+0.13}$ & $\pm$0.6 & $\pm$0.08 & $^{-0.03}_{+0.05}$ & \hspace{6pt}$^{-13.1}_{+\rm ND}$ & \hspace{10pt}$^{-15}_{+22}$ & $^{-0.11}_{+0.14}$ & $^{-0.25}_{+0.15}$ & \hspace{4pt}$^{-7}_{+7}$ & \hspace{4pt}$^{-3}_{+5}$ \\[1pt]
  \hline
  \end{tabular}
  \end{center}
\end{table*}

\subsection{The assumption of spherical symmetry}

\fastwind\ assumes a spherically symmetric star and wind. For very
high rotational speeds, \vrot, this may potentially be a problematic
assumption.  As it is well known, a high rotation rate leads to a
distortion of the stellar surface and, via the von Zeipel theorem, to
a decrease in flux and effective temperature from the pole to the
equator \citep[][(slightly) modified by Maeder (1999) for the relevant
case of shellular, i.e., radially dependent, differential rotation,
$\omega = \omega(r)$]{vonzeipel24}.  This so-called ``gravity''
darkening does not only affect the stellar parameters, but also the
wind \citep[e.g.][]{cranmer95,owocki96,petrenz00}.
Thus, it might be questioned in how far the derived properties (which
then depend on inclination angle and have a somewhat {\it local}
character) are representative for the {\it global} quantities (e.g.,
mass, luminosity and mass-loss rate) which refer to integral
quantities.

As shown by various simulations \citep{cranmer95,petrenz96,
petrenz00}, the difference between local and global quantities remains
small unless the star rotates faster than $\ga 60{\ldots}70$ \% of its
critical speed. Unfortunately, however, we cannot directly access the
actual rotational speed, \vrot, but only its projected value, \vsini.

Accounting for the average value of $< \sin i > = \pi/4$, we have
calculated the ratio of average to critical rotational speed, $\Omega
\approx \vsini /(<\sin i> \vcrit)$, and found that only two objects
lie above this value, namely N11-033 and N11-051, both with $\Omega
\approx 0.7$. All other objects lie well below $\Omega = 0.3$ (except
for N11-087 with $\Omega \approx 0.4$). Of course, we cannot exclude
that also the latter objects lie above the decisive threshold (if
observed pole-on), but the rather large sample size implies that such
a possibility should be actually present only for a minority of
objects. In conclusion, we suggest that the majority of our objects
is, if at all, only weakly affected by a distortion of surface and
wind. Note, e.g., that $\Omega = 0.3$ leads to a difference in \teff\
and \rstar\ between pole and equator of less than 5\%, i.e., of the
same order or less than our error estimates
(cf. Tab.~\ref{tab:errors}). Regarding the derived gravities and
masses, finally, we have applied a consistent centrifugal correction
anyhow (cf. Sect.~\ref{sec:gravities}).

For the two objects with large rotational speeds, on the other hand,
it is rather possible that we observe them almost equator-on, i.e.,
the observed profiles contain an intrinsic averaging over the complete
stellar disk and thus correspond, at least in part, to the global,
polar angle averaged values. \cite{howarth01} analyzed two galactic
fast rotators, HD\,149757 ($\zeta$\,Oph) and HD\,191423, accounting
for effects of non-sphericity and von Zeipel's theorem. Their study
showed that for these objects, rotating at $\Omega = 0.9$, differences
in effective temperature and radius between pole and equator can
amount to 20--30 \%. Interestingly, analysis of the same objects by
\citet{herrero02}, \citet{villamariz02}, and \citet{villamariz05}
using both hydrostatic and spherically symmetric expanding atmospheres
yielded (average) parameters that agreed to within the standard
deviation (5--10\%) with the results obtained by
\citeauthor{howarth01}. For stellar parameters it thus appears
justified to conclude that non-sphericity has a very small impact {\em
even} on stars with extreme \vsini.
But note also that with respect to global
mass-loss rates the situation is much more unsecure and more-D
simulations would be required to constrain their actual values. As
shown, e.g., by \cite{petrenz00}, the modified wind-momentum rate from
a rapidly rotating ($\Omega = 0.85$) B-supergiant might be
underestimated up to one magnitude if seen equator on.
As the wind properties of N11-031 and N11-051 are rather normal it
seems however unlikely that in these cases rotational effects
have this type of dramatic impact.

\section{Fundamental parameters}
\label{sec:fun_par}

\subsection{Effective temperatures}
\label{sec:teff}

\begin{figure}[t]
  \centering \resizebox{8.8cm}{!}{ \includegraphics{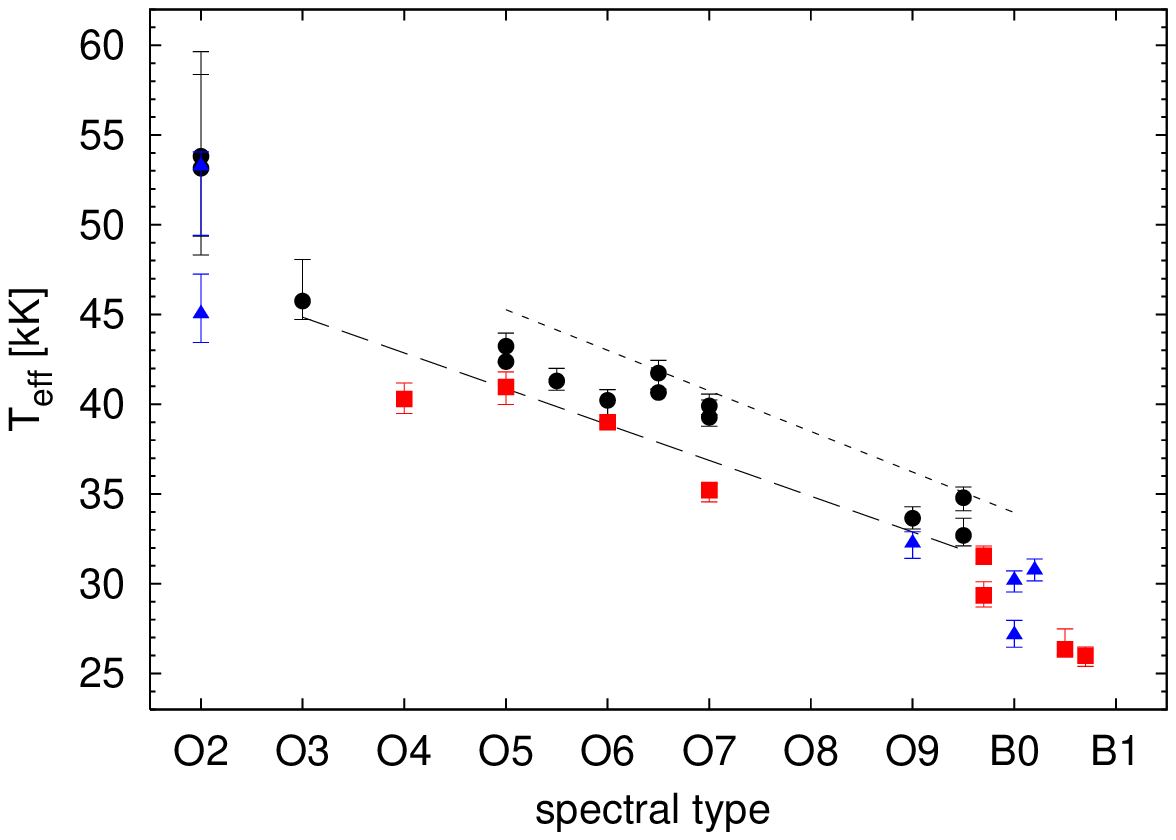}}
  \caption{Effective temperatures as a function of spectral type for
  the investigated LMC sample. Different luminosity classes are
  denoted using circles, triangles and squares for class V, III and
  I-II objects, respectively. Shown as a dashed line is the \teff\
  calibration of \cite{martins05a} for Galactic O-type dwarfs. The
  dotted line corresponds to the average effective temperature of the
  SMC dwarfs studied in \citetalias{mokiem06}. The LMC dwarfs are
  found to lie in between these two average scales, with a typical
  $\sim$2~kK offset from both the average Galactic and SMC relations.}
  \label{fig:teff}
\end{figure}

In Fig.~\ref{fig:teff} the distribution of the effective temperatures
of our programme stars as a function of spectral type is shown. The
different luminosity classes are denoted using circles, triangles and
squares, respectively, for class V, III and I-II objects. Similar as
in \citetalias{mokiem06} we see that for a given spectral type the
dwarfs are systematically hotter than the giants and supergiants. This
separation can be interpreted as the result of the reduced surface
gravities of the more evolved objects. A lower surface gravity results
in an increased helium ionisation \citep[e.g.][]{kudritzki83},
reducing the \teff\ needed for a given spectral type
\citep[e.g.][]{mokiem04}. A second reason is that the more evolved
objects have stronger winds. These denser winds induce an increased
line blanketing effect, further reducing the required temperature
\citep[e.g.][]{schaerer97}. \cite{massey05}, who also analysed a
sizeable sample of LMC stars, do not find evidence for the dwarfs
being hotter than the giants. Though, we note that their analysis only
contained two giant \teff\ determinations for spectral type later than
O2.

For comparison we also show in Fig.~\ref{fig:teff} as a dashed line
the observed \teff\ vs.\ spectral calibration for Galactic O-type
dwarfs from \cite{martins05a}. With a dotted line the average
temperature of the SMC dwarfs studied in \citetalias{mokiem06} is
shown. The LMC dwarfs are found to occupy the region in between these
two average temperature scales, with an average behaviour intermediate
to that of the Galactic and SMC dwarfs. We interpret this as the
result of the metallicity of the LMC that is lower than the Galactic
value and higher than in the SMC. As the amount of line blanketing in
a stellar atmosphere depends on metallicity the LMC objects have
temperatures in between that of objects in the other two galaxies.

\subsection{The \teff\ scale of O2 stars}
\label{sec:O2teff}

Our sample contains four O2 type stars. This spectral type was
introduced by \cite{walborn02a} and is assigned based primarily on the
ratios of selective emission lines of \niv\ and \niii. By modelling
these lines \cite{walborn04} have shown that indeed, as had been
hypothesised, the O2 spectra correspond to higher effective
temperatures. However, the correct treatment of nitrogen lines in
stellar atmosphere models is notoriously difficult. The relevant
ionisation stages of this atom represent much more complex ion models
compared to the relatively simple hydrogen and helium ions. Moreover,
for the higher ions of nitrogen the ionisation depends on the
extreme-UV radiation field, which may be affected by non-thermal
processes, such as shocks. Adding to the complexity is the fact that
the nitrogen abundance has to be treated as a free parameter, as many
early type stars show evidence of atmospheric abundance enhancements
\cite[e.g.][]{crowther02, bouret03}. In our analysis method we solely
model the hydrogen and helium lines and self consistently allow for
abundance enhancements by treating the helium abundance as a free
parameter. In principle our temperature determination should,
therefore, not be affected by such problems. {\em As a result of this,
our analysis can provide an independent confirmation for the hot
nature of O2 stars.}

In Fig.~\ref{fig:teff} we see that the objects with an O2 spectral
type indeed correspond to the hottest stars in our sample. With
exception of the giant N11-031, which based on its helium lines
we find to be cooler (see below), they have temperatures in
excess of 50~kK and, therefore, are significantly hotter than the O3
star at $\teff = 46$~kK. The error bars, though, are
considerable. Compared to the average error of $\sim$3 percent the O2
effective temperatures have an uncertainty of 5 up to 11 percent. This
large uncertainty can be explained by the weak or even absent neutral
helium lines in the spectra of these objects. As a result of this, our
fitting method has to predominately rely on the \heii\ line profiles
to determine the correct helium ionisation equilibrium. Based on a
single ionisation stage the determination of this equilibrium is more
uncertain, which explains the larger error bars. One should also be
wary for a possible degeneracy effect between \teff\ and \yhe\ that
can occur as a result of the missing neutral helium lines
\citepalias[see e.g.][]{mokiem05}. This, however, is not an issue as
the helium abundances derived for all O2 stars correspond to normal
values close to 0.10. 

It is also possible to question whether the fact that the \hei\
lines are so weak could result in a systematically higher \teff\
determination by the automated method compared to ``by eye'' fits. In
other words, would a ``by eye'' fit prefer a lower temperature
solution? For the O2 giant N11-026 this is a relevant question, as the
weak \heil4471 line, shown in Fig.~\ref{fig:fits_1} in the appendix,
is not fitted perfectly. Note, however, that due to the relative plot
scale the discrepancy is exaggerated and that the fit
quality is good and is comparible to that of the \heii\ lines. Still,
to assess whether the fit of this particular line could be improved,
i.e.\ to search for a solution predominantly using the \hei\
diagnostic, we ran test fits with increasing relative weight of
\heil4471. We found that a solution with an improved fit could be
obtained for an increase of the relative line weight with a factor of
five. This solution has a \teff\ lower by 3.7~kK and all other fit
parameters approximately equal to the solution with the low \heil4471
weight. 
The overal fit quality of the other lines has somewhat decreased.
As these lines are relatively strong, this is difficult to discern by
eye.
Also note that within the lower error estimate for \teff\
of 3.9~kK the two solutions agree. Similar results were also obtained
for the O2 stars BI~237 and BI~253, where a reduction of \teff\ by
3.8~kK and 5.4~kK was found, respectively, for an increase in the
relative weight of \heil4471 by a factor of six and two,
respectively. Consequently, based on our analysis we can only give a
{\em range} for the \teff\ scale of the O2 stars of $49-54$~kK, the
bounds being set by the temperatures based on the \heil4471 and
hydrogen/\heii\ diagnostics.

The upper end of our hydrogen/helium based O2 effective
temperature range is in good agreement compared to the average
effective temperature of 54~kK as determined by \cite{walborn04}.
As we already mentioned the giant N11-031 with a \teff\
lower by approximately 8~kK compared to the other O2 stars forms an
exception to this. Its effective temperature compares better to the
temperature of the O3 star N11-060. A close inspection of the spectra
of the two stars reveals why this is so. Both stars have exactly the
same equivalent width ratio of the \heil4471 and \heiil4541 lines,
which results in similar values for \teff. To also test whether
the relatively strong \heil4471 line could be dominating the fit,
forcing a relatively low \teff, we refitted the spectrum of N11-031
ignoring the neutral helium lines. This again resulted in an effective
temperature of 45~kK.

Interestingly, the nitrogen line analysis of \citeauthor{walborn04}
for N11-031 did result in a higher effective temperature of
55~kK. We are not sure whether this discrepancy is the result of
a systematic offset between the \nv~$\lambda$4603-20/\niv\ and
\heil4471/\heii\ temperature scales. A more recent comparison by one
of us (P.A.C.) of the spectral fit of \citeauthor{walborn04} to new
VLT data, however, showed relatively large discrepancies in the helium
line fits compared to the nitrogen line fits, indicating that this
could be the case. The discrepancy could also be due to differences in
fitting assumptions and approaches. \citeauthor{walborn04} adopted a
surface gravity of $\logg = 4.0$ and estimated a mass loss rate of
$\mdot = 1.0\times10^{-6}\,\msunyr$ from the \heiil4686 line. Our fit
indicates that the former parameter should be lower by 0.15~dex,
consequently lowering \teff. With respect to the mass loss rate we
find a value higher by a factor four. An increase in \mdot\ of this
magnitude can have a significant effect on the strength of different
nitrogen lines \citep[e.g.][]{crowther02} and, therefore, on the
derived effective temperature.

\begin{figure}[!t]
  \centering \resizebox{8.8cm}{!}{ \includegraphics{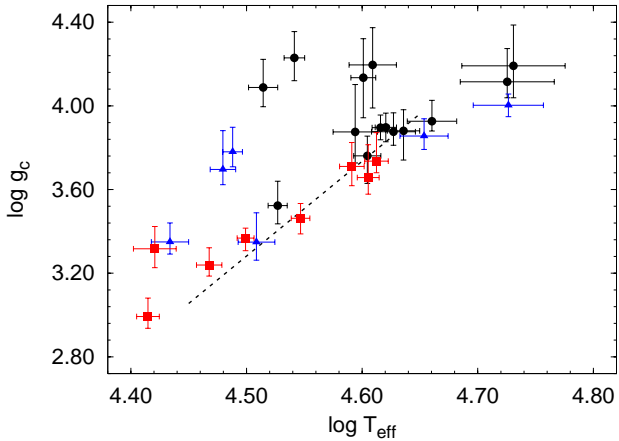}}
  \caption{The $\log \teff$ -- \loggc\ plane for the analysed LMC
  objects. Different luminosity classes are denoted using circles,
  triangles and squares for dwarfs, giants, and bright giants and
  supergiants, respectively. With exception of two objects the dwarfs
  are clearly separated from the luminosity class I-II objects. The
  giants seem to overlap with both the luminosity class V and I-II
  objects. Shown as a dotted line is the average $\log \teff$ --
  \loggc\ relation of the SMC O-type stars of luminosity class I-II-III
  from
  \citetalias{mokiem06}.}
\label{fig:logg}
\end{figure}

\cite{massey05} analysed a total of 11 O2 stars. They could determine
the effective temperature for three dwarfs and one giant, with $
47.0\,{\rm kK} \lesssim \teff \lesssim 54.5\,{\rm kK}$. For the
supergiants only lower limits of $\teff \gtrsim 42$~kK and higher
could be derived. These results are in agreement with our
findings. However, these authors note that the correlation with \teff\
for the O2--3.5 spectral types is not tight. In particular no good
agreement was found between the ratios of the \niii\ and \niv\
emission lines and the \hei\ and \heii\ lines. Consequently, a more
thorough investigation of the O2 stars based on both the nitrogen and
helium spectrum is necessary to resolve this degenerate class
adequately.

\subsection{Gravities}
\label{sec:gravities}

The distribution of our programme stars in the $\log \teff$ -- \loggc\
plane is presented in Fig.~\ref{fig:logg}. To calculate the surface
gravity corrected for centrifugal acceleration (\loggc) the method
discussed by \cite{herrero92} and \cite{repolust04} was
adopted. Different luminosity classes are denoted using circles,
triangles and squares for, respectively, class V, III and I-II
objects. In this figure we see that the dwarfs, with exception of two
objects, form a group clearly separated from the latter two groups. In
contrast to our findings in \citetalias{mokiem06} we do not find a
clear separation between the dwarfs and giants. Instead the latter
group of objects shows an overlap with both luminosity class V and
I-II objects.

Shown in Fig.~\ref{fig:logg} as a dotted line is the average $\log
\teff$ -- \loggc\ relation of the SMC O-type stars of luminosity
class I-II-III from
\citetalias{mokiem06}. The majority of the evolved LMC objects seem to
agree with this trend, that illustrates the evolutionary correlation
between effective temperature and surface gravity. However, note that
some evolved objects are at considerable distance from the average
relation. Thus, a calibration of the two parameters for a given
luminosity class should be taken with care \citep[see
also][]{repolust04}.

\begin{figure}[!t]
  \centering \resizebox{8.8cm}{!}{ \includegraphics{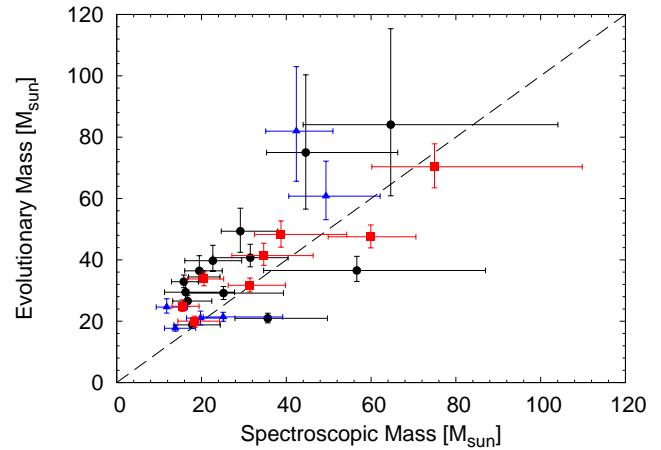}}
  \caption{Comparison of spectroscopic masses with masses derived from
  the evolutionary tracks of \cite{schaerer93}. The one-to-one
  correlation between the two mass scales is given by the dashed
  line. Symbols have the same meaning as in Fig.~\ref{fig:logg}. The
  objects with the highest evolutionary masses exhibiting a mass
  discrepancy correspond to the four O2 stars in our sample.}
\label{fig:mass}
\end{figure}

The comparison of spectroscopically determined masses (\Ms) with
masses obtained from evolutionary calculations is presented in
Fig.~\ref{fig:mass}. Using the same symbols as in Fig.~\ref{fig:logg}
dwarfs, giant and supergiants are distinguished. Evolutionary masses
(\Mev) were derived from the evolutionary tracks calculated for a
metallicity of $Z = 0.4 \, \zsun$ from \cite{schaerer93}. The errors on
these masses correspond to the maximum mass interval in the error box
spanned by the uncertainties in luminosity and effective temperature.

The tracks from \citeauthor{schaerer93} do not include the effects of
rotation. Consequently, this additional source of error is not
included. Calculations including \vrot\ show that in some cases one
can no longer assign an unambiguous $M(L, \teff)$. This is a result of
rotationally enhanced mixing or the unknown inclination angle, if the
star has a non-spherically symmetric distribution of \rstar\ and
\teff, causing complicated tracks including loops during the secular
redward evolution \citep{meynet05}.  Assessing the impact of \vrot\ on
the \Mev\ determination we showed in \citetalias{mokiem06} from a
comparison of masses derived from non rotating tracks to those
obtained from tracks calculated for $\vrot = 300 \,\kmsec$ that the
error in \Mev\ will not increase by more than approximately ten
percent.

In Fig.~\ref{fig:mass} we see that the majority of the objects are
located left of the one-to-one correlation, given by the dashed
line. The error bars of twelve objects do not even touch this
correlation. Consequently, we find a significant mass
discrepancy. Similar mass discrepancy problems have been discussed by
e.g.\ \cite{herrero92}, \cite{dekoter03}, and \cite{repolust04}. Most
of these classical problems were attributed to limitations in the
stellar atmosphere models \citep{herrero93} and to potential biases in
the fitting process \citepalias[see][]{mokiem05}. Here we cannot
explain the found discrepancy in such a manner. We will provide a more
thorough investigation and discussion in Sect.~\ref{sec:mdisc}.

\subsection{Helium abundances}
\label{sec:yhe}

For the SMC sample that we analysed in \citetalias{mokiem06} we found
a correlation between the helium surface abundance and surface
gravity. This could partly be explained by evolutionary effects. As
the surface gravity decreases when a star evolves away from the ZAMS,
objects with lower gravities would correspond to more evolved objects
and are more likely to have atmospheres enriched with helium.

\begin{figure}[t]
  \centering
  \resizebox{8.8cm}{!}{ \includegraphics{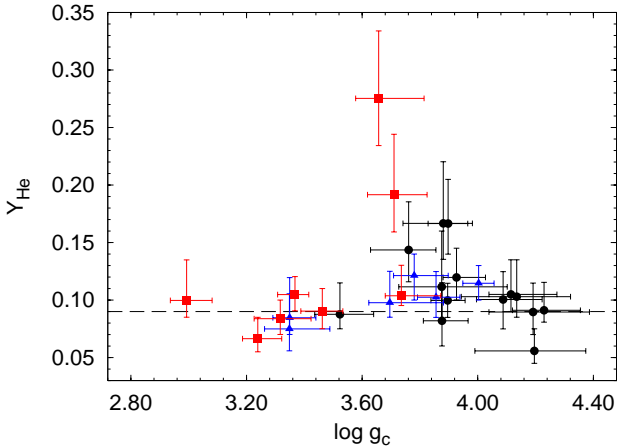}}
  \caption{Helium abundances as a function of surface gravity. Symbols
  have the same meaning as in Fig.~\ref{fig:logg}. The dashed line at
  $\yhe = 0.09$ is a measure for the initial helium abundance and
  corresponds to the mean of the helium abundances of the dwarfs with
  \yhe\ smaller than the sample average. Starting at the highest
  gravities an increase of the helium fraction is seen down to $\loggc
  \approx 3.6$.}
  \label{fig:yhe}
\end{figure}

To investigate whether the scenario discussed above also applies to
the current LMC sample, we plot the helium abundance as determined
with the automated method as a function of \loggc\ in
Fig.~\ref{fig:yhe}. Also shown as a dashed line is a measure of the
initial helium abundance. This value of $\yhe = 0.09$ was calculated
by averaging the surface helium abundances of the dwarf type objects
with a helium abundance smaller than the total sample
average. Compared to this measure for the initial helium abundance a
correlation between the surface gravity and helium enrichment can be
observed. Starting at the highest gravities, we find an increase in
the average helium abundance towards lower \loggc. Note that the two
objects exhibiting the largest helium fractions are supergiants.  The
increase can again be partially explained as an evolutionary effect.
In \citetalias{mokiem06} a similar correlation between average helium
abundance and gravity was found to exist down to the lowest gravities
investigated.

\begin{figure}[t]
  \centering \resizebox{8.8cm}{!}{\includegraphics{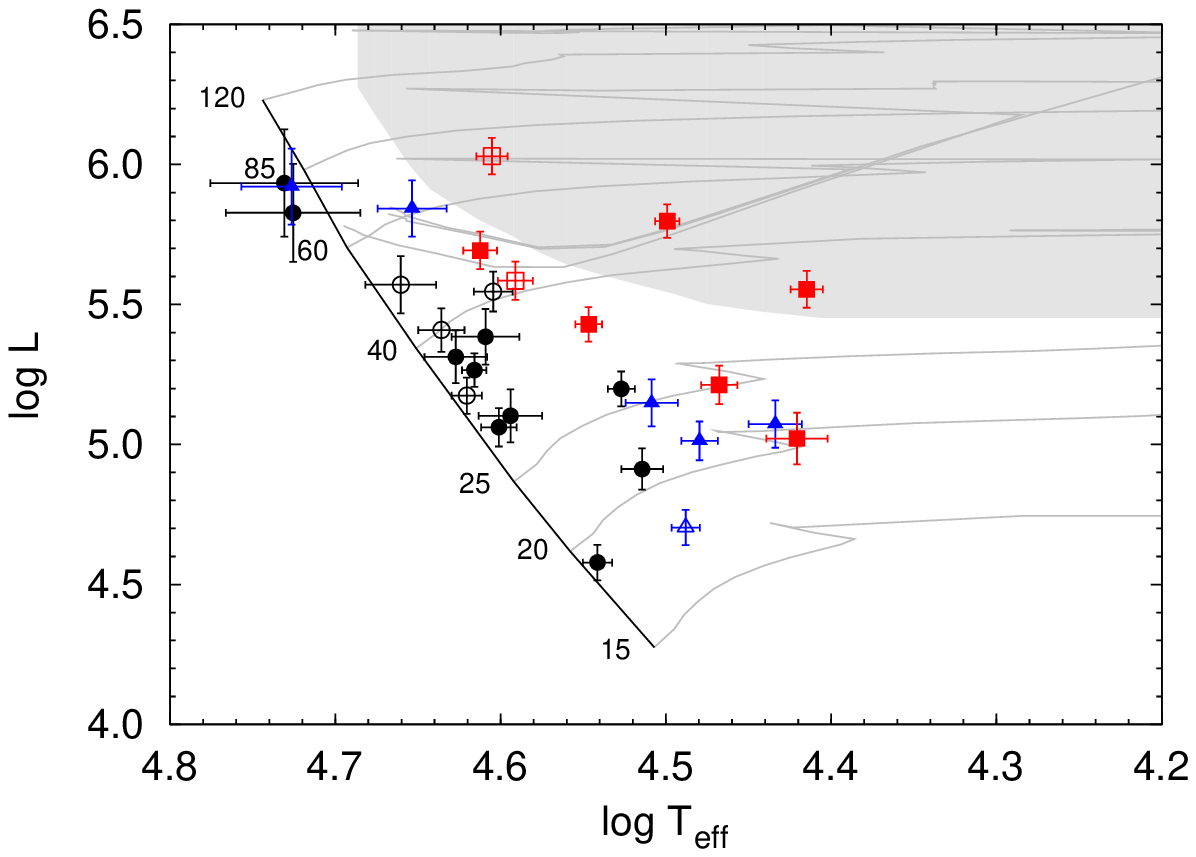}}
  \caption{Hertzsprung-Russell diagram for the LMC sample.  Symbols
  have the same meaning as in Fig.~\ref{fig:logg}. Over plotted as
  grey lines are the evolutionary tracks of \cite{schaerer93} for $Z =
  0.4\,\zsun$, with a black line representing the ZAMS. Open symbols
  indicate objects with $\yhe \geq 0.12$. The grey area correspond to
  the region in which the rotating evolutionary models of
  \cite{meynet05} predict a relative helium enhancement of at least
  0.01.}
\label{fig:hrd}
\end{figure}

Interestingly, in Fig.~\ref{fig:yhe} we see that for our LMC sample no
helium enrichment is found at $\logg \lesssim 3.6$. Not even the
supergiants show evidence of enrichment below this gravity. Why is
this so? In Fig.~\ref{fig:hrd}, where we present the HR-diagram for
our sample, the answer to this question is given. Shown as a grey area
in this figure is the region in which the rotating evolutionary models
of \cite{meynet05} predict a helium surface enrichment of at least ten
percent. The majority of the evolved objects are located outside this
region. Consequently, based on their specific evolutionary phase these
objects are not expected to show any enrichment. Also important is the
fact that we have selected the hottest objects from the N11
cluster. As a result of this our sample is biased and does not contain
the low gravity objects with a high luminosity, i.e.\ those more
likely to be enriched, as these have evolved into cool B-type stars.

Apart from the supergiants also four dwarfs are found to be
enriched. Figure~\ref{fig:hrd}, in which we have highlighted stars
with a helium abundance of at least 0.12 using open symbols, shows
that one these dwarfs, Sk~$-66$~18, is located relatively close to the
region in which enrichment is predicted. Consequently, rotationally
enhanced mixing is a possible explanation. The three remaining dwarfs,
N11-60, N11-065 and Sk~$-70$~69, in contrast lie relatively close to
the ZAMS. Therefore, ``normal'' mixing cannot explain their
enrichment. Instead, a possible explanation is given by chemically
homogeneous evolution. As this may also be linked to the mass
discrepancy, we will return to it in Sect.~\ref{sec:mdisc}.

\subsection{Microturbulence}
\label{sec:vturb}

\begin{figure}[t]
  \centering \resizebox{8.8cm}{!}{\includegraphics{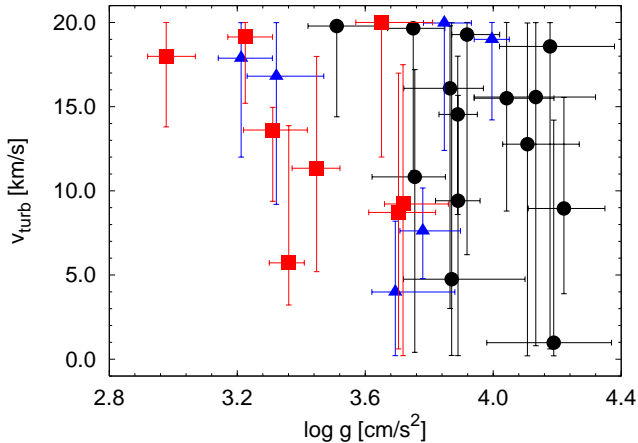}}
  \caption{Microturbulent velocities determined using line profile
    fits as a function of surface gravity. Symbols have the same
    meaning as in Fig.~\ref{fig:logg}.  For $\logg \lesssim 3.6$ a
    trend of increasing \vturb\ with decreasing \logg\ is visible. For
    larger gravities the uncertainties in the \vturb\ determinations
    are too large to discern any possible relation.}
\label{fig:vt_logg}
\end{figure}

The microturbulent velocities determined with the automated fitting
method are also given in Tab.~\ref{tab:results}. Though the error
estimates are relatively large (see Tab.~\ref{tab:errors}), we find
that for the current data set the \vturb\ determinations were
sufficiently accurate to reveal a weak correlation between this
parameter and the surface gravity. This is shown in
Fig.~\ref{fig:vt_logg}, where it can be seen that for $\logg \lesssim
3.6$ the average microturbulence recovered from the line profiles
increases systematically. The situation for $\logg \gtrsim 3.6$ is
less clear, as the error bars are on average larger and the values for
\vturb\ are more or less randomly distributed between 0 and
20~\kmsec. The reason for this is that for larger values of the
surface gravity the line profiles become intrinsically broader due to
the increased Stark broadening, making it more difficult to accurately
recover \vturb\ from the line profiles only.

Based on samples predominantly consisting of unevolved early B-type
Galactic stars other authors have also found a relation between
microturbulence and surface gravity, e.g.\ \cite{kilian91},
\cite{gies92} and \cite{daflon04}. More recently \cite{hunter06}
analysed a sample of early B-type stars in the Magellanic clouds and
also found a trend of increasing \vturb\ for decreasing \logg. To
derive the values for \vturb\ all these authors relied on
curve-of-growth techniques, which were applied to metal lines such as
those of \siiii\ and \oiii\ calculated using plane parallel
models. Consequently, our line profile based analysis is an
independent confirmation of the existence ({\em or requirement by lack
of a physical explanation or failures in the line broadening
mechanisms}) of microturbulence in the atmospheres of these type of
stars.

The physical mechanism explaining the observed microturbulence or its
relation to the surface gravity is still poorly understood.
\cite{kudritzki92} and \cite{lamers94} have argued that the observed
microturbulence might be the result of a stellar outflow, 
implying the existence of a velocity gradient in the photospheric
layers, which can mimic microturbulence-like desaturation effects.
\cite{smith98}, however, showed by applying a simple core-halo model
to the Galactic O9.7 supergiant HD~152003 that this effect would not
be sufficient to explain the observed \vturb. Indeed, our analysis
employing a unified photosphere and wind model confirms that the
microturbulence cannot be explained as an artifact of a transonic
velocity field.

A possible explanation for the gravity dependence could be related to
instabilities in the wind. These instabilities are reflected in the
large turbulent velocities ($\sim$100--200~\kmsec) necessary to fit
the wind lines in the intermediate and outer wind
\citep[e.g.][]{groenewegen89, haser98, evans04a}
and could be related to shocks due to the intrinsic line-driven
instability \citep{lucy83, owocki88,owocki99}. For a low surface
gravity the wind starts at larger Rosseland optical depth compared to
the high gravity case. Therefore, the line forming region of low
gravity objects contains a relatively large contribution originating
from (the base of) the wind. Consequently, this region could be
affected by the onset of the line-driven instability introducing
wind-turbulence into the line profiles.

\begin{figure}[t]
  \centering \resizebox{8.8cm}{!}{\includegraphics{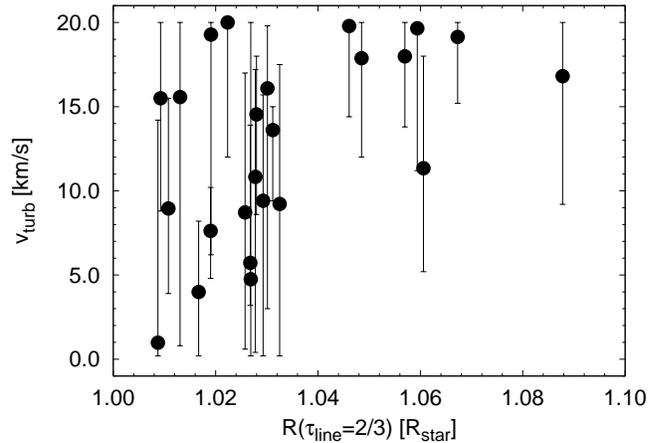}}
  \caption{Microturbulent velocity as a function of the location of
  the line forming region of \heil4471, which is defined as the
  location where the radial optical depth in the line core reaches a
  value of $\tau=2/3$. A weak trend is visible that suggests that
  for increasing extension of the atmosphere larger values of \vturb\
  are necessary to fit the line profiles.}
\label{fig:lfr_vt}
\end{figure}

Tentative support for the above described scenario may be implied by
Fig.~\ref{fig:lfr_vt}, where we show \vturb\ as function of the line
forming region of \heil4471 in units of the stellar radius. The
location of the line forming region is defined as the position at
which the radial optical depth in the line core reaches a value of
2/3. This figure shows that when the radial distance to this position
increases also the average microturbulent velocity increases. Though
the trend is weak, it does seem to indicate that when the atmosphere
becomes more extended, higher values of \vturb\ are necessary to
reproduce the line profiles. It may appear that this implies that the
line forming region enters in to the regime where wind turbulence
develops. However, for this statement no compelling evidence is
available, as we do not find any correlation between \vturb\ and the
distance between the line forming region and (for instance) the sonic
point. Moreover, for most objects at $R(\tau_{\rm line}=2/3) >
1.04$ values of \vturb\ close to the maximum allowed value are
found. Consequently, these values should be interpreted as lower
limits. Therefore, we can only conclude that our analysis points
towards a gradient in the turbulent velocity, possibly connected to a
link between microturbulence and wind instabilities, and suggest
further investigation in this direction.

\subsection{Wind parameters}
\label{sec:wind_param}

The wind parameters and their uncertainties determined using the
automated method are listed in Tabs.~\ref{tab:results} and
\ref{tab:errors}. Compared to our SMC analysis we find that we were
able to accurately determine these parameters for a significantly
larger number of objects. This is mainly the result of an on average
higher signal-to-noise ratio of the spectra as well as of denser winds
for the LMC objects compared to their SMC counterparts. In total we
determined 22 mass loss rates and 6 upper limits.  The upper limits
are defined (and can be identified in Tab.~\ref{tab:errors}) by an
error bar $-\log \mdot > 1.0$~dex.

To provide a meaningful comparison of the mass loss rates we place the
LMC objects in the modified wind momentum luminosity diagram. This
diagram shows as function of stellar luminosity the distribution of
the so-called modified wind momentum, which is defined as $\Dmom
\equiv \mdot \vinf R^{1/2}_{\star}$. Not only does this allow for an
assessment of the behaviour of \mdot\ within our sample, it also
provides a convenient method to compare the observed wind strengths to
the predictions of line driven wind theory. According to this theory
\Dmom\ is predicted to behave as
\begin{equation}
  \log \Dmom = x \log \left(\lstar/\lsun\right) + \log D_{\circ}~,
\end{equation}
where \lstar\ is the stellar luminosity \citep{kudritzki95,
puls96}. In this equation $x$ is the inverse of the slope of the
line-strength distribution function corrected for ionisation effects
\citep{puls00}.  The vertical offset $D_{\circ}$ is a measure for the
effective number of lines contributing to the acceleration of the
outflow.

\begin{figure}[t]
  \centering \resizebox{8.8cm}{!}{\includegraphics{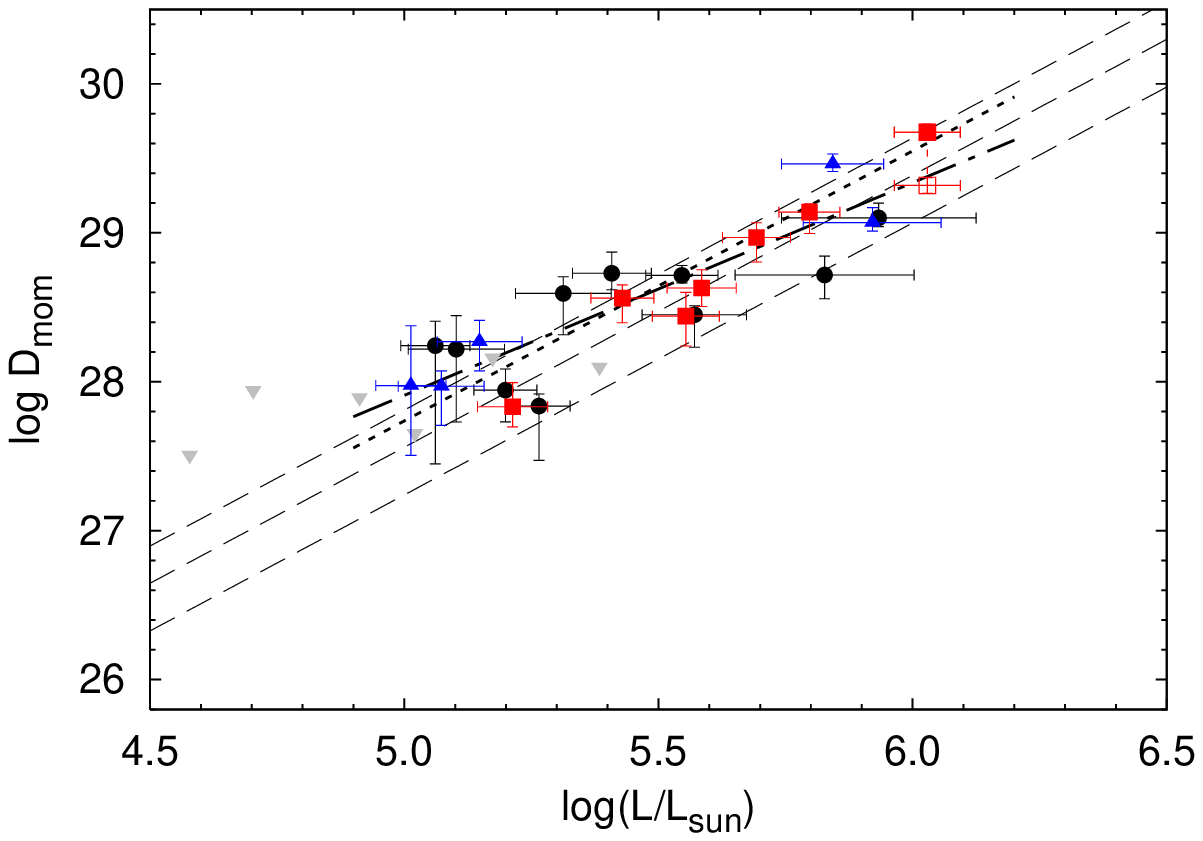}}
  \caption{Modified wind momentum (\Dmom) in units of ${\rm
  g\,cm\,s^{-2}}\,\rsun^{1/2}$ vs.\ luminosity. Symbols have the same
  meaning as in Fig.~\ref{fig:logg}.  Upper limits are shown as
  inverted triangles. 
  Dashed lines correspond to the predicted wind-momentum luminosity
  relations (WLR) from \cite{vink00, vink01}. The upper, middle and
  lower relation, respectively, correspond to predictions for
  Galactic, LMC and SMC metallicity.  The observed modified wind
  momenta show a strong correlation with luminosity with an average
  relation that lies in between the predicted Galactic and SMC
  WLR. This is quantified by the empirical WLR that was constructed
  for the LMC objects and that is shown by the dotted line. The open
  square corresponds to the wind momentum corrected for clumping of
  the supergiant Sk $-67$~166. Shown as a dashed-dotted is the
  empirical WLR obtained using this corrected \Dmom.}
\label{fig:wlr}
\end{figure}

In Fig.~\ref{fig:wlr} the distribution of the modified wind-momenta
for our programme stars are presented. Indicated using circles,
triangles and squares are objects of, respectively, luminosity class
V, III and I-II. Upper limits are shown as grey inverted triangles. A
clear correlation between $\lstar$ and $\Dmom$ can be observed in this
figure. Over an order of magnitude in $\lstar$ the average modified
wind-momentum decreases by approximately 1.5~dex. A comparison of the
behaviour of \Dmom\ with predictions is facilitated by the theoretical
WLRs calculated by \cite{vink00, vink01} that are shown as a set of
dashed lines. The upper, middle and lower of these predicted power laws
were calculated for, respectively, Solar, LMC and SMC
metallicity. Compared to these predictions we find that the LMC
wind-momenta are approximately confined between the theoretical WLR
for Galactic and SMC metallicity. In other words, compared to the
Galactic and SMC case, stars in the LMC have intermediate wind
strengths.

To further quantify the behaviour of LMC winds relative to that of
Galactic and SMC outflows we have fitted a power law to the observed
modified wind-momentum distribution, consistently accounting for both
the symmetric errors in \lstar\ and the asymmetric errors in \Dmom,
This yielded the following empirical WLR
\begin{equation}
  \log \Dmom =  (1.81 \pm 0.18) \log \left(\lstar/\lsun\right)
               + (18.67 \pm 1.01)~.
\end{equation}
The theoretical LMC relation from \citeauthor{vink00} is given by
$x=1.83$ and $\Dmom = 18.43$. The error bars of the
theoretical and empirical relations are in agreement. More
importantly, in Fig.~\ref{fig:wlr} the empirical relation, shown as a
dotted line, is found to lie between the predicted Galactic and SMC
relations. These predictions have been found to be in good agreement
with the observed Galactic WLR (\citealt{repolust04},
\citetalias{mokiem05}) and observed SMC WLR
\citepalias{mokiem06}. {\em Consequently, our empirical LMC WLR is
quantitative evidence for the fact that massive stars in this system
have mass loss rates intermediate between those of massive stars in
the Galaxy and SMC.}

The differences between the empirical and theoretical LMC WLR at the
start and end of the observed luminosity range are, respectively, 0.17
and 0.16~dex. In Fig.~\ref{fig:wlr} this seems to imply a systematic
offset between the two relations. However, we note that these
differences are still smaller than the typical uncertainty in \Dmom\
of 0.2~dex. More importantly, no correction was applied for the
possibility that the winds of our sample stars are, in contrast to our
assumption, not smooth but structured. In recent years evidence
has been mounting for this so-called clumping in the winds of O- and
early B-type stars. In particular spectroscopic modelling of UV
(resonance) lines \citep{crowther02, hillier03, bouret03, massa03,
martins04, martins05b, bouret05, fullerton06} seems to suggest the
existence of clumping factors in the range of 10-100, implying
corresponding reductions of \mdot\ by factors 3 up to 10. 
More recently, \cite{puls06} also found clumping factors of the order
5 to 10 (normalized to the unknown clumping properties in the
outermost, radio-emitting wind) from the analysis of \ha, infrared,
millimetre and radio fluxes.
In the present study we try to account for possible wind clumping
effects by correcting the mass loss rates of stars with \ha\ in
emission.  \cite{markova04} and \cite{repolust04} argue that the mass
loss rates of these stars could be overestimated as a result of the
fact that \ha\ emission lines are formed over a relatively large
volume where clumping might have set in. In contrast, for stars with
\ha\ in absorption the line is formed relatively close to the stellar
surface, where clumping effects are negligible. Based on the
comparison of dwarfs and supergiants in their Galactic sample
\citeauthor{repolust04} derived a numerical correction factor of 0.44 for the
mass loss rates of supergiants with \ha\ in emission. 

We have applied the clumping correction to the super giant Sk $-67$
166, which has a \ha\ emission profile. In Fig.~\ref{fig:wlr} its new
wind momentum is indicated using an open symbol. Using this value the
following empirical WLR is obtained\footnote{The small errors in the
relevant parameters of Sk $-67$ 166, which dominates the WLR at very
high luminosity, cause the significant difference between Eqs.\,2 and
3. Note that our least square fitting does not account for the
correlation of both quantities (due to \rstar). In view of the well
known distance these effects are likely small compared to the Galactic
case \cite[see][]{markova04,repolust04}.}
\begin{equation}
  \log \Dmom = (1.43 \pm 0.17) \log \left(\lstar/\lsun\right) + (20.77
               \pm 0.97)~.
\end{equation}
In Fig.~\ref{fig:wlr} we see that for $\log \lstar/\lsun \gtrsim 5.3$
the new WLR compares better to the \citeauthor{vink01} relation. For
lower luminosities the situation is less clear. Due to the large
uncertainties, this range has a relatively low weight in the
fit. Consequently, a discrepancy for low \lstar\ is less significant
than the good agreement obtained for the higher luminosities. For this
reason it is difficult to investigate the existence of a ``weak wind
problem'' for stars at $\log \lstar/\lsun \lesssim 5.3$, first
reported by \cite{bouret03}. This study, as well as later studies
\citep{hillier03, evans04b, martins04, martins05b} report a steepening
in the WLR relation relative to predictions starting at about the
above mentioned \lstar, leading to an over prediction of the wind
strength by up to a factor 100 at $\log \lstar/\lsun \sim 4.5$.  Our
LMC results do not appear to confirm this break between observations
and seem to follow the predictions down to $\log \lstar/\lsun \approx
5.0$. In a forthcoming paper we will present a comprehensive overview
of the observed WLR relations in our Galaxy and the Magellanic Clouds
as well as a thorough discussion of the successes and failures of the
theory of radiation driven winds in predicting the WLR, including
possible causes for the weak wind problem (Mokiem et al.\ in
preparation).

\section{The mass discrepancy}
\label{sec:mdisc}

\begin{figure}[t]
  \centering
  \resizebox{8.8cm}{!}{ \includegraphics{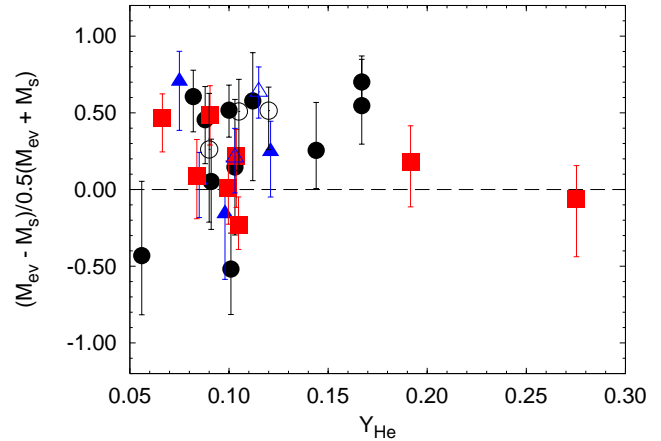}}
  \caption{Mass discrepancy as a function of helium abundance for the
  LMC sample. Evolutionary masses used to calculate the discrepancy
  were derived from the non-rotating tracks of \cite{schaerer93}.
  Symbols have the same meaning as in Fig.~\ref{fig:logg}. The open
  symbols correspond to stars with $\teff \geq 45$~kK.}
  \label{fig:massdisc}
\end{figure}

In Fig.~\ref{fig:massdisc} we investigate the mass discrepancy of our
LMC targets as a function of the helium surface abundance. On the
vertical axis a measure for this discrepancy is shown, which is scaled
to the mean of the evolutionary and spectroscopic mass. This ensures
that positive and negative discrepancies follow the same linear
scale. For non-enriched stars, i.e.\ $\yhe \lesssim 0.10$,
approximately three times as many objects lie above the one-to-one
correlation than below it. Therefore, {\em in contrast to our finding
for the equivalent SMC case \citepalias{mokiem06} a significant mass
discrepancy is found for our sample of non-enriched LMC stars}. The
reason for this is unclear. As we stated before our analysis employs
state-of-the-art atmosphere models, and is in principle not hampered,
as were previous studies, by potentially unoptimised fits. As stated,
in our SMC dataset, analysed in an identical manner, no evidence was
found for a mass discrepancy for $\yhe < 0.10$.

\cite{massey05} also study the \Ms\ vs. \Mev\ problem in a set of LMC
stars. For objects hotter than 45~kK they find a mass discrepancy that
is even stronger than what we find. This behaviour could be the result
of an underestimate of the photospheric line pressure in this high
temperature regime. In Fig.~\ref{fig:massdisc} we have highlighted the
objects with $\teff \geq 45$~kK using open symbols. Although their
mass discrepancy is considerable they do not stand out as a separate
group.To further illustrate this, note that the hottest supergiant in
our sample Sk~$-67$~166 ($\teff = 40$~kK) at $\yhe = 0.28$ has a \Ms\
of 75~\msun\, which is in very good agreement with its \Mev\ of
70~\msun. Consequently, though we can not explain the mass discrepancy
we do not anticipate that it is connected to a flawed treatment of the
photospheric radiation pressure that manifests itself at metallicities
as high as that of the LMC environment (but not yet at values typical
for the SMC).

If we accept the analysis of stellar and photospheric parameters, the
presence of a mass discrepancy may point to an oversimplified picture
of the evolution of massive stars used to determine \Mev\ and/or,
possibly, to a breakdown of the assumption of a spherically symmetric
atmosphere.  One obvious simplification may be that we have used
tracks for non-rotating stars. One of the effects introduced by
rotation is a wide bifurcation in the evolutionary tracks
\citep{maeder87}. Stars rotating faster than roughly half the surface
break-up velocity will essentially follow tracks representative of
homogeneous evolution.  \cite{langer92} showed that as a result of
this the $M/L$-ratio will be a monotonically decreasing function for
increasing helium enrichment. Consequently, stars evolving along
homogeneous tracks are expected to be increasingly under-massive for
increasing helium abundance. One would therefore derive a positive
mass discrepancy if the evolution of a rapidly rotating star was
incorrectly described using non-rotating or modestly rotating model
tracks.

So, can rapid rotation be used to explain the observed mass
discrepancy?  Let us first focus on the supergiants. For these stars
the mass discrepancy problem appears absent, though we note two
exceptions at $\yhe = 0.07$ and $\yhe = 0.09$.  This absence of signs
of a distinct mass discrepancy was also found in our study of SMC
bright giants and supergiants \citepalias{mokiem06}. The conclusion
seems to be that the supergiants follow non-rotating or modestly
rotating evolutionary tracks. This implies that, apparently, all
supergiant targets that we have selected in both the SMC
\citepalias{mokiem06} and LMC happen to have started out their
evolution with low or modest initial rotational velocities. As the
initial rotational velocity distribution derived in
\citetalias{mokiem06} for the NGC\,346 cluster implies that only some
$5-15$ percent of stars are expected to evolve along homogeneous tracks
this should not be alarming. One supergiant, however, might not
fit this picture: Sk -67 166. This source does not feature a mass
discrepancy, but does show an increased helium abundance. Given its
position in the HRD this is quite difficult to explain in term of evolution
without rotation. A scenario accounting for a more delicate interplay
between mass loss and rotation may be required to explain its
properties \cite[see][]{herreroIAUS215}.

Now let us turn to the dwarfs and giants.  For our SMC sample we found
a correlation between the helium surface enrichment and the mass
discrepancy for class V and III stars, with \Ms\ being systematically
smaller than \Mev\ for $\yhe \gtrsim 0.11$. Interestingly, this
dependence was only found for dwarfs and giants (see above).  It was
suggested that this behaviour was the result of rotationally enhanced
mixing, enriching the atmospheres with primary helium. The dwarfs and
giants at $\yhe > 0.11$ displayed in Fig.~\ref{fig:massdisc} also
systematically suffer from a positive mass discrepancy. Consequently,
this scenario also seems to be a good explanation for the situation in
the LMC sample. However, we realise that these seven objects are a
relatively small sample and that the typical uncertainties in \yhe\
are 0.03. This statement should therefore be seen as a working
hypothesis.

For helium enriched dwarfs and giants the above scenario seems a
logical one, as helium enrichment early on in the evolution must imply
efficient mixing. For the dwarfs and giants that do not show excess
helium in their surface layers such a clue or indication is not
present (unless they evolve left of the main sequence as do chemically
homogeneous stars, see Sect.\,\ref{sec:O2}).  In principle these non
\yhe-enriched stars could rotate sufficiently rapid to cause a mass
discrepancy, but not to the extent of chemically homogeneous
evolution. However, if so, a very significant fraction of the stars
should have started their evolution at super-critical rotation. At
least for the stars studied in the SMC cluster NGC\,346
\citepalias{mokiem06} this is not the case.

\subsection{Chemically homogeneously evolving O2 stars}
\label{sec:O2}

Three out of the four O2 stars are found to lie to the left of the
ZAMS in Fig.~\ref{fig:hrd} and exhibit relatively large mass
discrepancies (see open symbols in Fig.~\ref{fig:massdisc}). These are
N11-026, BI~237 and BI~253. For a discussion of the effective
temperature of the fourth O2 object, N11-031 ($\teff \sim 45$\,kK), we
refer to Sect.~\ref{sec:teff}. Even though the hottest three ($\teff
\sim 53-54$\, kK) are not significantly enriched in helium it is
tempting to speculate on a possible (near) homogeneous evolution of
these objects. This would not only provide an explanation for their
mass discrepancy, it would also explain their peculiar location in the
HRD. The possibility that these are true ZAMS stars is very
exciting. However, due to the short evolutionary time scales in this
part of the HR-diagram, it would also be unlikely.  Based on the
non-rotating model tracks of \cite{schaerer93} we estimate that a 60~\msun\
star will within 2~Myr evolve away from the ZAMS to a location in the
HRD at $\teff \approx 45$~kK. This is well beyond the error bars on
the parameters of these objects.
Given the fact that N11-026 and BI~253 are thought to be associated
with the LH~10 and 30~Doradus clusters, respectively, which have ages
of approximately 3 (see Sect.~\ref{sec:agesLH}) and 2~Myr
\citep{dekoter98}, a normal evolutionary scenario appears unlikely.
Based on its large radial velocity the field star BI~237 probably is a
runaway star \citep{massey05}, therefore it is also likely to be
relatively old.

The problems with a reconciliation of the hottest three O2 stars with
fully homogeneous evolution and an age of at least 2 Myr, are {\em i)}
that their surface helium abundances are not significantly enriched,
and {\em ii)} that their rotational velocities are not extreme. The
latter issue need not be ``a smoking gun'' considering the possibility
that we may see them relatively pole-on and the fact that their
\vsini\ values, ranging between 110-190 \kmsec, are above the sample
average. Concerning the first point, it appears that we must concede
to the possibility that stars may evolve along tracks similar to those
for homogeneous evolution while in fact they are not {\em fully}
homogeneous -- i.e. the near surface layers are not yet strongly
affected by the mixing that must occur deeper in. Having said this, we
do point to the fact that the error bars on \yhe\ do allow for
relatively large ages {\em even within the hypothesis of fully
homogeneous evolution}. Within the error bars the maximum \yhe\ is
$\sim$ 0.13 for all three stars. It takes fully homogeneously evolving
stars of 60 and 40~\msun\ about $\sim 1.5$ and $\sim$2.0~Myr,
respectively, to build up this amount of helium enrichment
(S.\,-C. Yoon, private communication). This is close to the derived
cluster ages of LH~10 and 30~Doradus. For reference, if the star would
evolve along near-homogeneous tracks they would be older.

We tentatively conclude that the HRD position, helium abundance, and
rotational velocity of the three hottest O2 stars in our sample could
be consistent with (near-)homogeneous evolution. Evidence for
efficient rotation-induced mixing during the main sequence phase of O
stars has also been presented by \cite{lamers01}, based on the
chemical abundance pattern of ejected circumstellar nebulae.

\begin{figure}[t]
  \centering
  \resizebox{8.8cm}{!}{ \includegraphics{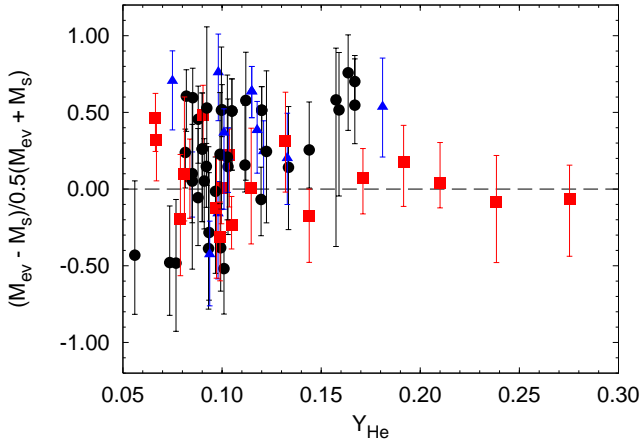}}
  \caption{Mass discrepancy as a function of helium abundance for the
  combined samples from \citetalias{mokiem05} (Galactic),
  \citetalias{mokiem06} (SMC) and the current LMC sample.  Symbols
  have the same meaning as in Fig.~\ref{fig:logg}.  Note the good
  correspondence between \Ms\ and \Mev\ for the bright giants and
  supergiants (square symbols).}
  \label{fig:tot_massdisc}
\end{figure}

\subsection{Large sample trends}

To firmly establish our findings with respect to the mass discrepancy
problems, we compare in Fig.~\ref{fig:tot_massdisc} spectroscopic and
evolutionary masses for all stars that have been analysed using our
automated fitting method so far. Presented in this figure are the
combined Galactic \citepalias{mokiem05}, SMC \citepalias{mokiem06} and
the current LMC samples, corresponding to a total of 71 O- and early
B-type stars.

\begin{figure*}[t]
   \centering
   \caption{The FLAMES field for the star forming region N11 with the
   associations LH9 (south of centre) and LH10 (north of centre). The
   FLAMES targets are identified using circles and their star
   identification is given.  From \cite{evans06}.  [Included as separate JPG file in 
   astro-ph submission.]}
   \label{fig:N11field}
\end{figure*}

The figure confirms our two main findings. First, the good agreement
between \Ms\ and \Mev\ for the supergiants is clearly visible. This is
an encouraging finding and we believe that this means that the
improvements in the stellar atmosphere models, evolutionary
calculations and spectral analysis techniques have finally resolved
the long standing mass discrepancy as found by \cite{herrero92}. The
correspondence between the spectroscopic and evolutionary mass scale,
in particular for increasing helium enrichment is striking. The reason
for this is probably related to the fact that the class I-II objects
are found in the region of the HR-diagram in which stars are not
expected to undergo extreme evolutionary phases, such as homogeneous
evolution.  Consequently, the enriched bright giants and supergiants
have evolved along relatively simple evolutionary tracks, which (in
this respect) appear well understood.

Our second finding, i.e. that of a correlation of mass discrepancy
with the helium abundance, is also corroborated by the large sample in
Fig.~\ref{fig:tot_massdisc}. For $\yhe > 0.10$ all dwarfs and giants,
except for one dwarf, are found above the $\Ms = \Mev$ line. In all
fairness, given the typical uncertainty of 0.03 in \yhe, from a
statistical point of view the region $0.09 < \yhe < 0.12$ should be
regarded with care. For larger helium abundance, however, the
correlation can be regarded to be statistically significant. A total
of 11 objects show a positive mass discrepancy, with only a single
counter example. Moreover, the magnitude of the discrepancy seems to
be related to the amount of enrichment. This strongly points to
efficient mixing in the main sequence phase, leading to
(near-)chemically homogeneous evolution.

\section{The evolutionary status of N11}
\label{sec:age}

In this section we explore the evolutionary status of the N11
field. We will first briefly outline the current understanding of N11
with respect to the OB associations LH9 and LH10 and in particular the
sequential evolutionary link between the two. Based on the analysis of
the 22 stars in our sample associated with these clusters we will then
estimate their ages and discuss whether they are compatible with a
sequential star formation scenario.

\subsection{LH9 and LH10 in N11}

\begin{figure*}[t]
   \centering
   \resizebox{16cm}{!}{\includegraphics{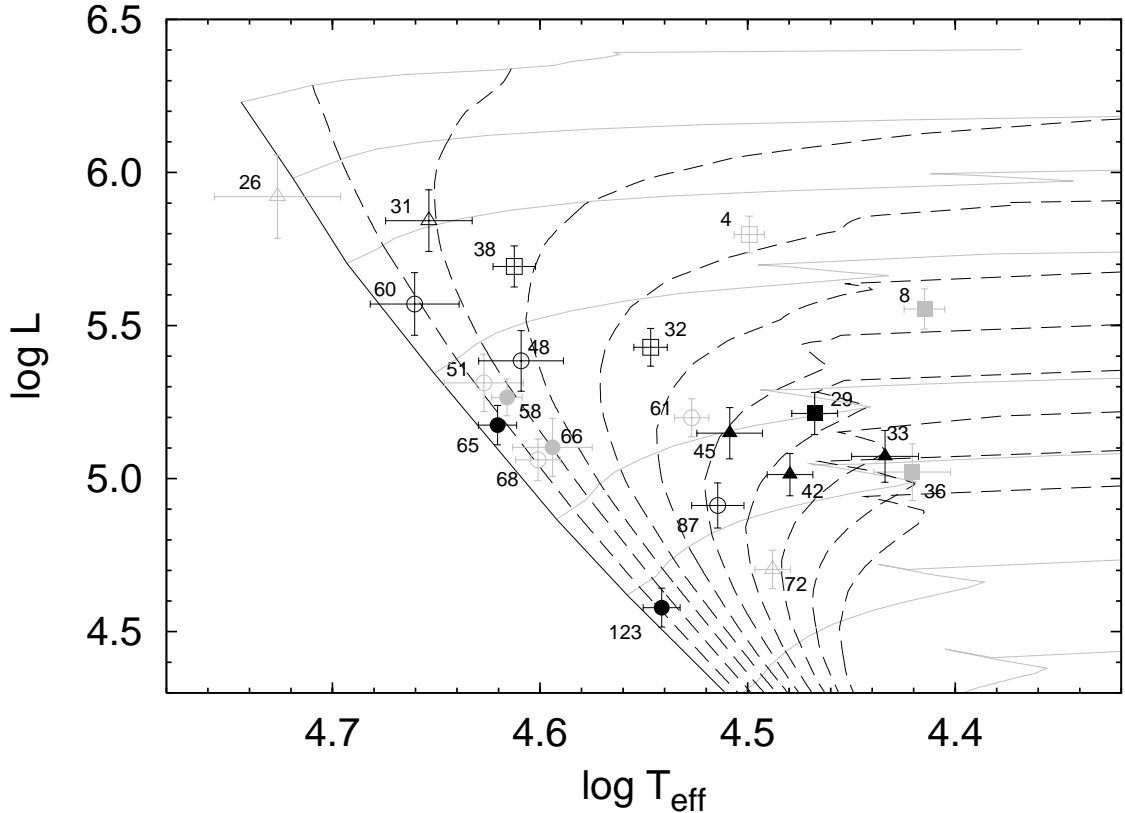}}
   \caption{Comparison of the programme stars located in N11 with
   isochrones derived from the evolutionary tracks of
   \cite{schaerer93}. The symbols used to denote dwarf, giants and
   bright- and supergiants are, respectively, circles, triangles and
   squares. Filled and open symbols are used to distinguish stars
   associated with LH9 and LH10, respectively. Black symbols refer to
   objects located within a radius of two arc minutes of the LH9 and
   LH10 core, whereas the grey symbols correspond to stars found
   outside this radius. Labels correspond to the N11 identification
   numbers given in Tab.~\ref{tab:data}. Isochrones (dashed lines)
   are shown from one up to ten million years with 1~Myr
   intervals. For reference evolutionary tracks from \cite{schaerer93}
   are drawn as grey lines, where for clarity purpose only the
   blueward part of the evolution for the most massive tracks is
   shown. The ZAMS corresponding to these tracks is given by the black
   solid line.}  \label{fig:hrd_age}
\end{figure*}

N11 is an intricate giant \hii\ region containing several massive star
forming complexes. The largest of these are the OB associations LH9
and LH10 \citep{lucke70}. An image of the region, including the
adopted star identifications, is presented in Fig.~\ref{fig:N11field}
\citep{evans06}. Several studies have suggested that the star forming
activity in these associations are linked. In particular the study by
\cite{parker92} has provided a key to understanding of the structure and
formation of LH9 and LH10. Their analysis of the stellar content of
N11 revealed the presence of several O3-O5 stars and possible ZAMS
stars in LH10. In contrast the earliest spectral type associated with
LH9 was found to be O6. They also found
the slope of the initial mass function of LH10 to be significantly
flatter than that of LH9, indicating that the former contains a higher
ratio of high mass to low mass stars. Combined with the fact that the
reddening of LH10 is larger than that of LH9, it was concluded that
LH10 is the younger of the two clusters. Based on these findings the
authors propose an evolutionary link between the two, where star
formation in LH10 could possibly be triggered by the stellar winds and
supernovae of massive stars in LH9.

Further evidence for a sequential star formation scenario was given by
\cite{walborn92}, who found a dual structural morphology of N11
analogous to that of the 30~Doradus \hii\ region. In the latter
substantial evidence suggests the presence of {\em current star
formation} in regions surrounding the central star cluster (e.g.\
\citealt{walborn87, hyland92}, also see \citealt{walborn97}). This
secondary burst seems to be set off approximately 2~Myr after the
initial star formation took place, possibly initiated by the energetic
activity of the evolving cluster core. \citeauthor{walborn92} argue
that this is very similar to N11 where the stellar content of LH9 and
LH10 also suggests an age difference of $\sim$2~Myr \citep[also
see][]{walborn99}. The process in N11, however, would be advanced by
$\sim$2~Myr, classifying it as an evolved 30~Doradus analogue, though
less massive.

More recently, several bright IR sources showing characteristics of
young stellar objects were discovered in the N11B nebula surrounding
LH10 by \cite{barba03}. These objects are probably intermediate mass
(pre-)main-sequence Herbig Ae/Be stars belonging to the same
generation as do the LH10 objects (the pre-main-sequence evolutionary
timescales of intermediate mass stars being longer than that of
massive OB stars). \citeauthor{barba03} also found that the massive
stars in LH10 have blown away their ambient molecular material and are
currently disrupting the surface of the parental molecular cloud
material surrounding LH10 (i.e. the material in N11B).

\subsection{The ages of LH9 and LH10}
\label{sec:agesLH}

To estimate the ages of our programme stars we compare their location
in the HR-diagram to theoretical isochrones in Fig.~\ref{fig:hrd_age}.
The luminosity classes V, III and I-II are represented using,
respectively, circles, triangles and squares. To differentiate between
stars associated with LH9 and LH10, respectively, filled and open
symbols are used. Membership is defined on the basis of minimum
distance to either the LH9 or LH10 cluster core. This, therefore,
should be taken with some care: the cores are about 4 arcmin or 60
parsec apart, which can be traversed in 2 Myr if the proper motion of
the star is some 30~\kmsec. Runaway O- and B-type stars have typical
velocities of 50--100~\kmsec, therefore, it is entirely possibly that
(a few) individual objects are assigned to the wrong association.
For position reference see Fig.~\ref{fig:N11field}. Isochrones in
Fig.~\ref{fig:hrd_age} are shown as dashed lines for one up to ten
million years with 1~Myr intervals and were derived from the
evolutionary tracks of \cite{schaerer93}. Note that these tracks do
not account for the effects of rotation.

The distribution of the stars in Fig.~\ref{fig:hrd_age} is such that
they can be separated in a group of objects younger than three million
years and a group of objects older than this age. We also see that the
oldest objects are predominantly found in LH9, whereas LH10 contains
the largest fraction of young objects, suggesting that the clusters
can indeed be separated in terms of age. At first sight it seems that
they can be characterised by an age of $\sim$7~Myr and $\sim$2~Myr,
respectively. However, a significant number of objects in both
clusters are found to be several million years younger and older than
these preliminary ages. A possible explanation for this large age
scatter could be ``contamination'' by field stars. To assess this
possibility we have assigned grey symbols to all objects outside a
radius of two arc minutes from the cluster cores. Disregarding these
objects reduces the age scatter significantly; still a number of stars
appear to contradict with the notion of two coeval populations.

\begin{figure}[t]
  \centering
  \resizebox{8.8cm}{!}{\includegraphics{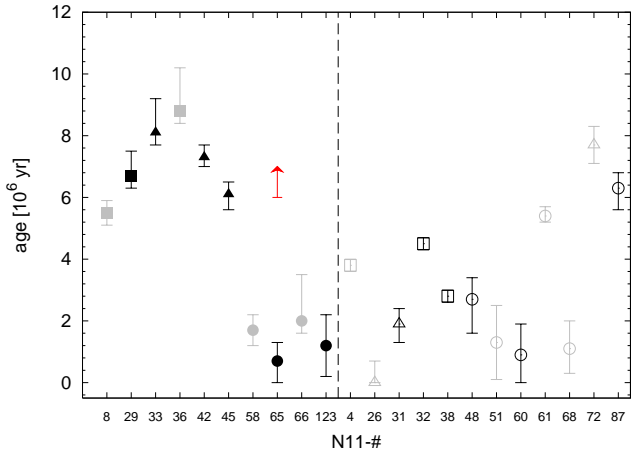}}
  \caption{Individual age estimates for stars in N11 based on the
  non-rotating tracks of \cite{schaerer93}. Symbols have the same
  meaning as in Fig.~\ref{fig:hrd_age}. The left and right part of the
  diagram, respectively, contain stars associated with LH9 and LH10
  and are separated using a dashed line. The upward pointing arrow
  denotes a lower age limit for N11-065 determined from its surface
  helium abundance using chemically homogeneous evolutionary models
  \citep{yoon06, yoon05}.}
  \label{fig:age_indv}
\end{figure}

To investigate the age distributions in more detail the individual age
estimates are shown in Fig.~\ref{fig:age_indv}. LH9 and LH10 objects,
which are shown using the identical symbols as in
Fig.~\ref{fig:hrd_age}, are placed in, respectively, the left and
right part of the diagram and are separated using a dashed line.
First concentrating on the LH9 objects we see that they are
near-coeval with exception of four dwarfs. Of these the two non-core
members are located at a distance of approximately six arc minutes
from the central concentration. Therefore, it is probable that they
are either spatially not related to N11 or that possibly their
formation was triggered more recently by the stellar activity in
LH9. The third dwarf N11-065 occupies a location very close to the
ZAMS and in principle would be the youngest member of LH9. However, we
also find a considerable helium enrichment for this object (\yhe =
0.17). Combined with the fact that N11-065 has a large mass
discrepancy, this suggests that it might be evolving chemically
homogeneously. Consequently, a more appropriate age estimate should be
derived from tracks appropriate for this kind of evolution. Adopting
such tracks from \citet[][also see \citealt{yoon05}]{yoon06} we
derive a lower limit of 6~Myr for the age of N11-065 based on the
surface helium abundance. In Fig.~\ref{fig:age_indv} this estimate is
indicated using an upward pointing arrow, and is in good agreement
with the bulk of the LH9 stars.

The dwarf N11-123 exhibits no chemical peculiarities and its location
on the sky places it in the central concentration of LH9, suggesting
that it should have been formed in the burst of star formation that
formed the cluster. However, given the fact that N11-123 is in
principle the only star deviating from the coeval nature of LH9 we
suspect that it is like N11-058 and N11-066 part of the periphery of
the cluster and that its relatively close position to the core is due
to a projection effect. Consequently, we conclude that the central
concentration of LH9 is coeval with an age of
$\sim$7.0~$\pm1.0$~Myr. The error bar is of the same order of
magnitude as the 0.5--1.0~Myr introduced by uncertainties in the
evolutionary tracks due to effects of (relatively modest) rotation
\citepalias[see][]{mokiem06}.

The right part of Fig.~\ref{fig:age_indv} shows that the stars in
the central concentration of LH10 have ages ranging from one up to
approximately six million years. Despite this large scatter it is
clear that the majority of the stars are younger than 4.5~Myr and that
only one object (N11-087) in the cluster core is older than this
age. An explanation for the large age of the latter object might be
that it formed in LH9 and, over time, migrated towards LH10. As
explained at the start of this section, this is a possibility given
that the distance from the centre of LH9 to the current position of
N11-087 is only three arc minutes.  Considering the error bars on the
age determinations and the additional uncertainty of 0.5--1.0~Myr
introduced by rotation \citepalias[see][]{mokiem06}, we finally
estimate an age best describing LH10 of $\sim$3.0~$\pm1.0$~Myr.

\subsection{Sequential star formation?}
If the starbursts in N11 are sequential the formation of LH10 should
be induced by supernova explosions and/or stellar winds in LH9. Given
the age difference and distance between the two clusters we find that
both means of triggering are possible. In the first case the age
difference of approximately four million years is compatible with the
time of $\sim$3~Myr it takes before the most massive stars end their
life in a supernova \citep[e.g.][]{schaller92}. The time needed for
the supernova shock to cross the distance of $\sim$3 arc minutes
between LH9 and LH10, corresponding to approximately 40 parsec at the
distance of the LMC, is only $\sim$$10^5$ years \citep[see
e.g.][]{falle81}; hence it is a possible scenario. If we consider
triggering through stellar winds, the time scales are also compatible.
\cite{garcia-segura96}, for instance, using hydrodynamical simulations
have shown that a 60~\msun\ star can create a wind-driven bubble in
the interstellar medium of $\sim$50~pc during its main sequence
lifetime of 3~Myr.  Consequently,
this scenario seems to be appropriate as well and in agreement with
the determined time scales.

In view of the above we conclude that a sequential scenario for LH9
and LH10 seems very likely. A combination of supernovae and stellar
winds from stars in LH9 may have initiated star formation in LH10.

\section{Summary and conclusions}
\label{sec:sum}

We have analysed a sample of 28 massive OB-type stars located in
the LMC. For a homogeneous and consistent treatment of the data we
employed the automated fitting method developed by \cite{mokiem06},
which combines the genetic algorithm based optimisation routine
\pikaia\ \citep{charbonneau95} with the fast non-LTE unified stellar
atmosphere code \fastwind\ \citep{puls05}. The sample is mostly drawn
from the targets observed within the context of the VLT-FLAMES survey
of massive stars \citep[][]{evans05}. In total 22 of these stars
are located in the LH9 and LH10 clusters within the giant \hii\ region
N11. This region is believed to have been the scene of
sequential star formation, with the stellar activity in LH9 igniting
secondary starbursts in different associations in N11. Our main
findings are summarised below.
\begin{enumerate}
  \item[\it i)] The effective temperature per spectral sub-type of the
    LMC stars is found to be intermediate between that of Galactic and
    SMC O- and early B-type stars, with the LMC objects being,
    respectively, cooler and hotter by typically $\sim$2~kK.

  \item[\it ii)] Based on the helium and hydrogen lines it was
    possible to determine the effective temperatures, though with
    relatively large error bars, of the four O2 stars in our
    sample. Three of these are found to be hotter by more than 3
    to 7~kK (see Sect.\,\ref{sec:O2teff}) compared to the O3 star in
    our sample, suggesting that O2 stars indeed represent a hotter
    subgroup within the O-type class. However, we note that for 
    the one ON2 star a relatively low \teff\ of 45~kK was obtained,
    indicating that the \niv\ and \niii\ classification lines
    \citep{walborn02a} are not fully compatible with the helium lines
    traditionally used for classification.

  \item[\it iii)] The spectroscopically determined masses of the dwarf
    and giant stars in our set of programme stars are found to be
    systematically smaller than those derived from non-rotating
    evolutionary tracks. For helium enriched dwarfs and giants,
    i.e. those having $\yhe > 0.11$, we find that {\em all} show this
    mass discrepancy. The same was found in an analysis of SMC stars
    using the same methods \citep{mokiem06}. We interpret this as
    evidence for efficient rotationally enhanced mixing leading to the
    surfacing of primary helium and to an increase of the stellar
    luminosity.

  \item[\it iv)] The bright giants and supergiants do not show any mass
    discrepancy, regardless of the surface helium abundance. This also
    is consistent with the finding for Galactic and SMC class I-II
    objects studied with the same methodology \citep{mokiem05,
    mokiem06}. 
    This suggests that shortly after birth all these stars must have
    rotated at less than about 30 to 40 percent of the surface break-up
    velocity.

  \item[\it v)] A weak correlation is found between microturbulent
    velocity and surface gravity. More extended atmospheres (i.e.\
    lower gravity stars) require a relatively large \vturb\ to fit the
    lines. The reason for this relation is unclear, however, it does
    not seem to be connected to the lines being formed closer to the
    sonic point of the wind flow in low gravity stars.

  \item[\it vi)] From a comparison of modified wind momenta \Dmom\ we
    find that the wind strengths of LMC stars are weaker compared to
    Galactic stars, and stronger compared to SMC stars. Comparing the
    derived \Dmom\ as a function of luminosity with predictions for
    LMC metallicities by \cite{vink01} yields good agreement in the
    entire luminosity range that was investigated ($5.0 < L/\lsun <
    6.1$).

  \item[\it vii)] We have determined an age of, respectively,
    $\sim$7.0$\pm1.0$~Myr and $\sim$3.0$\pm1.0$~Myr for the clusters
    LH9 and LH10. The age difference and relative distances are in
    good agreement with a sequential star formation scenario, in which
    stellar activity in LH9 triggered the formation of LH10. 
\end{enumerate}

\begin{acknowledgements}
  We would like to thank Sung-Chul Yoon and Alexander van der Horst
  for constructive discussions, and George Meynet for providing us
  with the evolutionary models of the Geneva group. M.R.M.\
  acknowledges financial support from the NWO Council for Physical
  Sciences. S.J.S.\ acknowledges the European Heads of Research
  Councils and European Science Foundation EURYI (European Young
  Investigator) Awards scheme, supported by funds from the
  Participating Organisations of EURYI and the EC Sixth Framework
  Programme. A.H.\ and F.N.\ acknowledge support from the Spanish MEC
  through project AYA2004-08271-C02. JSV acknowledges RCUK for his
  Fellowship.  Spectral fits were calculated using the LISA computer
  cluster at SARA Computing \& Networking Services.
\end{acknowledgements}

\bibliographystyle{aa}

\begin{small}
\bibliography{6489}
\end{small}

\clearpage

\appendix
\section{Fits and comments on individual objects}

\label{sec:fits}
The observed spectra shown in this section were corrected for radial
velocities. If not noted differently the lines that were fitted are
the hydrogen Balmer lines \ha, \hg\ and \hd; the \hei\ singlet line at
4387~\AA; the \hei\ triplet lines at 4026, 4471 and 4713~\AA; and the
\heii\ lines at 4200, 4541 and 4686~\AA. Over plotted are the best
fit spectra, unless noted differently. For a discussion of the
line weighting scheme adopted in our fitting procedure we refer to
\citetalias{mokiem05} (especially to their Table 3.).

\begin{figure*}[t]
  \centering
  \resizebox{17cm}{!}{
  \includegraphics{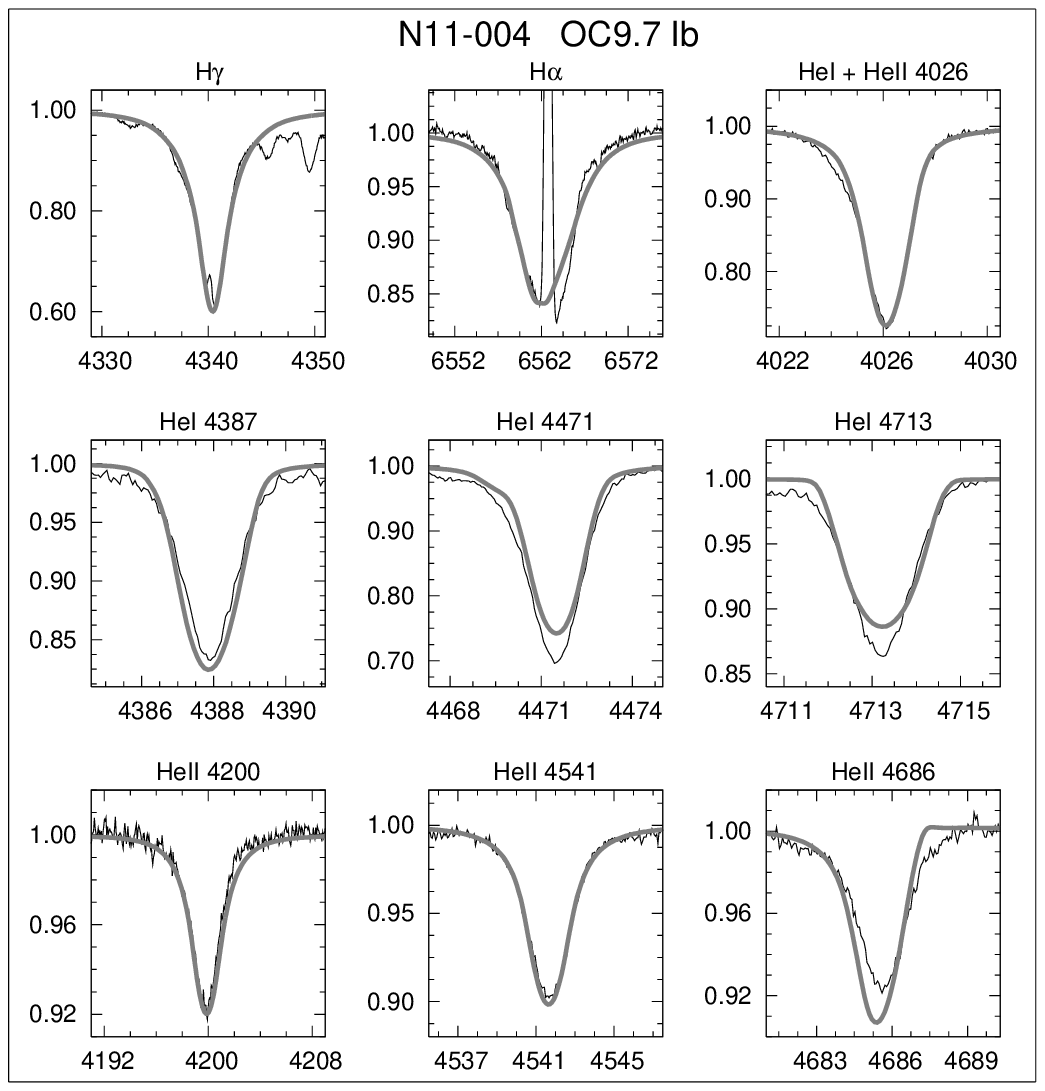}
  \includegraphics{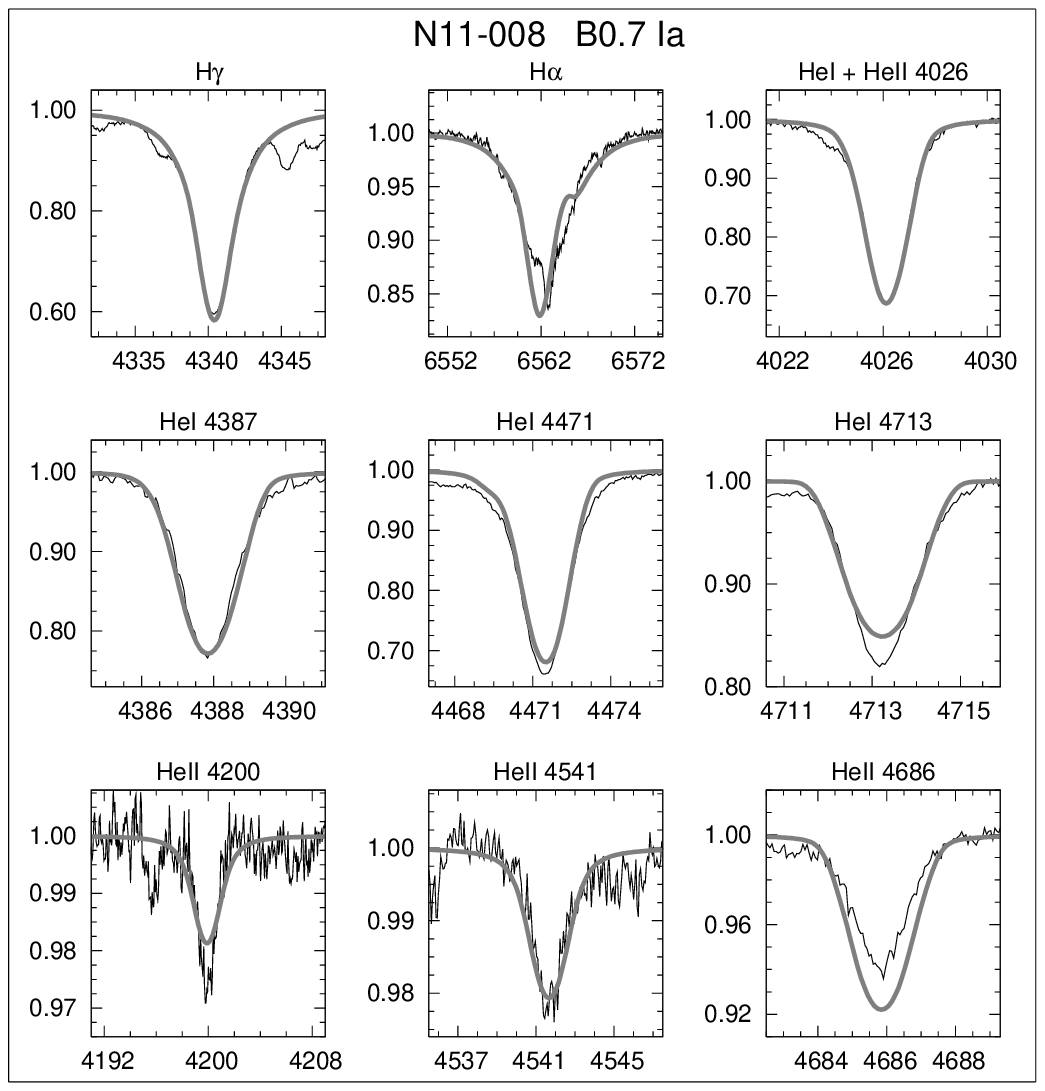} 
                     }

  \resizebox{1cm}{!}{ }

  \resizebox{17cm}{!}{
    \includegraphics{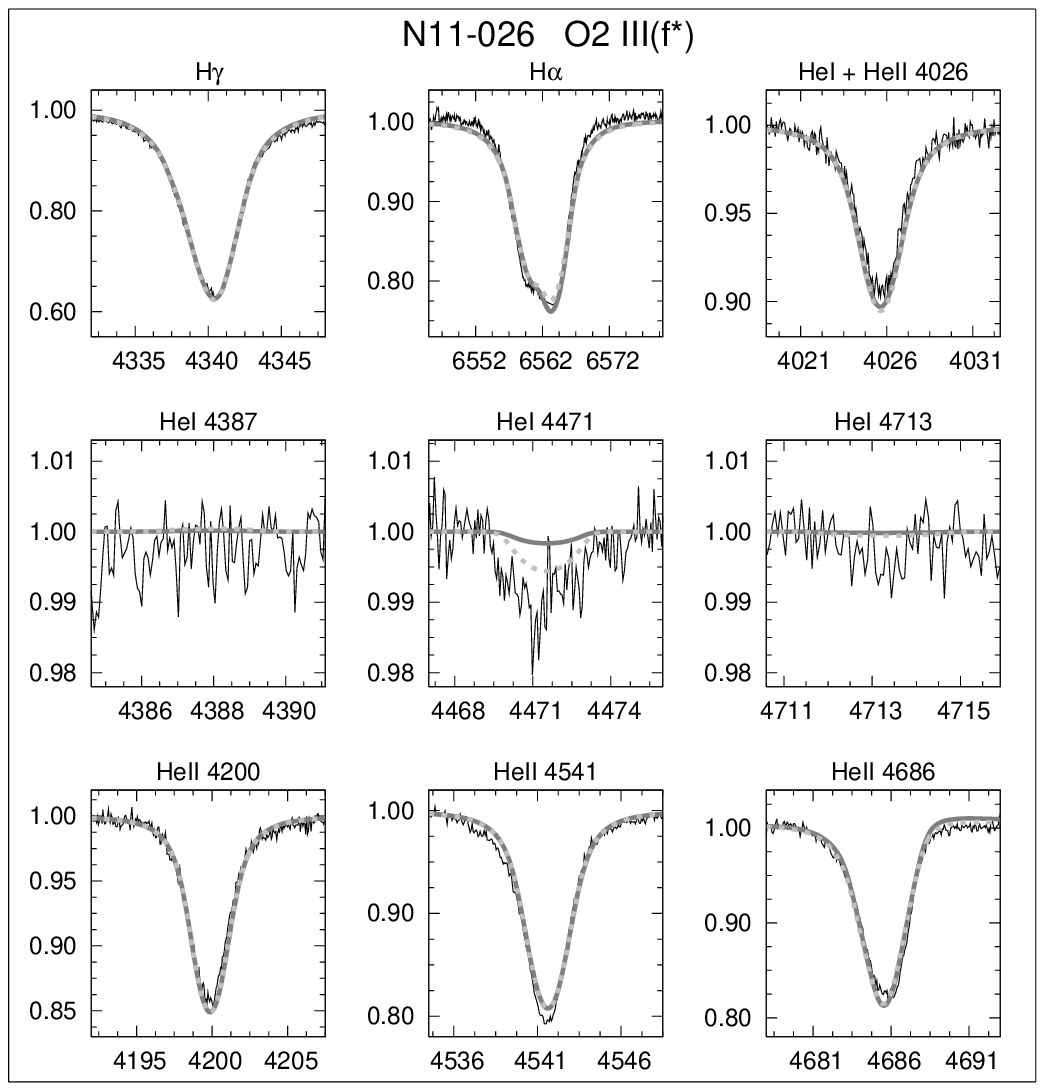}
    \includegraphics{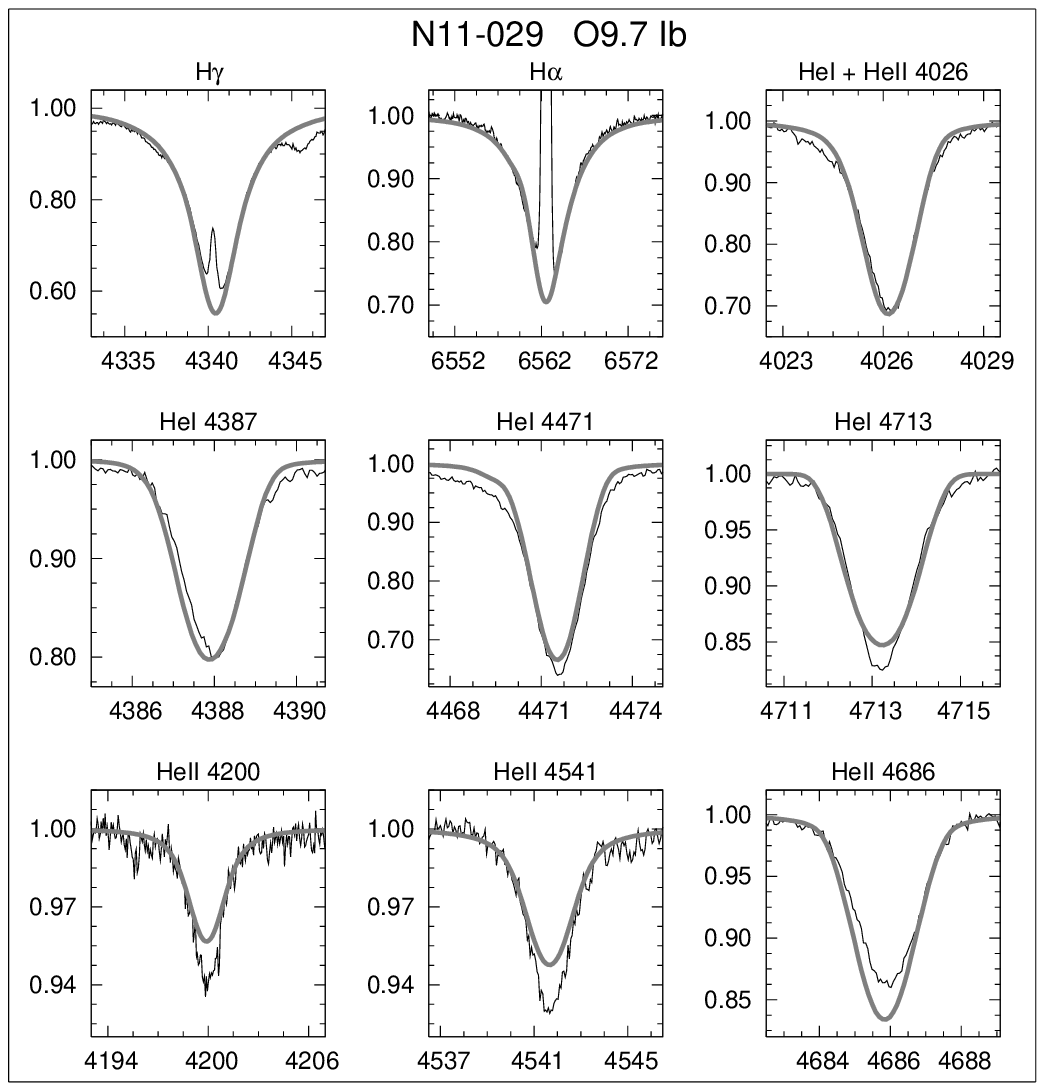}
  }
  \caption{Comparison of the observed line profiles of N11-004, -008,
  -026 and -028 with best fitting synthetic line profiles obtained
  using the automated fitting method (grey lines). Wavelengths are
  given on the horizontal axis in \AA. The vertical axis gives the
  normalised flux. Note that this axis is scaled differently for each
  line. The dotted line profiles for N11-026 correspond to the
  best fit obtained adopting a five times larger relative weight for
  the \heil4471 line. This best fit has a \teff\ lower by 3.7~kK and
  all other fit parameters approximately equal compared to the 
  fit for the nominal \heil4471 weight.}
  \label{fig:fits_1}
\end{figure*}

\begin{figure*}[t]
  \centering
  \resizebox{17cm}{!}{
  \includegraphics{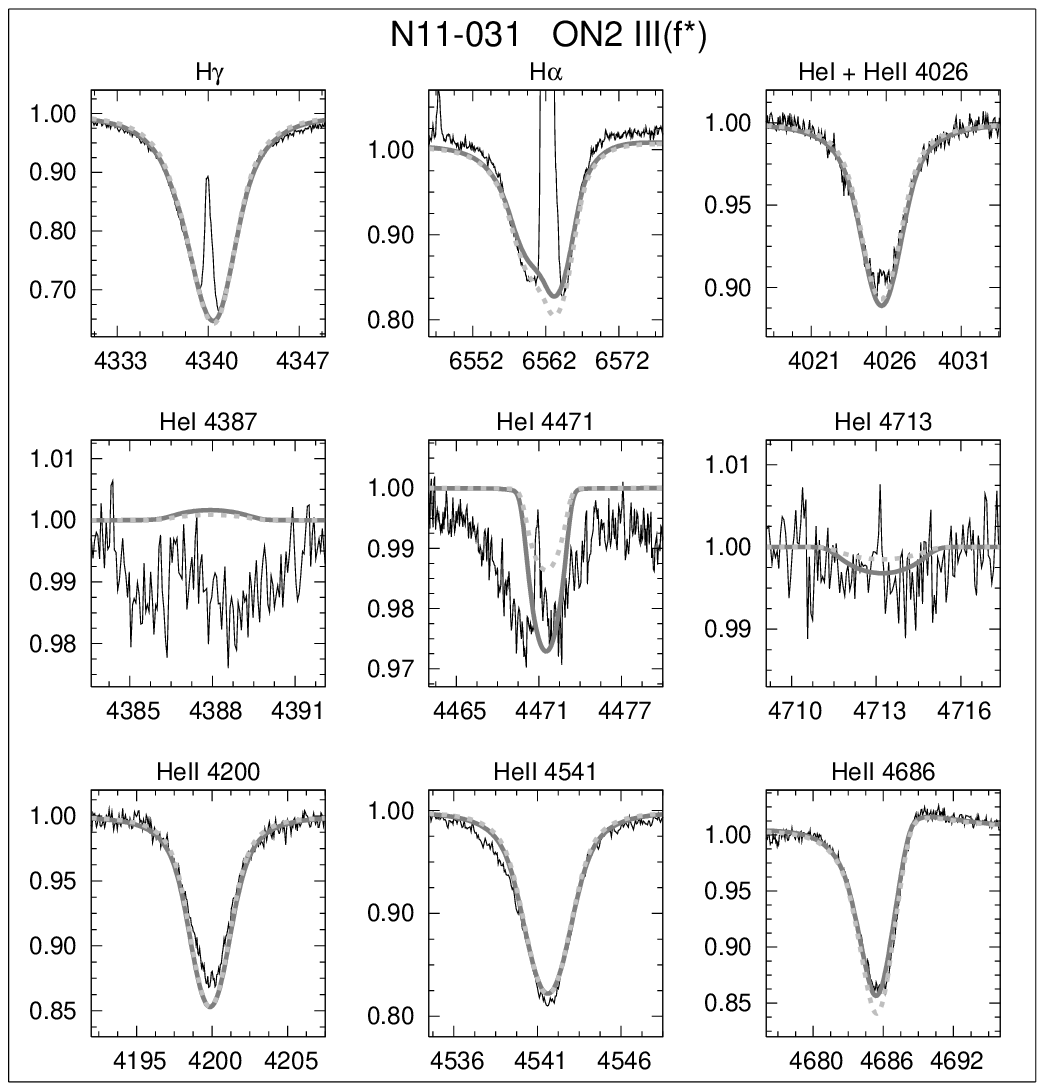}
  \includegraphics{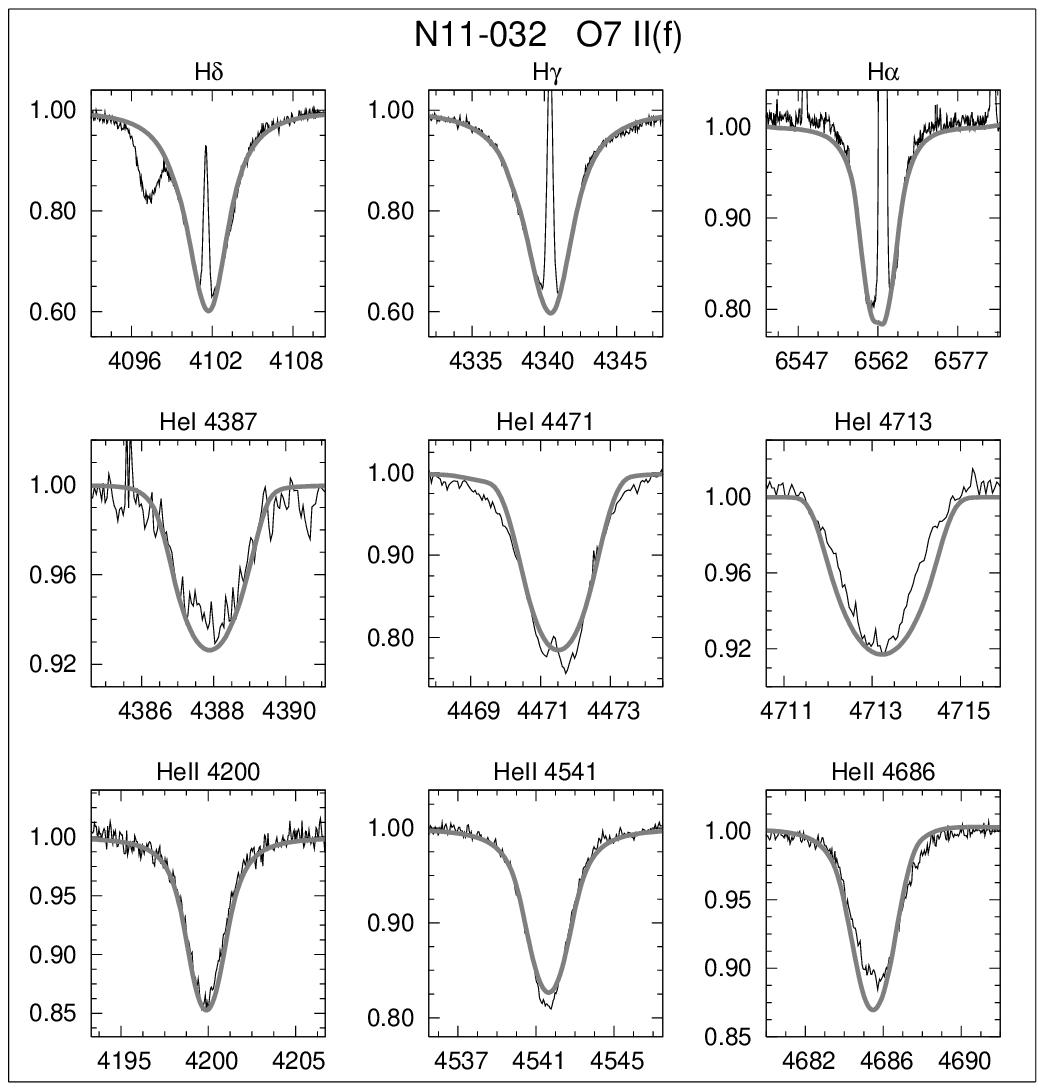} 
                     }

  \resizebox{1cm}{!}{ }

  \resizebox{17cm}{!}{
  \includegraphics{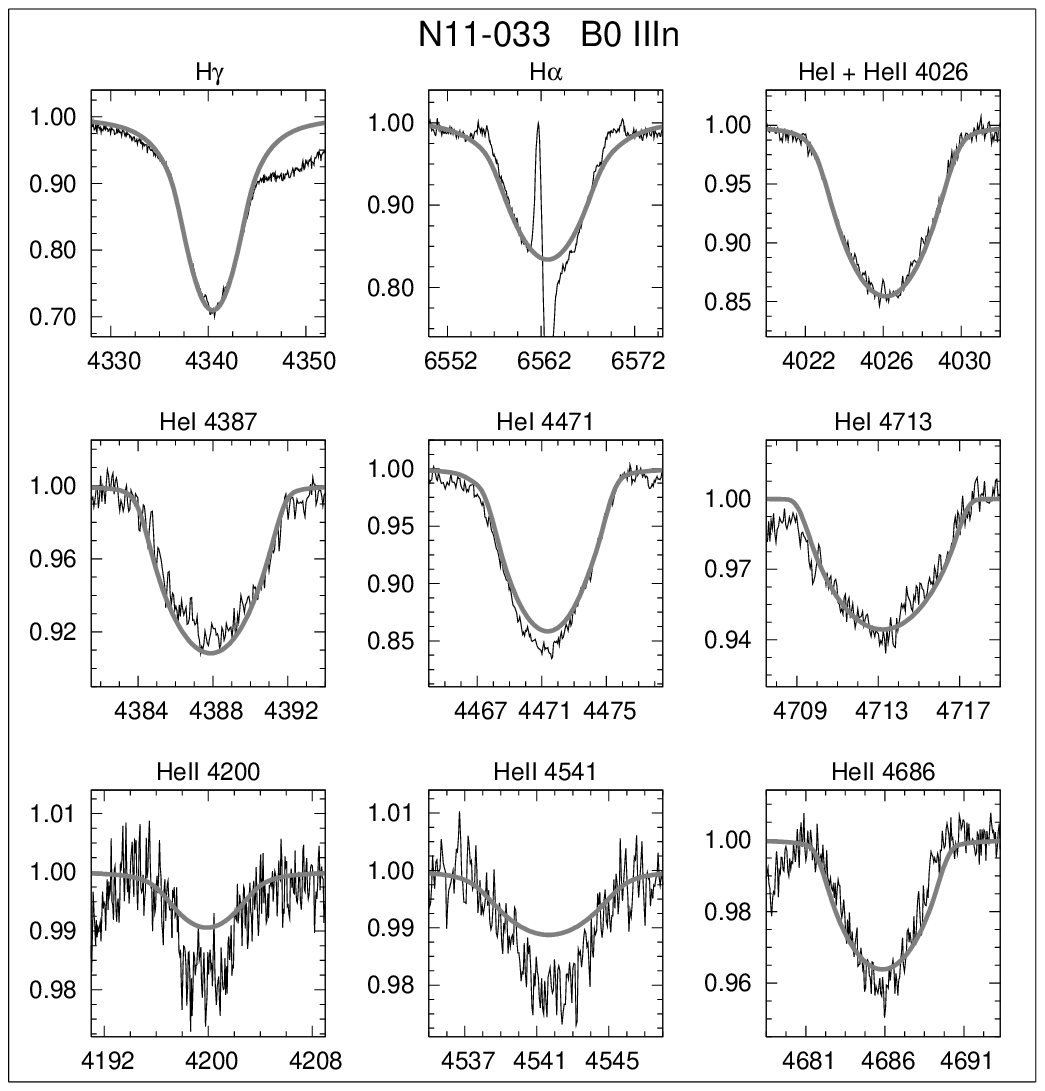}
  \includegraphics{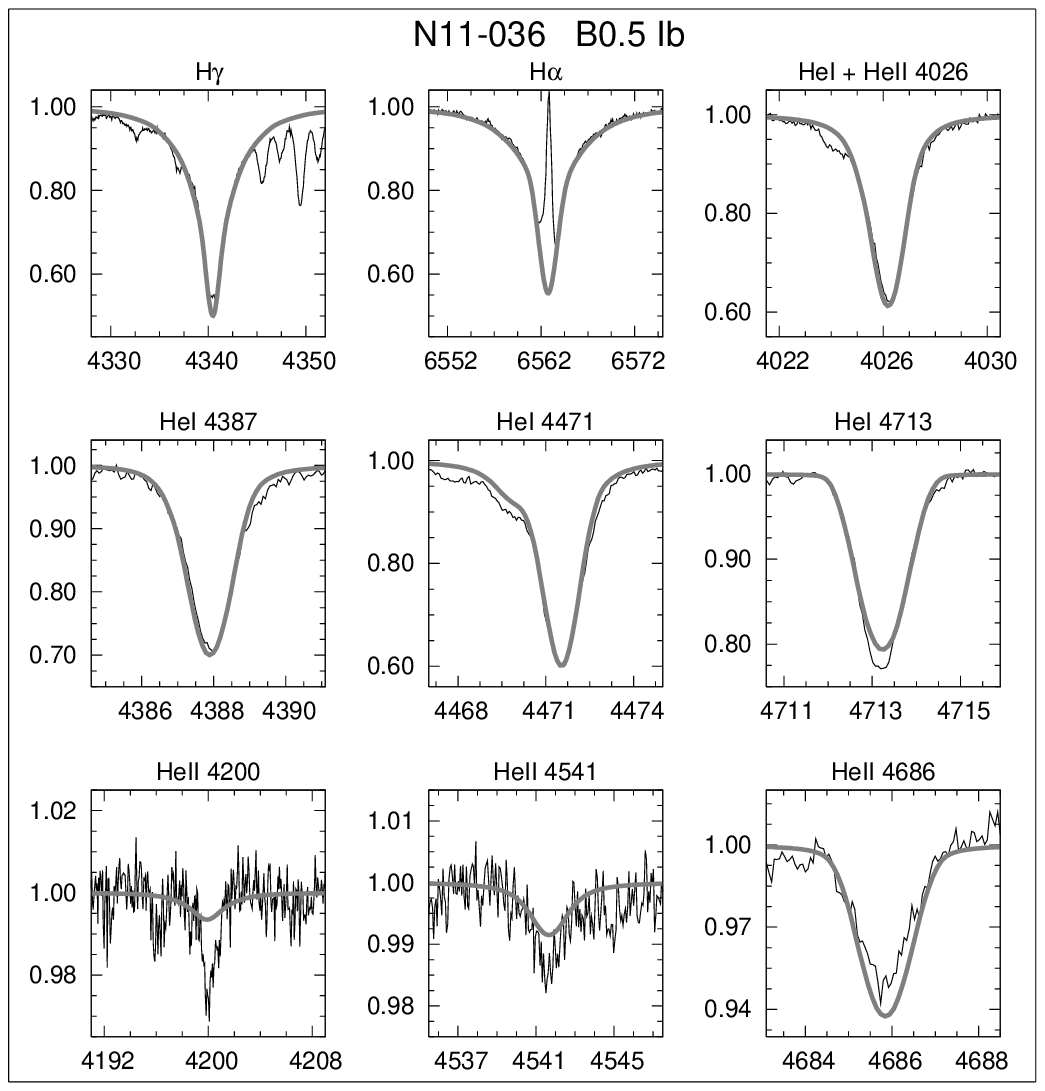}
  }

  \caption{Same as Fig.~\ref{fig:fits_1}, however, for N11-031, -032,
  -033 and -036. Shown as dotted profiles for N11-031 is the effect of
  and a 2.2~kK increase in \teff.}
  \label{fig:fits_2}
\end{figure*}

\begin{figure*}[t]
  \centering
  \resizebox{17cm}{!}{
  \includegraphics{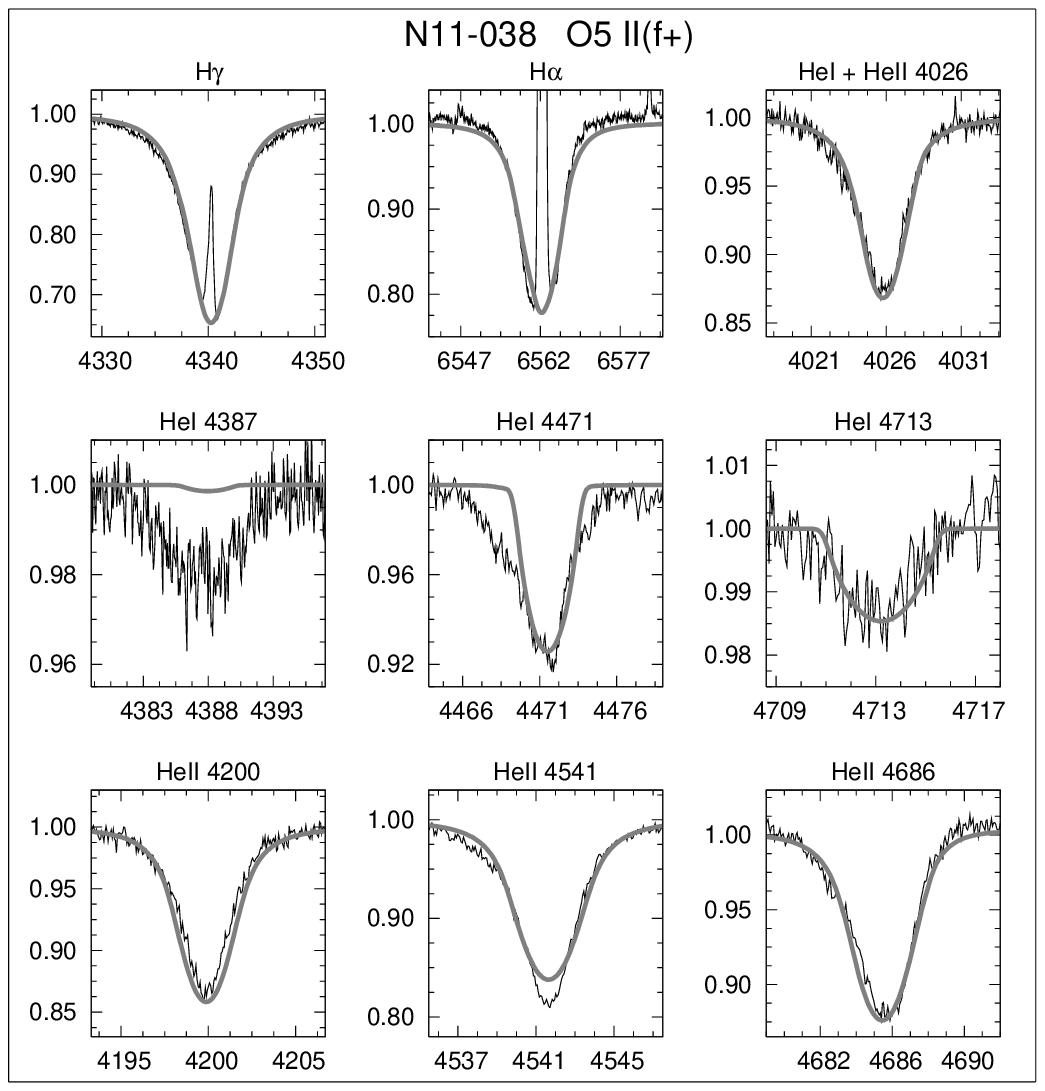}
  \includegraphics{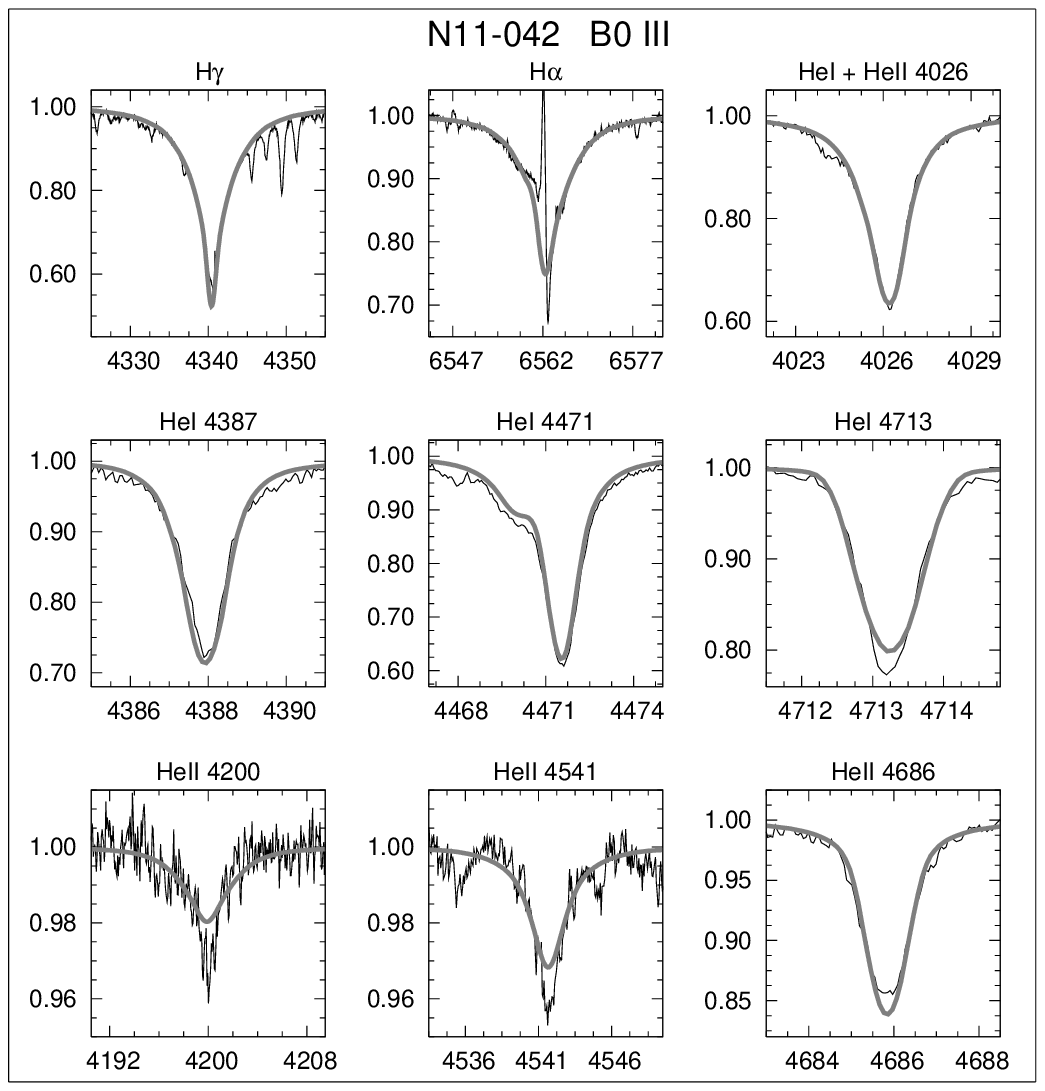} 
                     }

  \resizebox{1cm}{!}{ }

  \resizebox{17cm}{!}{
  \includegraphics{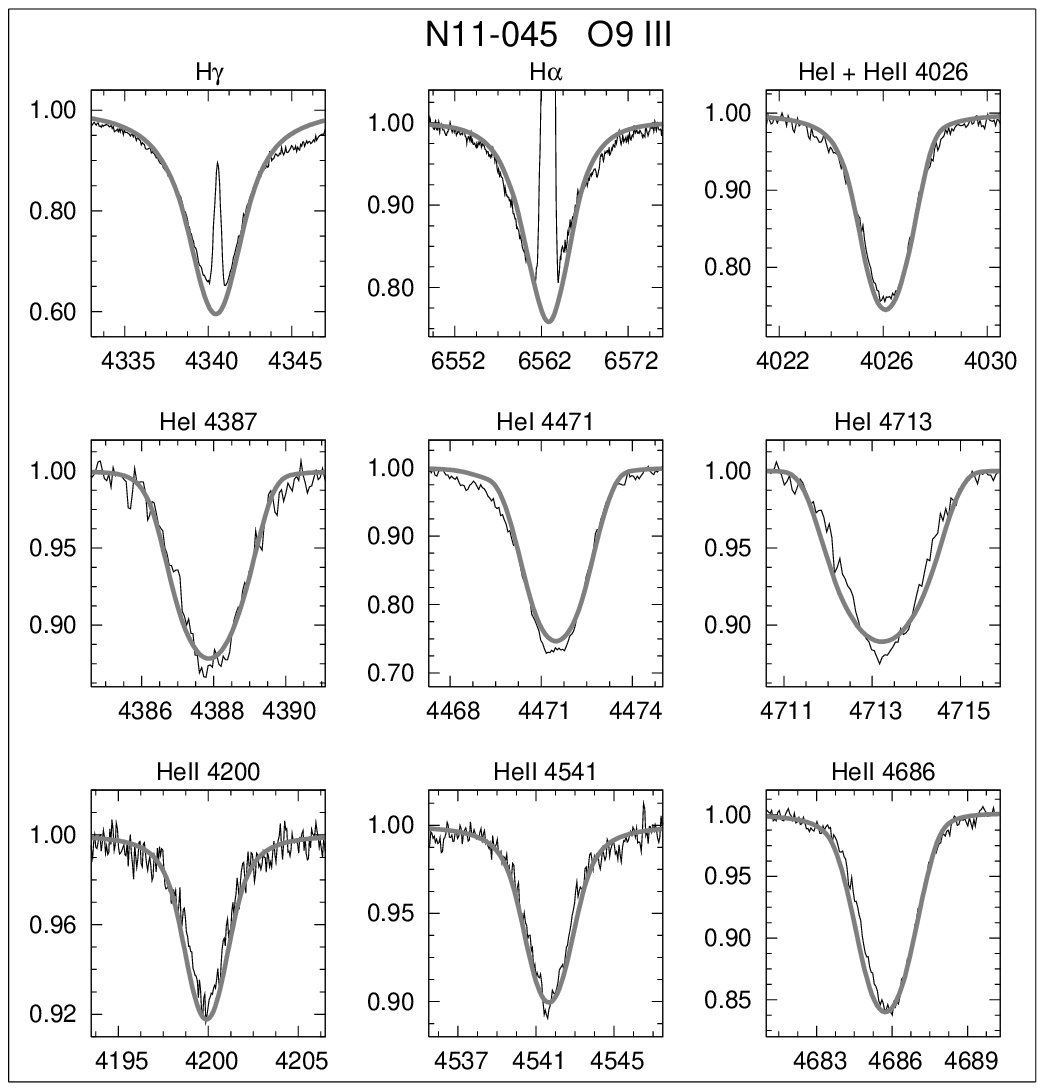}
  \includegraphics{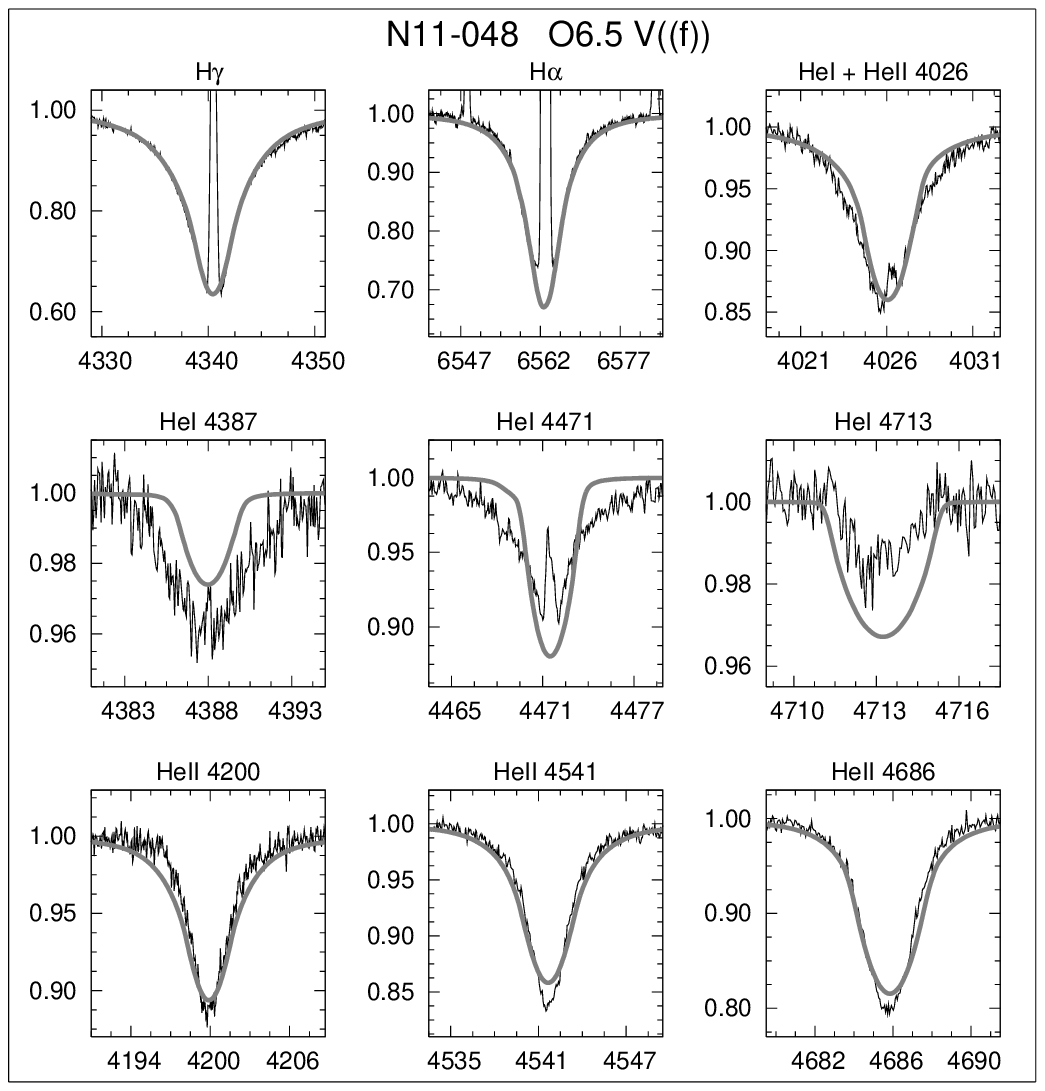}
  }

  \caption{Same as Fig.~\ref{fig:fits_1}, however, for N11-038, -042,
  -045 and -048.}
  \label{fig:fits_3}
\end{figure*}

\begin{figure*}[t]
  \centering
  \resizebox{17cm}{!}{
  \includegraphics{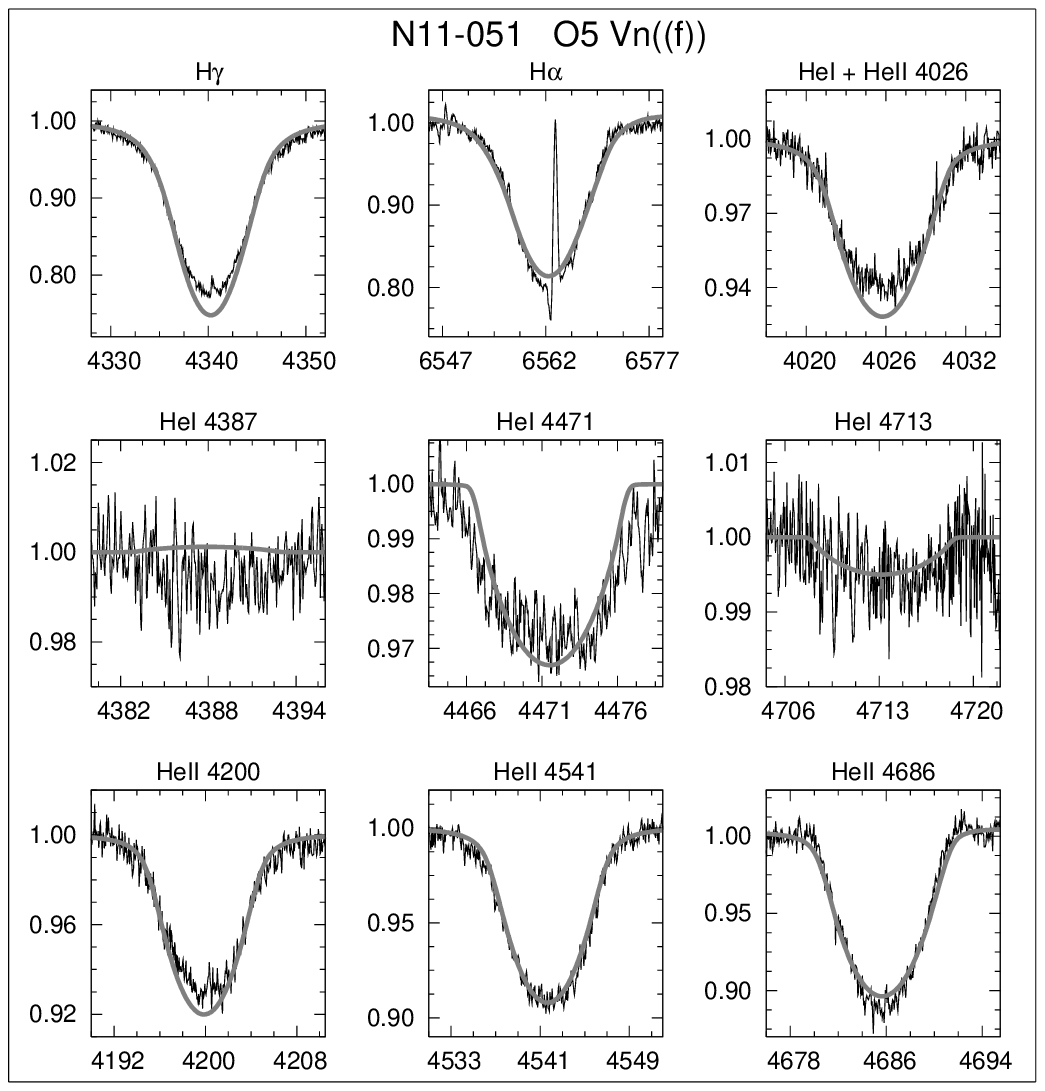}
  \includegraphics{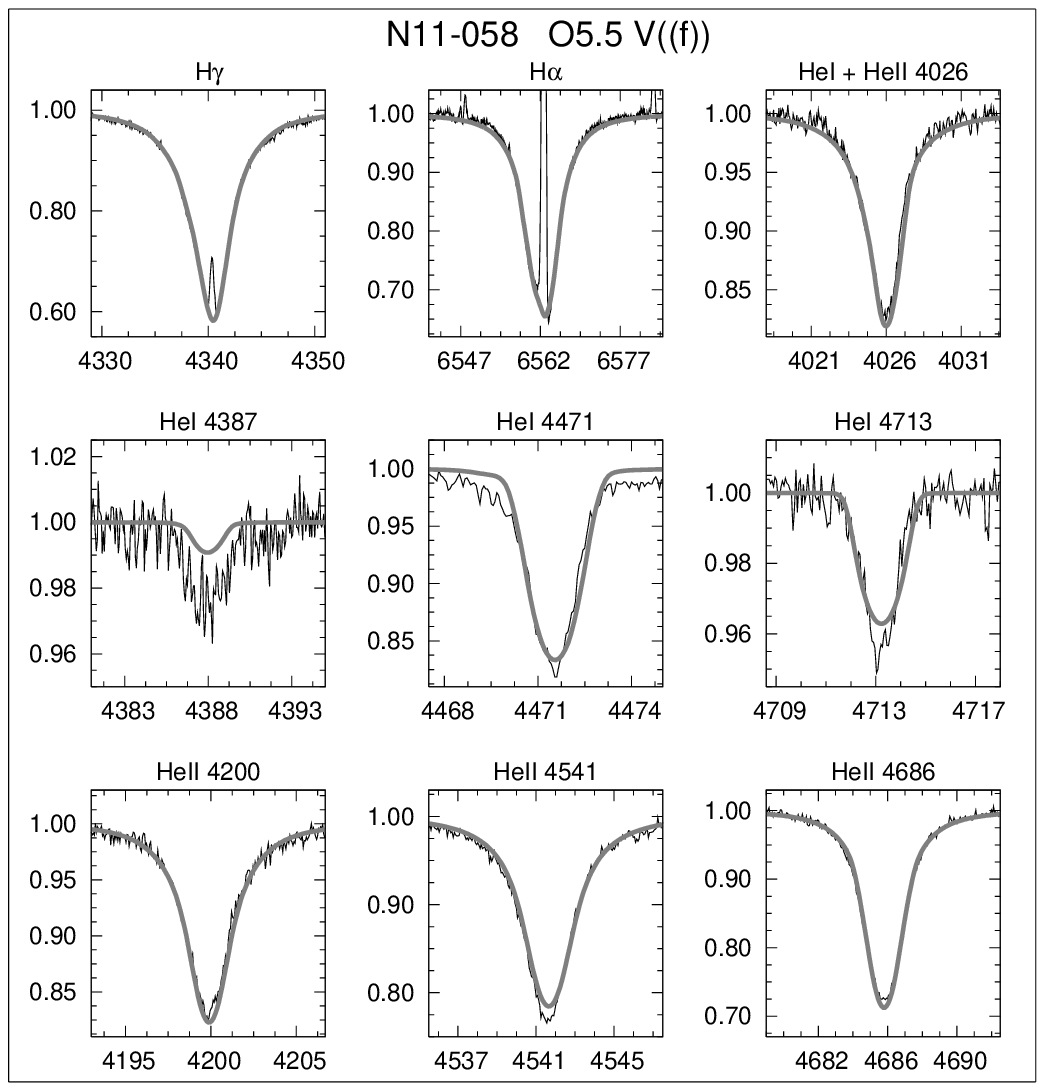}
  }

  \resizebox{1cm}{!}{ }

  \resizebox{17cm}{!}{
  \includegraphics{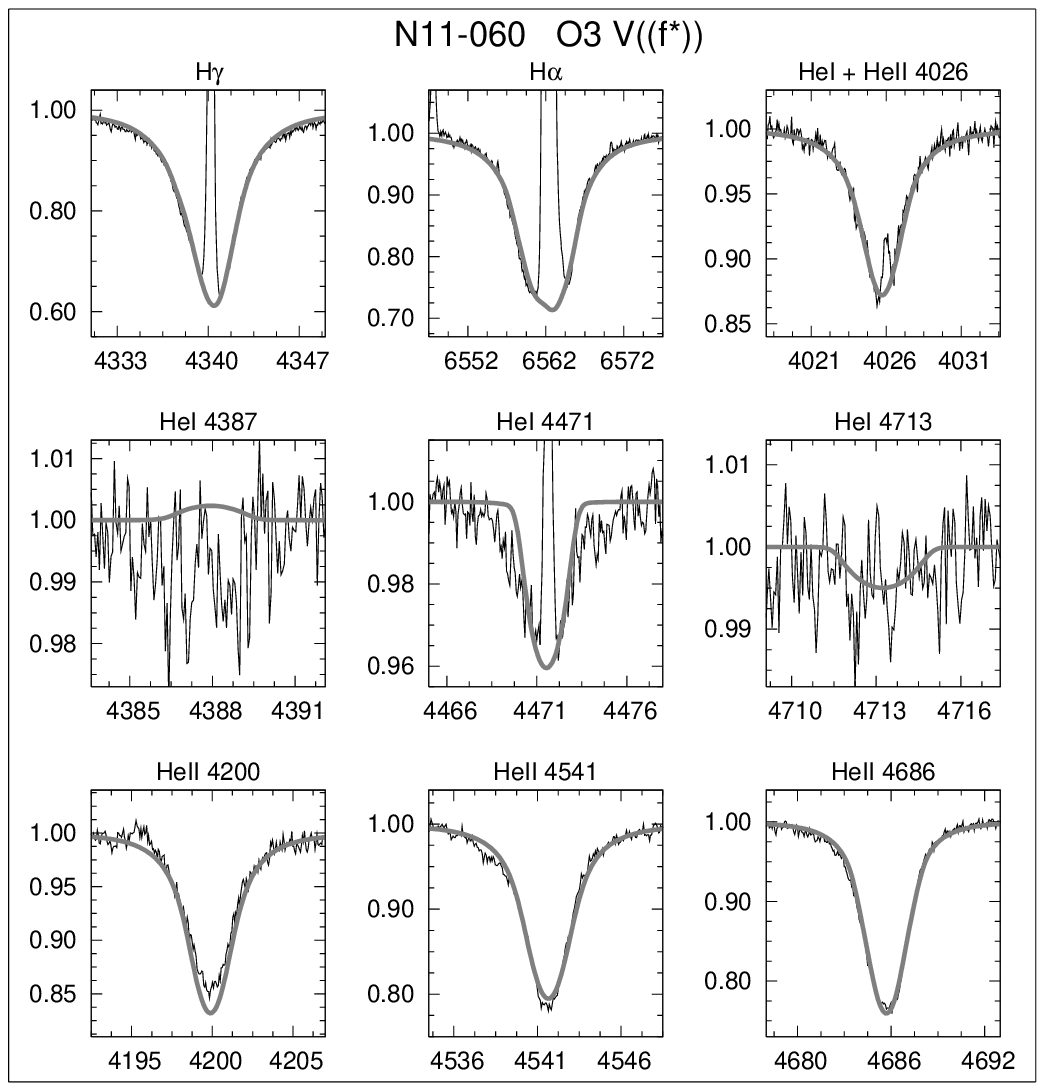}
  \includegraphics{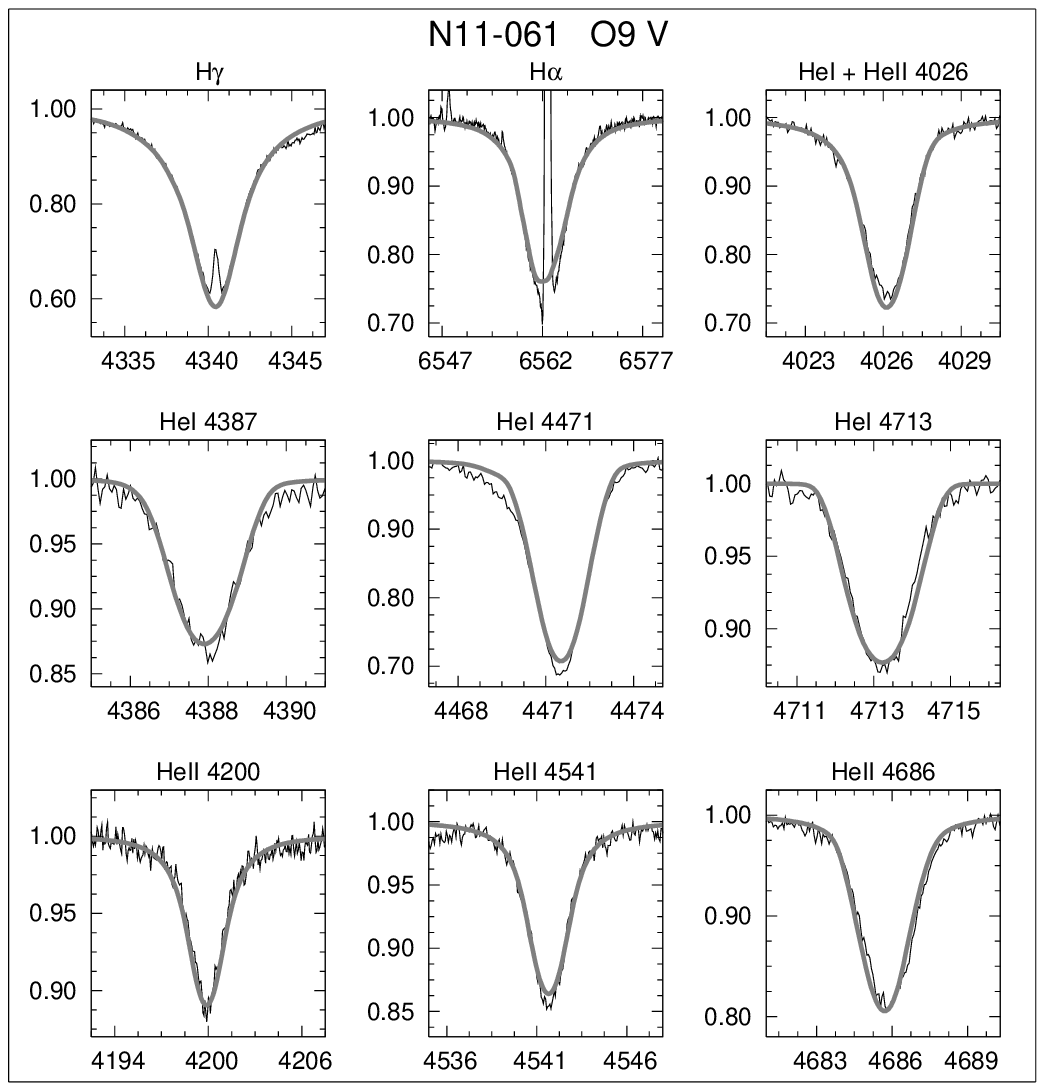}
  }

  \caption{Same as Fig.~\ref{fig:fits_1}, however, for N11-051, -058,
  -060 and -061.}
  \label{fig:fits_4}
\end{figure*}

\begin{figure*}[t]
  \centering
  \resizebox{17cm}{!}{
  \includegraphics{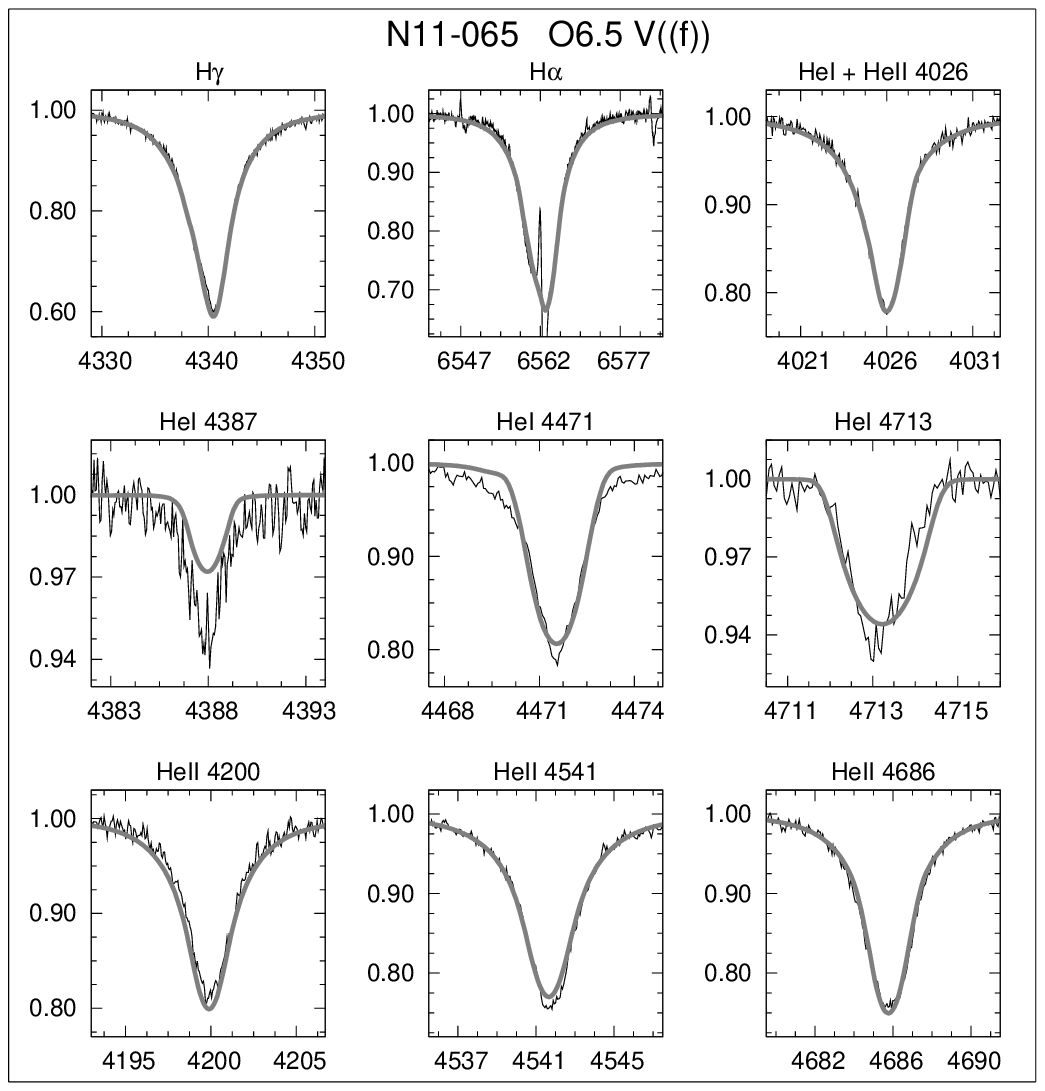}
  \includegraphics{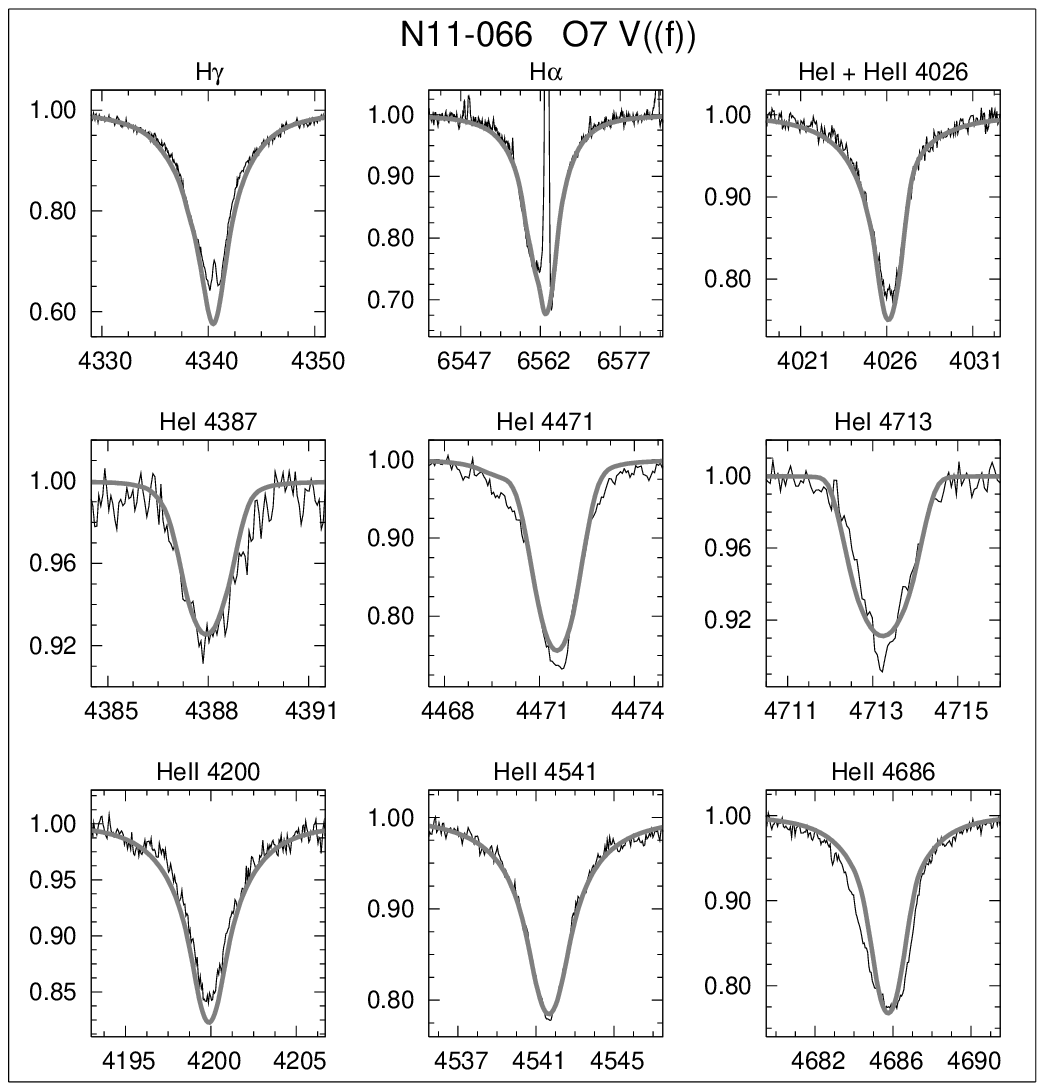}
  }

  \resizebox{1cm}{!}{ }

  \resizebox{17cm}{!}{
  \includegraphics{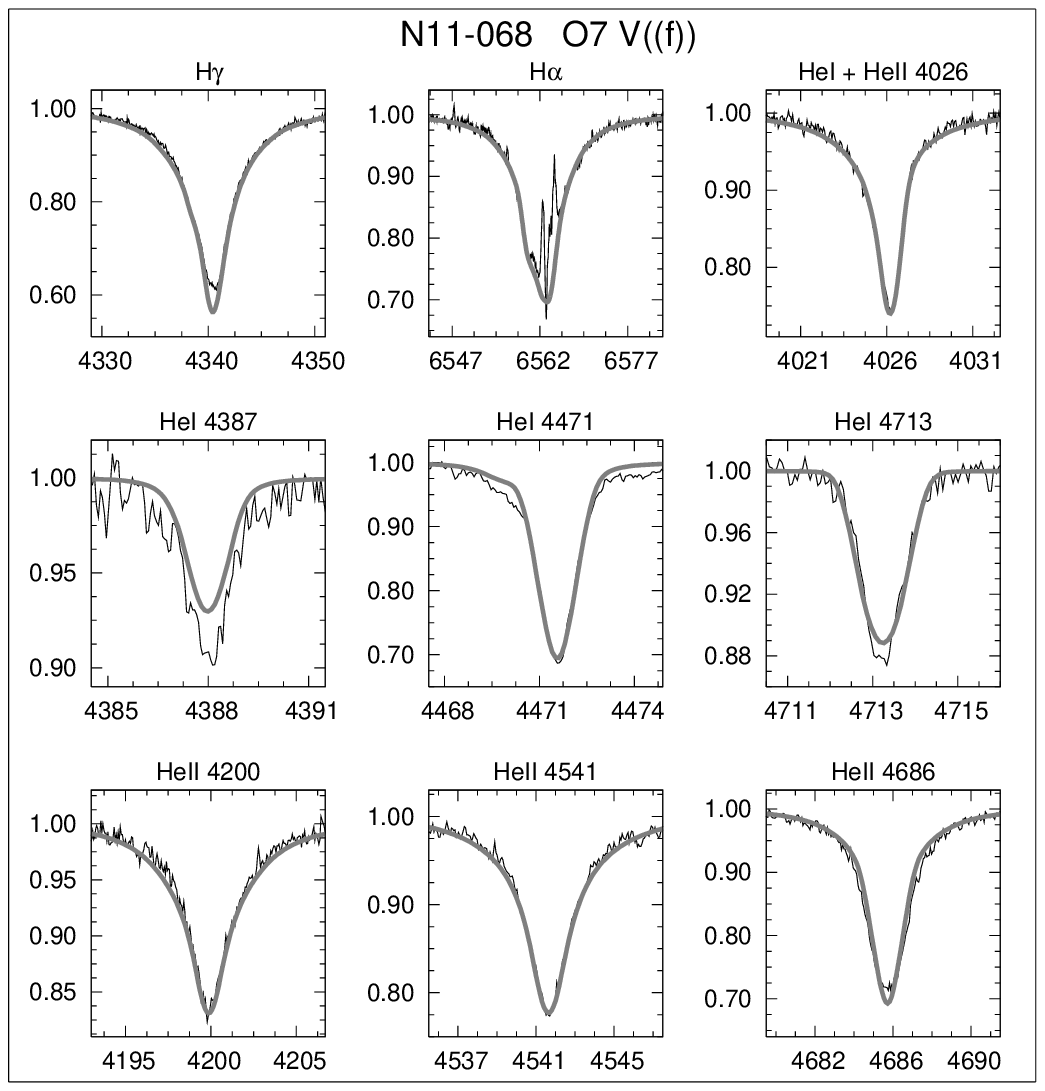}
  \includegraphics{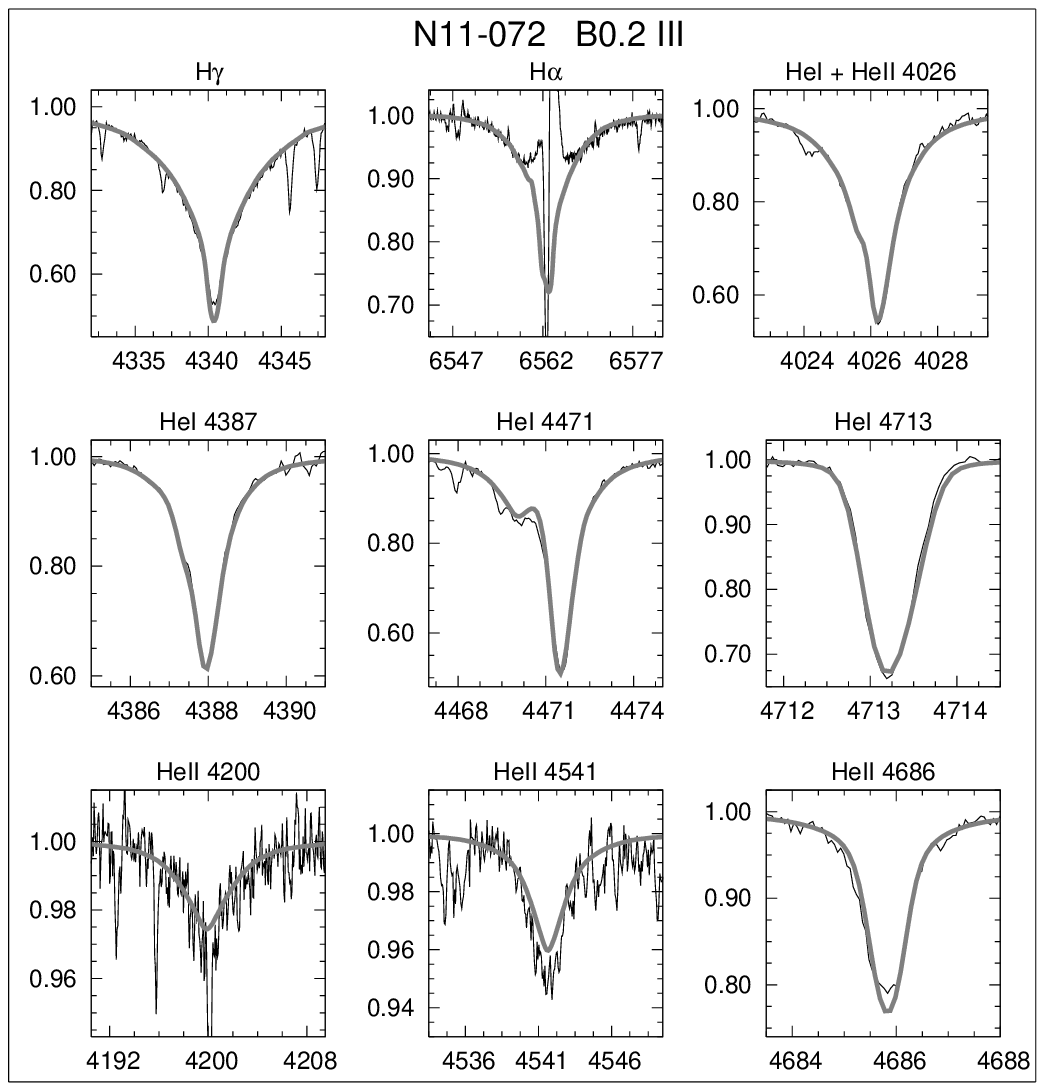}
  }

  \caption{Same as Fig.~\ref{fig:fits_1}, however, for N11-065, -066,
  -068 and -072.}
  \label{fig:fits_5}
\end{figure*}

\begin{figure*}[t]
  \centering
  \resizebox{17cm}{!}{
  \includegraphics{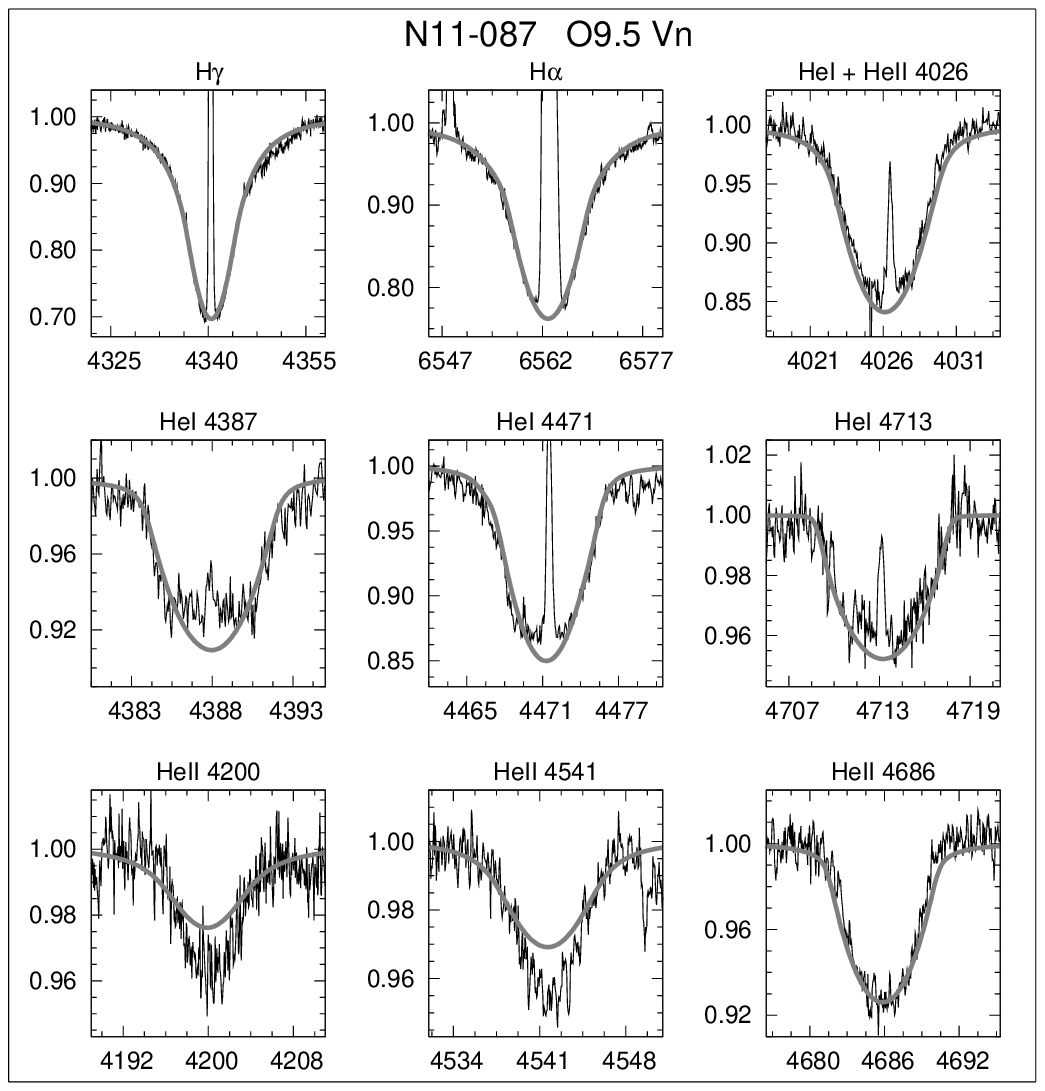}
  \includegraphics{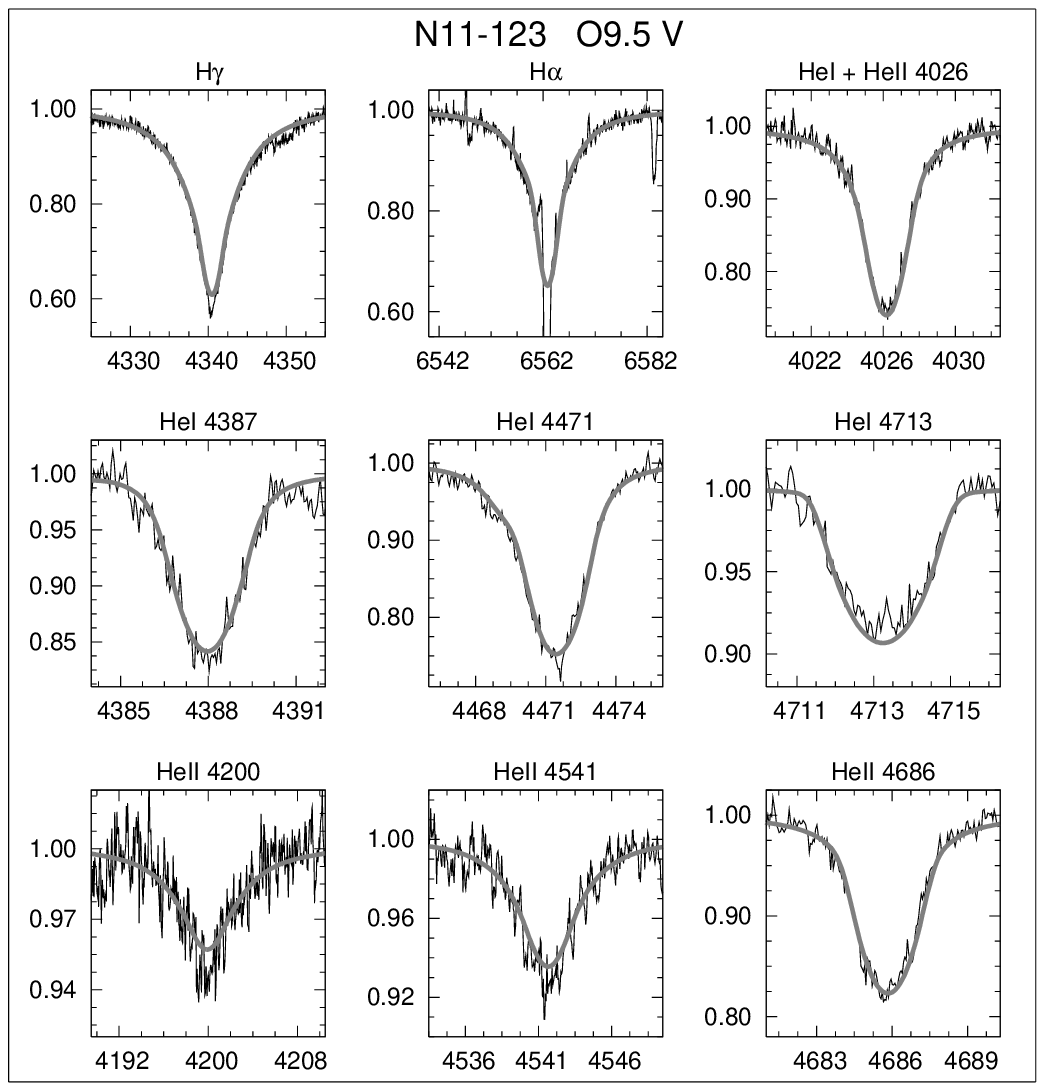}
  }

  \caption{Same as Fig.~\ref{fig:fits_1}, however, for N11-087 and
  -123.}
  \label{fig:fits_6}
\end{figure*}

\begin{figure*}[t]
  \centering
  \resizebox{17cm}{!}{
  \includegraphics{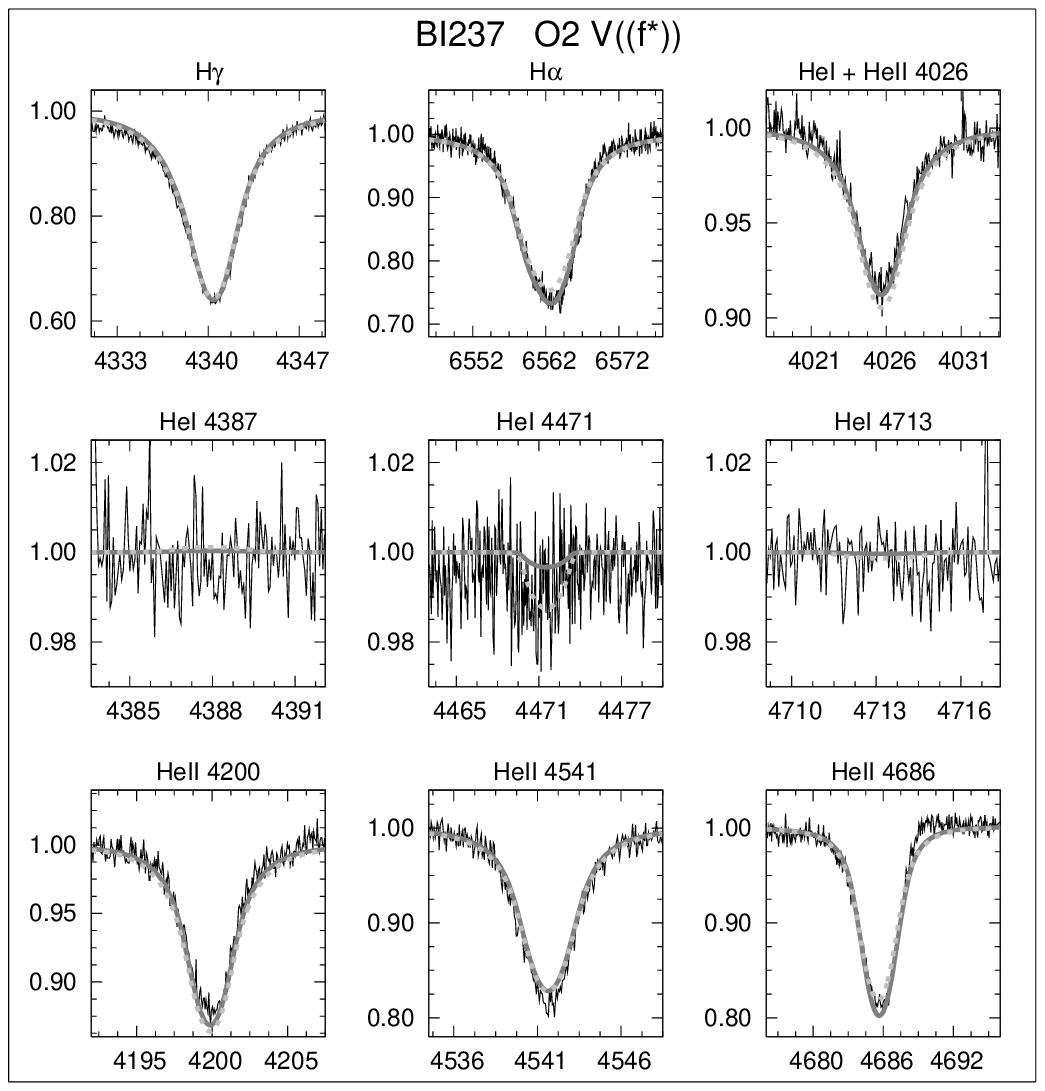}
  \includegraphics{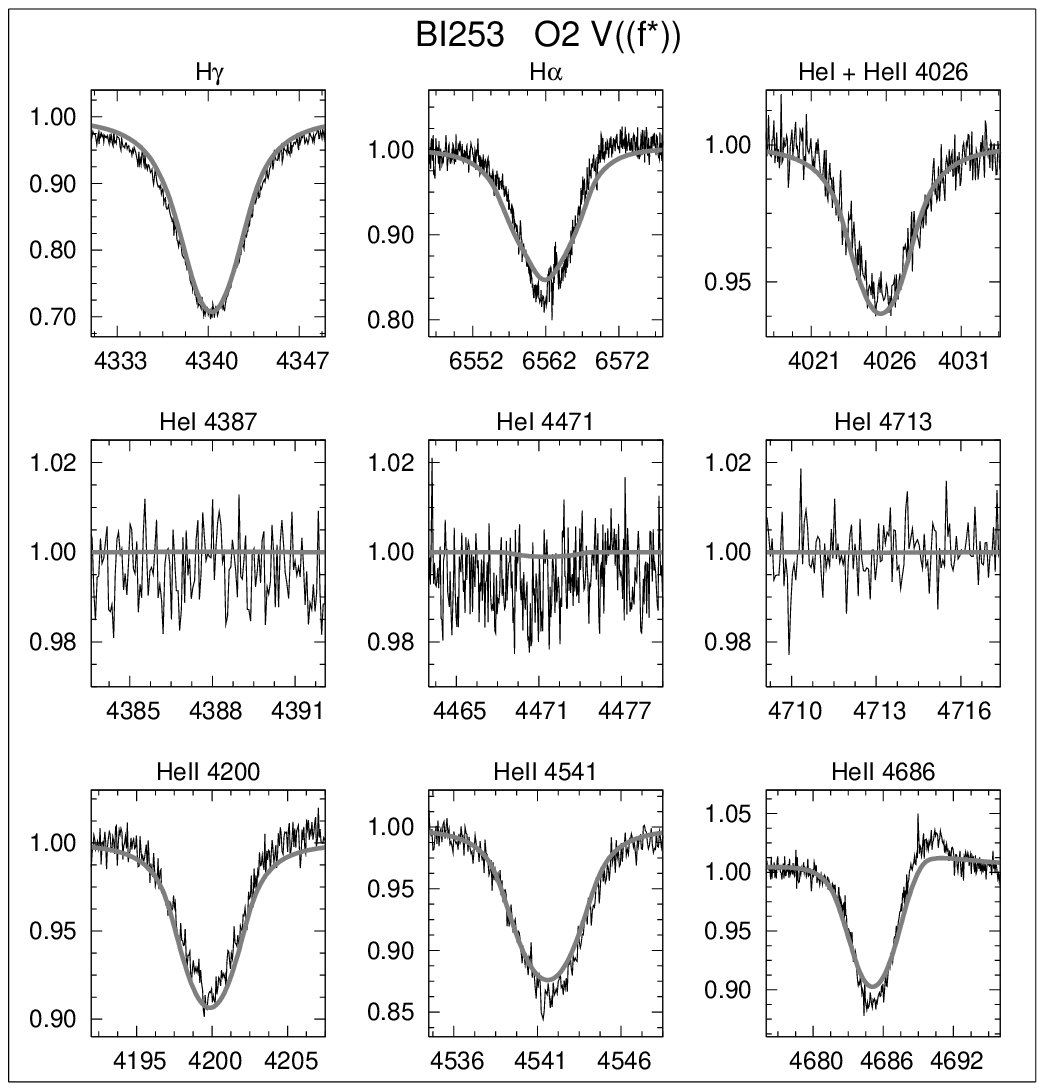}
  }

  \resizebox{1cm}{!}{ }

  \resizebox{17cm}{!}{
  \includegraphics{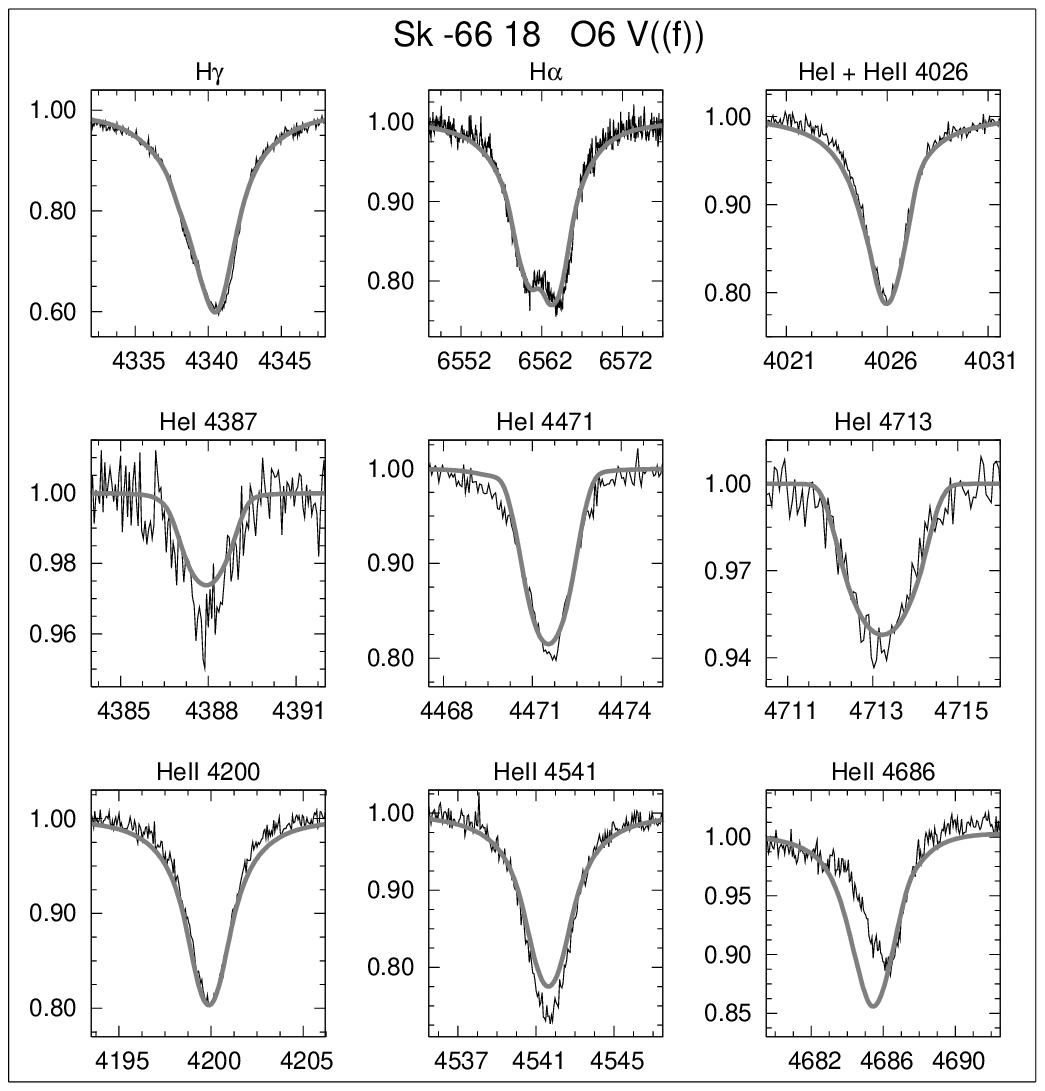}
  \includegraphics{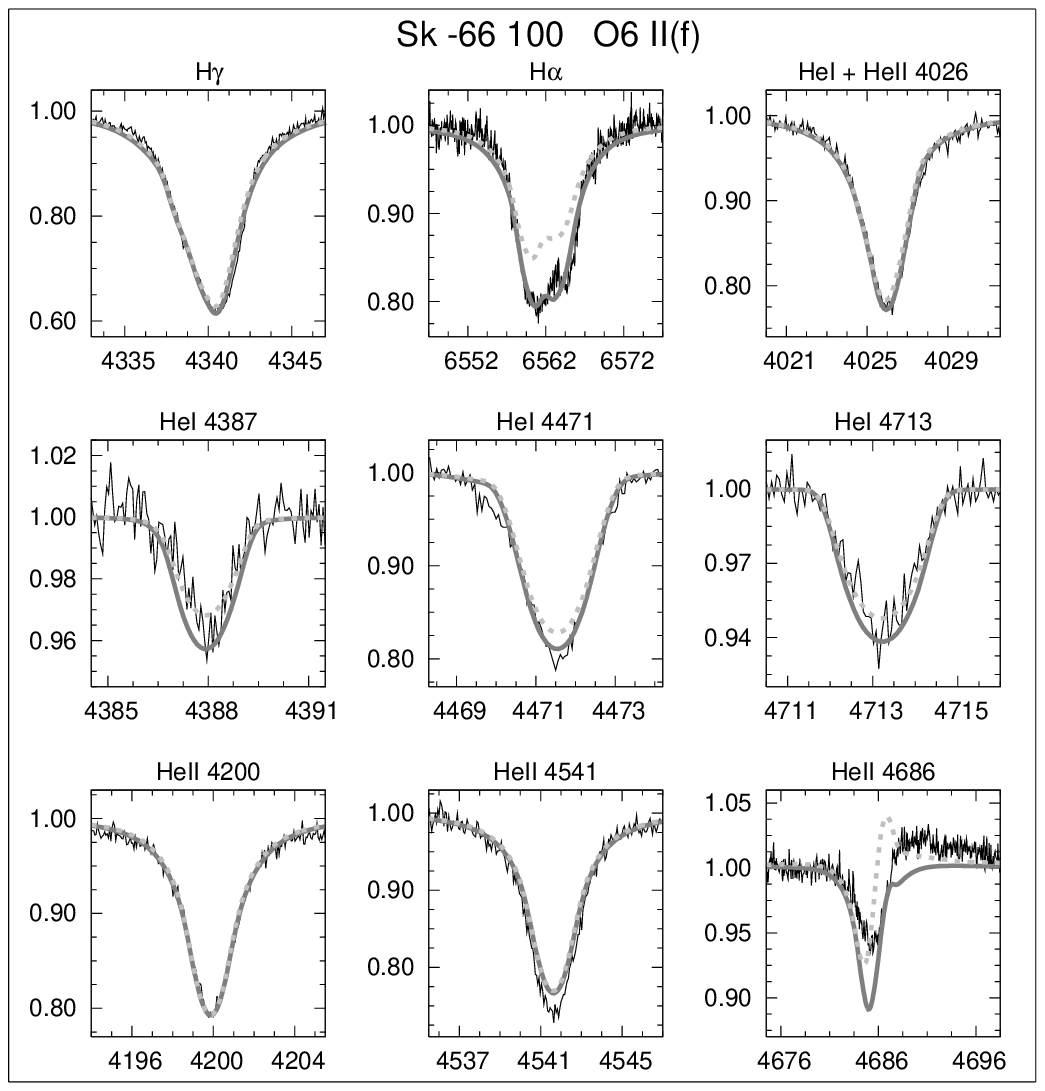}
  }

  \caption{Same as Fig.~\ref{fig:fits_1}, however, for BI~237, BI~253,
  Sk~$-66$~18 and Sk~$-66$~100. The dotted profiles for BI~237
  correspond to a model calculated with a \teff\ reduced by 3.8~kK
  compared to the best fit value. For Sk~$-66$~100 the dotted profiles
  show the effect of an increase in \mdot\ by 0.12.}
  \label{fig:fits_7}
\end{figure*}

\begin{figure*}[t]
  \centering
  \resizebox{17cm}{!}{
  \includegraphics{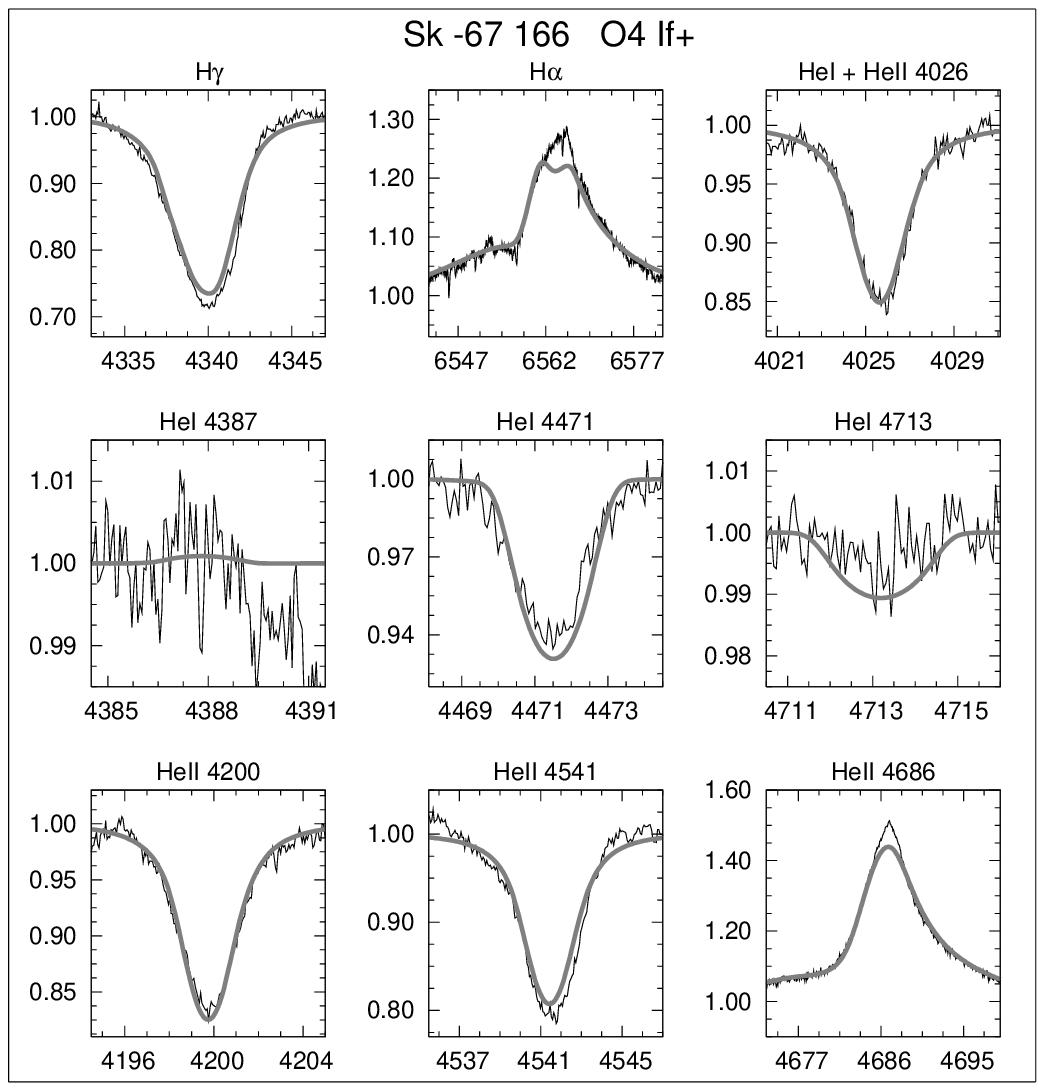}
  \includegraphics{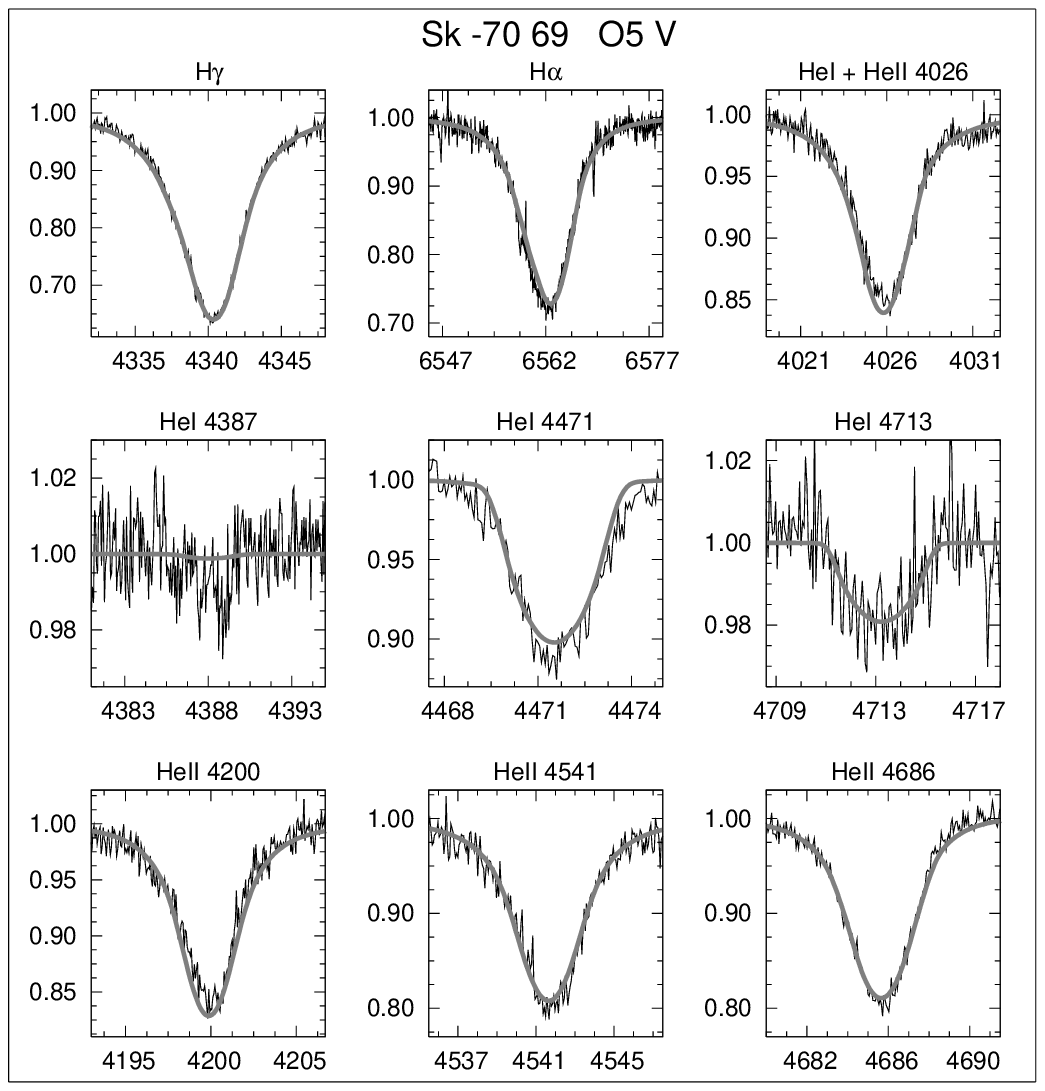}
  }

  \caption{Same as Fig.~\ref{fig:fits_1}, however, for Sk~$-67$~166
  and Sk~$-70$~69.}
  \label{fig:fits_8}
\end{figure*}

\paragraph{N11-004} (Fig.~\ref{fig:fits_1})
With exception of \heil4471 good fits for all lines of this OC9.7
supergiant were obtained. The under prediction of the \heil4471 line is
possibly connected to the so-called generalised dilution effect
(\citealt{voels89}, see also \citealt{repolust04}).

\paragraph{N11-008} (Fig.~\ref{fig:fits_1})
All lines are reproduced well. Note that given the weakness of
\heiil4686 and lower weight assigned to this line, the reproduction of
its profile can be considered good.

\paragraph{N11-026} (Fig.~\ref{fig:fits_1})
A high effective temperature of 53~kK was needed to fit the spectrum
of this O2 giant. Note that even though on its relative plot
scale the very weak \heil4471 line is not fitted perfectly, the
quantitative fit quality is good and is comparible to that of the
\heii\ lines. Still, trying to assess whether the fit could be
improved we ran test fits with increasing relative weight of the
\heil4471 line. It turned out that an increase of a factor of five was
necessary to obtain a solution with an improved \heil4471 line fit,
which is shown in Fig.~\ref{fig:fits_1} as a set of dotted lines.
This best fit has a \teff\ lower by 3.7~kK and all other fit
parameters approximately equal compared to the regular fit. 
The overal fit quality of the other diagnostic lines is somewhat less,
though this is difficult to discern by eye because these lines are
relatively strong (one has to zoom in to the 0.5 to 1 percent level,
as we do for \heil4471).
Note, however, that the error estimates for \teff\ are
relative large and within the lower error estimate of 3.9~kK (cf.\
Tab.~\ref{tab:errors}) the solutions agree. Also note that even for
the lower \teff\ solution this O2 star is still considerably hotter
than an O3 star (see Sect.~\ref{sec:teff}).  

\paragraph{N11-029} (Fig.~\ref{fig:fits_1})
A good fit was obtained. The slight underestimation of the core
strength of the \heiil4200 and \heiil4541 lines is, given the relative
weakness of these lines, not significant.

\paragraph{N11-031} (Fig.~\ref{fig:fits_2})
To fit the spectrum of this O2 star a relatively low effective
temperature of 45~kK was necessary. \cite{walborn04} derived a \teff\
of 55~kK for this object. However, their determination of this
parameter was based on the analysis of the \ion{N}{iv-v} ionisation
adopting a fixed gravity of $\logg = 4.0$. Based on the analysis of
the helium ionisation we exclude a \teff\ higher than
$\sim$47~kK. This is illustrated by the dotted lines, which correspond
to a model calculated for an effective temperature higher by the upper
error estimate we derived for \teff. For even higher temperatures the
\heil4471 would, in contradiction to the observations, disappear
completely (also see Sect.~\ref{sec:teff}).

\paragraph{N11-032} (Fig.~\ref{fig:fits_2})
The \hei\ and \heii\ blend at 4026~\AA\ was not observed for this
object. In the final fit, therefore, also the \hd\ Balmer line is
shown.

\paragraph{N11-033} (Fig.~\ref{fig:fits_2})
This B0 giant is fast rotator. To reproduce the observed line profiles
a projected rotational velocity of 256~\kmsec\ was necessary.  In the
presented fit the strength of \heiil4541 seems to be
under predicted. However, given the relative weakness of this line this
small discrepancy is negligible. Note that the sharp change in the
central part of the \ha\ profile is the result of an over subtraction of the
core nebular feature and was not taken into account in the fitting
procedure.

\paragraph{N11-036} (Fig.~\ref{fig:fits_2}) 
The line profiles of N11-036 could be reproduced quite accurately
because of its relatively slow rotation.

\paragraph{N11-038} (Fig.~\ref{fig:fits_3})
The relatively poor reproduction of \heil4387 in this O5 bright giant
is the result of the low weight assigned to this line based on the
spectral type of N11-038. A good fit to the \heil4471 could not be
obtained due to its peculiar line profile shape. Possibly this
triangular profile is related to macroturbulence.

\paragraph{N11-042} (Fig.~\ref{fig:fits_3})
A relatively low projected rotational velocity was found for this B0
giant allowing for a nearly perfect fit.

\paragraph{N11-045} (Fig.~\ref{fig:fits_3})
A good fit was obtained for this O9 giant.  Despite the good fit
quality a problem is apparent when \Ms\ and \Mev\ are compared in
Tab.~\ref{tab:results}. The spectroscopic mass is found to be less
than half its evolutionary equivalent.

\paragraph{N11-048} (Fig.~\ref{fig:fits_3})
Based on the fact that the \heiil4686 is the strongest line in its
spectrum \cite{parker92} classified this star as a Vz star. The fit
parameters, however, place this object at a considerable distance from
the theoretical ZAMS in Fig.~\ref{fig:hrd}. We also find that the
spectrum can not be reproduced accurately.  The width of the neutral
helium lines is systematically unpredicted, whereas the width of the
\heii\ is over predicted. This and the fact that the helium abundance
that is recovered from the spectrum is rather low ($\yhe = 0.06$) are
indicative of a possible binary nature.

\paragraph{N11-051} (Fig.~\ref{fig:fits_4})
The spectrum of the O5 dwarf N11-051 shows very broad lines,
indicating fast stellar rotation. To obtain the final fit a projected
rotational velocity of 333~\kmsec\ was required.

\paragraph{N11-058} (Fig.~\ref{fig:fits_4})
A good fit was obtained for all lines of this O5.5 dwarf. The small
under prediction of the \hei4387 line strength is the result of the
relatively low weight assigned to this line for the spectral type of
this star.

\paragraph{N11-060} (Fig.~\ref{fig:fits_4})
Note that the \heil4471 line of this O3 dwarf has a strength that is
comparable to the strength of this line in the spectrum of the O2
giant N11-031. As the effective temperature we find for these two
objects are also comparable, the helium spectrum suggest a spectral
type of O3 for both objects.

\paragraph{N11-061} (Fig.~\ref{fig:fits_4})
All lines of this O9 dwarf are reproduced correctly. The mass loss
could be reliably determined using the automated method at a rate of
$2.1 \times 10^{-7}\,\msun$. Note that the value of 1.8 obtained for
the wind acceleration parameter $\beta$ is rather high for a dwarf
type object.

\paragraph{N11-065} (Fig.~\ref{fig:fits_5})
A relatively high helium abundance of 0.17 was recovered from the
spectrum. The good fit quality obtained for all lines, however, does
not indicate an overestimation of this parameter. A strong mass
discrepancy is found with \Ms\ being smaller than \Mev\ by
approximately a factor of two.

\paragraph{N11-066} (Fig.~\ref{fig:fits_5})
Apart from a slight under prediction of the width of \heiil4686 all
lines of this O7 dwarf could be fitted with good accuracy. Like the
previously discussed object, again a large mass discrepancy is found
with a ratio of \Mev\ to \Ms\ of 1.8.

\paragraph{N11-068} (Fig.~\ref{fig:fits_5})
The final fit reproduces the observed line profiles very
accurately. No further comments are required.

\paragraph{N11-072} (Fig.~\ref{fig:fits_5})
The star has a very sharp lined spectrum. The best fit required a
\vsini\ of 14~\kmsec. Note that given the lower error estimate, given
in Tab.~\ref{tab:errors}, and the spectral resolution of the data this
value can also be interpreted as an upper limit.

\paragraph{N11-087} (Fig.~\ref{fig:fits_6})
To fit the spectrum of N11-087 a \vsini\ of 276~\kmsec\ was
needed. The final fit shows a good reproduction of the line profiles
with a slight under prediction of the \heiil4541 line. Given the
signal-to-noise ratio of the spectrum and the relative strength of
this line, we do not believe this to be significant.

\paragraph{N11-123} (Fig.~\ref{fig:fits_6})
Like the previously discussed object N11-123 is also classified as an
O9.5 dwarf. The effective temperature needed to fit the spectrum is,
however, higher by more than 2~kK. This is the result of a \logg\ that
is higher by $\sim$0.2 dex.

\paragraph{BI 237} (Fig.~\ref{fig:fits_7})
A good fit was obtained for this O2 dwarf star. The \heil4471 line is
hardly visible. Fortunately, the hydrogen and \heii\ lines provide the
automated fitting method with enough information to determine an
effective temperature. The error estimates on this parameter, however,
are considerable (see Tab.~\ref{tab:errors}). The effect of lowering
\teff\ to the lower error estimate of 49~kK is illustrated by the
dotted profiles.

BI 237 was recently also analysed by \cite{massey05}. Compared to this
study we find large differences for \teff\ ($+5$~kK), \logg\
($+0.2$~dex) and \mdot\ ($-0.4$~dex). Our solution for higher
effective temperature and surface gravity can be explained by the fact
that the ionisation structure of the atmosphere is set by both these
parameters (also see \citetalias{mokiem05}). The reduced mass loss
rate we find is the result of the increased value for $\beta$ ($+0.4$)
obtained by the automated method.

\paragraph{BI 253} (Fig.~\ref{fig:fits_7})
The star has as an identical spectral type as the previously discussed
object. The photospheric parameters determined from the spectrum are
also very similar to the parameters of BI~237. Note that in contrast
to this agreement BI~253 is found to have a much denser wind. Its mass
loss rate is larger by a factor 2.5 and reflects the fact that
compared to BI~237 the wind lines \ha\ and \heiil4686 are more filled
in.
In their analysis \cite{massey05} could only determine a lower
estimate for \teff\ of 48~kK, which is consistent with our lower error
estimate. The other parameters determined by these authors are in fair
agreement with our analysis. An exception to this is the surface
gravity for which \citeauthor{massey05} estimate a value lower by
0.3~dex, which is the result of the lower \teff\ used for their fit.

\paragraph{Sk $-$66 18} (Fig.~\ref{fig:fits_7})
With exception of the \heii\ lines at 4541~\AA\ and 4686~\AA\ we
obtained a near perfect fit. The helium abundance $\yhe = 0.14$
indicates an enriched atmosphere.

\paragraph{Sk $-$66 100} (Fig.~\ref{fig:fits_7})
was previously analysed by \cite{puls96}. Compared to this study we
find similar parameters with exception of a lower \teff\ which is the
result of the inclusion of blanketing and an increased helium
abundance of $\yhe = 0.19$. The final fit shows that with this large
value all lines could be reproduced quite accurately. The slight
mismatch of the \heiil4686 line can be improved by increasing \mdot\
to its upper error estimate. The effect of this increase of 0.12~dex
on the line profiles is shown using dotted lines.

\paragraph{Sk $-$67 166} (Fig.~\ref{fig:fits_8})
This supergiant has both \ha\ and \heiil4686 in emission indicating
the presence of a dense stellar outflow. To correctly reproduce the
line profiles a very high mass loss rate of $9.3 \times 10^{-6}\,
\msunyr$ was required. This value is in good agreement with the
findings of \cite{crowther02} who studied this objects using the model
atmosphere code \cmfgen\ \citep{hillier98} in the optical, UV and
far-UV. Compared to these authors we also find that the other fit
parameters are in good agreement. An exception to this is the helium
abundance. \cite{crowther02} adopted a fixed value of $\yhe = 0.2$
whereas our automated fitting method was able to self consistently
determine a value of $\yhe = 0.28$.

\paragraph{Sk $-$70 69} (Fig.~\ref{fig:fits_8})
To fit the spectrum of this O5 dwarf Sk~$-$70~69 a helium abundance of
$\yhe = 0.17$ was required. Note that, similar to N11-065, which is
also strongly enriched with helium, the spectroscopic mass is found to
be much smaller than the evolutionary mass.

\end{document}